\documentclass[11pt]{iopart}
\usepackage[english]{babel}
\pdfoutput=1
\usepackage{wasysym}
\usepackage{booktabs}
\usepackage{amssymb}
\usepackage{pifont}
\usepackage{amsbsy}
\usepackage{verbatim}
\usepackage{graphicx}
\usepackage{epstopdf}
\usepackage{color}
\usepackage{sidecap}
\usepackage{bm}% bold math
\usepackage[colorlinks,bookmarks=false,citecolor=blue,linkcolor=red,urlcolor=blue]{hyperref}
\usepackage{cite}

\catcode`@=11 % If we need private TeX macros
\renewcommand\tableofcontents{%
  \section*{\contentsname}%
  \@starttoc{toc}%
}
\catcode`@=12% '@' is no more a character

\def\be{\begin{equation}}
\def\ee{\end{equation}}

\def\nm{\newmoon}
\def\fm{\fullmoon}
\def\T{\rule{0pt}{.5cm}}
\def\B{\rule[-.3cm]{0pt}{0pt}}

\begin{document}

\setlength{\parindent}{0pt}

\title{Bound states and entanglement in the excited states of 
quantum spin chains}

\author{Jan M\"olter$^1$, Thomas Barthel$^{2,1}$, Ulrich Schollw\"ock$^1$, 
and Vincenzo Alba$^1$}

\address{$^1$ Department of Physics and Arnold Sommerfeld
Center for Theoretical Physics, Ludwig-Maximilians-Universit\"at
M\"unchen, D-80333 M\"unchen, Germany}

\address{$^2$ Laboratoire de Physique Th\'{e}orique et Mod\`{e}les 
statistiques, Universit\'{e} Paris-Sud and CNRS, UMR 8626, 91405 Orsay, 
France}

\date{\today}

%%%%%%%%%%%%%%%%%%%%%%%%%%%%%%%%%%%%%%%%%%%%%%%%%%%%%%%%%%%%%%%%
\begin{abstract} 

We investigate entanglement properties of the excited states of 
the spin-$\frac{1}{2}$ Heisenberg ($XXX$) chain with isotropic antiferromagnetic 
interactions, by exploiting the Bethe 
ansatz solution of the model. We consider eigenstates obtained from 
both real and complex solutions (``strings'') of the Bethe equations. 
Physically, the former are states of interacting magnons, whereas 
the latter contain bound states of groups of particles. We first focus on the 
situation with few particles in the chain. Using exact 
results and semiclassical arguments, we derive an upper bound $S_{MAX}$ 
for the entanglement entropy. This exhibits an intermediate behavior between 
logarithmic and extensive, and it is saturated for highly-entangled states. 
As a function of the eigenstate energy, the entanglement entropy is organized 
in bands. Their number depends on the number of blocks of contiguous Bethe-Takahashi 
quantum numbers. In presence of bound states a significant reduction in the 
entanglement entropy occurs, reflecting that a group of bound particles behaves 
effectively as a single particle. Interestingly, the associated entanglement 
spectrum shows edge-related levels. At finite particle density, the semiclassical 
bound $S_{MAX}$ becomes inaccurate. For highly-entangled states $S_A\propto L_c$, 
with $L_c$ the chord length, signaling the crossover to extensive entanglement. 
Finally, we consider eigenstates containing a single pair of bound particles. 
No significant entanglement reduction occurs, in contrast with the few-particle 
case. 
\end{abstract}

\maketitle

%%%%%%%%%%%%%%%%% INTRODUCTION %%%%%%%%%%%%%%%%%%%%%%%%%%%%%%%%%
\section{Introduction}
\label{intro}

In recent years the study of entanglement measures and other 
entanglement-related quantities have become of paramount importance 
in order to understand the behavior of low-dimensional quantum many-body 
systems~\cite{amico-2008,eisert-2009,calabrese-2009,cc-rev}. Considering a 
system $S$ in a pure state $|\Psi\rangle$, and its spatial bipartition 
into two blocks $A$ and $B$, as $S=A\cup B$, one can decompose 
$|\Psi\rangle$ as (Schmidt decomposition)  
\begin{equation}
|\Psi\rangle=\sum\limits_i e^{-\xi_i/2}|\psi_i^A\rangle\otimes 
|\psi_i^B\rangle,
\label{sch_dec}
\end{equation}
with $|\psi_i^A\rangle$ and $|\psi_i^B\rangle$ two orthonormal bases for 
the subsystems Hilbert spaces ${\mathcal H}_A$ and ${\mathcal H}_B$. 
Here the so-called entanglement spectrum (ES) levels $\{\xi_i\}$~\cite{kauke-1999,
li-2008,calabrese-2008,regnault-2009,nienhuis-2009,bray-ali-2010,fidkowski-2010,
lauchli-2010,thomale-2010,yao-2010,prodan-2010,pollmann-2010,kargarian-2010,
turner-2010,papic-2011,fidkowski-2011,deng-2011,dubail-2011,schliemann-2011,
hughes-2011,regnault-2011,alba-2011,qi-2012,lundgren-2012,poilblanc-2012,
dechiara-2012,alba-2012,lepori-2013,laeuchli-2013,chung-2013,lundgren-2013,
kolley-2013,chandran-2013,grover-2013,assaad-2013,petrescu-2014,luitz-2014,
luitz-2014a,luitz-2014b,lundgren-2014,schliemann-2014,udagawa-2014} 
are related to the eigenvalues $\{\lambda_i\}$ of the reduced density matrix 
for part $A$, $\rho_A\equiv\textrm{Tr}_B(|\Psi\rangle\langle\Psi|)$, as 
$\xi_i\equiv-\log(\lambda_i)$. A proper measure of the quantum entanglement 
between $A$ and $B$ is the so-called von Neumann entropy (entanglement entropy) 
$S_A$, which is defined as 
\begin{eqnarray}
\label{vn_ent}
S_A\equiv-\sum_i\lambda_i\log(\lambda_i).
\end{eqnarray}
It is now well established that the {\it ground-state} entanglement entropy 
contains universal information about one-dimensional quantum many-body 
systems~\cite{amico-2008,eisert-2009,calabrese-2009}. A spectacular example 
is provided by critical systems described by a conformal field theory (CFT), 
for which it has been proven that $S_A$ obeys the scaling behavior~\cite{Holzhey,
Vidal,cc-04}  
\begin{eqnarray}
\label{vn_cft}
S_A=\frac{c}{3}\log\Big[\frac{L}{\pi}\sin\Big(\frac{\pi L_A}{L}\Big)\Big]+c'_1. 
\end{eqnarray}
Here $c$ is the central charge of the conformal field theory, $c_1'$ a 
non-universal constant, while $L$ and $L_A$ are the sizes of the full system 
and of part $A$, respectively. Universal information can also be extracted from 
the entanglement between many disjoint blocks~\cite{fps-08,cg-08,cct-09,ip-09,fc-10,cct-11,
ch-04,ffip-08,rt-06,atc-09,atc-11,f-12,coser-2013}, or from the scaling corrections 
of the von Neumann entropy~\cite{ccen-10,ce-10,cc-10,ccp-10,fc-10b,xa-11,cmv-11}.  

In contrast, although many results are available in the literature~\cite{alba-2009,
alcaraz-2011,pizorn-2012,berganza-2012,wong-2013,storms-2013,berkovits-2013,essler-2013,
nozaki-2014,ramirez-2014,ares-2014,huang-2014,palmai-2014}, much less is known about 
entanglement properties of  excited states. Remarkably, for eigenstates obtained from 
primary fields of a CFT, which correspond to low-energy excitations, a scaling 
law similar to~\eref{vn_cft} has been obtained in Ref.~\cite{alcaraz-2011,berganza-2012}. 
Some results are available for exactly-solvable and free models. A complete classification 
of entanglement behaviors in the excited states of the $XX$ chain~\cite{lieb-1961,
barouch-1970,barouch-1971,barouch-1971a,mccoy-1971} 
is provided in Ref.~\cite{alba-2009} (see also~\cite{ares-2014}). Two main classes of 
eigenstates are present, displaying logarithmic (similar to~\eref{vn_cft}, with an effective 
``central charge''), or the extensive behavior $S_A\propto L_A$, respectively. The 
two scenarios are related to the local properties of the eigenstates, namely the 
distribution of occupied momenta in the Fermi sea that is obtained after mapping the $XX$ 
chain onto a free-fermion model. Interestingly, a logarithmic entanglement scaling is 
also observed in some high-energy eigenstates of the $XXZ$ chain. This has been 
shown in Ref.~\cite{alba-2009} using a numerical method based on the algebraic Bethe ansatz 
solution of the model~\cite{kor-book}. Yet, due to technical reasons, Ref.~\cite{alba-2009} 
focuses only on eigenstates obtained from {\it real} solutions of the so-called Bethe 
equations, and for small anisotropy (i.e., near the non-interacting $XX$ chain limit).  

Although it is interesting {\it per se}, the study of entanglement in excited 
states arises naturally in the context of out-of-equilibrium dynamics after a quantum 
quench. Formally, at any time after the quench the state of the system can be obtained 
as a time-dependent superposition of the eigenstates of the post-quench Hamiltonian. 
Understanding entanglement properties in individual eigenstates could be helpful  
in devising an effective truncation scheme to study real-time dynamics, especially 
in integrable models~\cite{barmettler-2010,gritsev-2010}. On the numerical side, this 
could give insights on the simulability of quench problems using DMRG (Density Matrix 
Renormalization Group) or other MPS (Matrix Product States) based algorithms~\cite{vidal-2004,daley-2004,
feiguin-2004,schollwoeck-2005,schollwoeck-2011}. Similarly, as excited states are 
necessary ingredients in constructing the Gibbs ensemble, a detailed knowledge of 
their entanglement properties could be useful for studying finite-temperature systems. 
\paragraph{Outline of the results.}
In this paper we investigate the entanglement entropy of the eigenstates of the 
spin-$\frac{1}{2}$ isotropic Heisenberg ($XXX$) chain. This is defined by the 
Hamiltonian 
\begin{equation}
{\mathcal H}=J\sum_{i=1}^{L}\frac{1}{2}(S_i^{+}S_{i+1}^{-}+S_{i}^-
S_{i+1}^+)+S_{i}^zS_{i+1}^z,  
\label{xxx_ham}
\end{equation}
where $S_i^{+,-,z}$ are the raising and lowering spin operators $S_i^{\pm}
\equiv(\sigma_i^x\pm i\sigma_i^y)/2,S_i^z\equiv\sigma_i^z/2$, and  
$\sigma_i^{x,y,z}$ the Pauli matrices. We set for convenience $J=1$ 
in~\eref{xxx_ham}. Periodic boundary conditions are used, i.e., sites $L+1$ 
and $1$ are identified. The $XXX$ chain is exactly solvable, and  
its eigenstates and eigenenergies can be constructed via the Bethe 
ansatz~\cite{bethe-1931,taka-book,kor-book}. The Hilbert space of~\eref{xxx_ham} 
is spanned by $2^L$ basis states, which can be generated starting from the 
ferromagnetic state $\left|\Omega\right.\rangle\equiv\left|\uparrow\uparrow
\uparrow\cdots\uparrow\right.\rangle$, flipping $M$ of the spins, with 
$M\in[0,L]$. Since it is conventional in the Bethe ansatz literature, we refer 
to flipped spins as ``particles''. The total magnetization $S^z_{T}\equiv
\sum_{i=1}^L S_i^z=L/2-M$ is a conserved quantity for~\eref{xxx_ham}, implying that
the eigenstates of the $XXX$ chain can be labeled by $S^z_T$ (equivalently by $M$).

In contrast with Ref.~\cite{alba-2009}, here we consider eigenstates of the $XXX$ 
chain obtained from both real and {\it complex} (``strings'') solutions of the Bethe 
equations (rapidities). Physically, while the former correspond to interacting 
magnons, complex rapidities signal the presence of many-particle bound states. Notice that the 
intriguing physics of bound states has received constant attention, both 
theoretically~\cite{hanus-1963,wortis-1963,fogedby-1980,schneider-1981,kohno-2009,
ganahl-2012} and experimentally~\cite{haller-2009,fukuhara-2013}. 
Finally, while Ref.~\cite{alba-2009} focuses mostly on eigenstates of the $XXZ$ 
chain exhibiting logarithmic entanglement entropy, here we characterize the crossover 
to extensive entanglement, which is expected in highly-entangled eigenstates 
of~\eref{xxx_ham}. 

We first focus on the situation with few particles in the chain, restricting ourselves 
to $M=2$ and $M=3$, i.e., high-magnetization sectors of~\eref{xxx_ham}. For any  eigenstate 
in the two-particle sector ($M=2$), its entanglement entropy and entanglement spectrum 
are constructed analytically exploiting the Bethe ansatz solution of~\eref{xxx_ham}. This, 
in particular, allows us to derive an upper bound $S_{MAX}$ for the entanglement entropy. 
Its generalization to arbitrary $M$ is obtained using semiclassical arguments as 
\begin{eqnarray}
\label{ub_intro}
S_{MAX}(M)=-M(\omega\log(\omega)+(1-\omega)\log(1-\omega)),   
\end{eqnarray}
where $\omega$ is the ratio $\omega\equiv L_A/L$. Clearly, for $M=1$, $S_{MAX}$ 
is the entanglement entropy of a free particle. For $M=2$ the bound~\eref{ub_intro}  has 
been also discussed in Ref.~\cite{berkovits-2013}. The linear dependence $\propto M$ 
in~\eref{ub_intro} reflects that in the semiclassical approximation the interactions 
(scattering) between particles can be neglected, as expected. Interestingly, for 
$L_A/L\ll 1$ (small blocks) one has $S_{MAX}\approx -(ML_A/L)\log(L_A/L)$, which is an 
``intermediate'' behavior between a logarithmic and an extensive growth. Notice that~\eref{ub_intro} 
is similar to the entanglement entropy of a ferromagnet~\cite{popkov-2005,popkov-2010,castro-alvaredo-2011,castro-alvaredo-2012}. 
In the sector with $M=2$ and for real rapidities we find that $S_{MAX}(M)$ is 
saturated for any $L_A$, meaning that $S_A\approx S_{MAX}$ (apart from ${\mathcal 
O}(1/L)$ terms) for some highly-entangled eigenstates of~\eref{xxx_ham}.

We also investigate the behavior of the half-chain entanglement entropy as a function of 
the eigenstate energy. For $M=2$ and real rapidities, a two-band structure appears, 
with eigenstates in the same band showing similar entanglement. In the higher band  
we have $S_A\approx S_{MAX}$. This band structure is understood in terms of the two Bethe-Takahashi 
quantum numbers, which in the Bethe ansatz formalism are used to identify the eigenstates 
of~\eref{xxx_ham}~\cite{taka-book}. Specifically, the lower and upper band correspond to 
the two quantum numbers being next to each other or far apart, respectively.

For states with real rapidities only, these features are generic in the vanishing density 
regime $M/L\to 0$ at fixed $M$: In practice, we numerically observe that the bound~\eref{ub_intro} is 
almost saturated. Moreover, the entanglement entropy is organized in 
bands. Higher entanglement bands correspond to increasing number of blocks of contiguous 
Bethe-Takahashi quantum numbers. For a given $M$, the number of bands is hence obtained 
as the number of integer partitions $p(M)$ of $M$. Physically, a small (large) number of 
blocks of Bethe-Takahashi numbers reflects the particles having small (large) relative 
quasi-momentum. All these findings are numerically confirmed by studying the full set of 
eigenstates of~\eref{xxx_ham} (for finite $L$) in the sector with $M=3$. 

The scenario is strikingly different in presence of bound states. For bound states 
of two particles ($2$-string solutions of the Bethe equations) a substantial 
decrease in the entanglement entropy is observed, as compared 
with eigenstates obtained from two real rapidities. This reflects a reduction in 
the number of degrees of freedom, since the bound state can be treated as an 
effective single particle. Nevertheless, we observe that $S_A>S_{MAX}(M=1)$,    
signaling that the bound states are extended objects. This is quantitatively understood in 
terms of the bound-state  extension $\ell_b$, which is obtained analytically from the 
Bethe ansatz solution of~\eref{xxx_ham}, and is an increasing function of the bound-state 
energy. As a consequence, while high-energetic bound states with $\ell_b/L\approx 1$ behave 
almost as two unbound particles, low-energy ones with $\ell_b/L\ll 1$ can be treated as 
a single particle. Clearly, larger values of $\ell_b\lesssim L$ are associated with higher  
entanglement. Interestingly, since in the large $L$ limit $\ell_b$ changes continuously as 
a function of the bound-state energy, no bands are observed in the entanglement entropy, 
in contrast with the case with two real rapidities. 

The presence of bound states has striking effects also at the level of the half-chain 
entanglement spectrum (ES). Specifically, we show that the ES exhibits edge-related levels. 
The corresponding entanglement eigenfunctions (i.e., the eigenvectors of the reduced density 
matrix) are exponentially localized at the boundary between the two subsystems (edge-states). 
Similar features, i.e., reduction of the entanglement entropy, edge-related ES levels, and 
absence of entanglement bands, are observed at fixed $M>2$, in the limit $L\to\infty$. 

This scenario breaks down at finite particle density $\rho\equiv M/L$. 
A striking change is that the upper bound~\eref{ub_intro} is no longer saturated. However, 
we numerically observe that in highly-entangled eigenstates $S_A\propto L_c$, where 
$L_c=(L/\pi)\sin(\pi L_A/L)$ is the so-called chord length. Since $L_c\approx L_A$ for 
$L_A\ll L$, this signals the crossover from~\eref{ub_intro} to the volume law $S_A\propto L_A$. 
We also investigate how the entanglement entropy depends on the Bethe-Takahashi quantum numbers. 
Similarly to the situation with few particles, we find that a 
large number of Bethe number blocks tends to correspond to highly-entangled eigenstates of the $XXX$ chain. 
Finally, we focus on the eigenstates with $\rho\approx 1/2$ containing a single two-particle 
bound state. We show that no significant entanglement reduction occurs (at least for finite 
chains), in contrast with the few-particle case.

%%%%%%%%%%%%%%%%%%%%%%% BETHE ANSATZ APPROACH %%%%%%%%%%%%%%%%%%%%%%%%%%%%%%
\section{Survey of Bethe ansatz results for the $XXX$ chain}
\label{ba_approach}

We start reviewing some aspects of the Bethe ansatz solution of~\eref{xxx_ham} 
(see~\cite{bethe-1931,kor-book,taka-book,karbach-1997} for more details) that 
are relevant for this work. In section~\ref{bethe_equations} the generic form of 
the eigenstates of~\eref{xxx_ham} and the so-called Bethe equations are introduced. 
Their solutions, both real ($1$-strings) and complex ($n$-strings, with $n\in{
\mathbb N}$, $n>1$), are discussed in~\ref{real_ex} and~\ref{string_ex}, within the 
framework of the string hypothesis~\cite{taka-book}. In particular, we introduce 
the so-called Bethe-Takahashi equations. Since it is important in order to 
understand entanglement properties of the $XXX$ chain, we detail the structure 
of the Bethe-Takahashi quantum numbers, which identify the solutions of the 
Bethe-Takahashi equations, and the eigenstates of~\eref{xxx_ham} thereof. 
Finally, focusing on complex solutions of the Bethe-Takahashi equations, we 
define the bound state length $\ell_b$.

%%%%%%%%%%%%%%%%%%%%%%%%%%%%%%%%%%%%%%%%%%%%%%%%%%%%%%%%%%%%%%%%%%%%%%%%%%%
\subsection{Bethe equations and wavefunctions}
\label{bethe_equations}

The generic eigenstate of~\eref{xxx_ham} in the sector with $M$ particles 
can be written as 
\begin{equation}
\label{ba_eig}
|\Psi_M\rangle=\sum\limits_{1\le x_1<x_2<\dots<x_M\le L}A_M(x_1,x_2,
\dots,x_M)|x_1,x_2,\dots,x_M\rangle,
\end{equation}
where the sum is over the positions $\{x_i\}$ of the particles, and $A_M(x_1,
x_2,\dots,x_M)$ is the eigenstate amplitude corresponding to particles 
at positions $x_1,x_2,\dots, x_M$. $A_M(x_1,x_2,\dots, x_M)$ is given as 
\begin{equation}
\label{ba_amp}
A_M(x_1,x_2,\dots,x_M)\equiv\sum\limits_{{\mathcal P}\in S_M}\exp\Big[i
\sum\limits_{j=1}^Mk_{{\mathcal P}_j}x_j+i\sum\limits_{i<j}\theta_{{
\mathcal P}_i{\mathcal P}_j}\Big].
\end{equation}
Here the outermost sum is over the permutations $S_M$ 
of the so-called quasi-momenta $\{k_1,k_2,\dots,k_M\}$. The two-particle 
scattering phases $\theta_{m,n}$ are defined as 
\begin{equation}
\label{s_phases}
\theta_{m,n}\equiv \frac{1}{2i}\log\Big[-\frac{e^{ik_m+ik_n}-2e^{ik_m}+1}
{e^{ik_m+ik_n}-2e^{ik_n}+1}\Big].
\end{equation}
The energy associated to the eigenstate~\eref{ba_eig} is  
\begin{equation}
\label{ba_ener}
E=\sum\limits_{\alpha=1}^M(\cos(k_\alpha)-1).
\end{equation}
The quasi-momenta $\{k_\alpha\}$ are obtained by solving the so-called 
Bethe equations  
\begin{equation}
\label{ba_eq}
e^{ik_\alpha L}=\prod\limits^M_{\beta\ne\alpha}\Big[-\frac{1-2e^{
ik_\alpha}-e^{ik_\alpha+ik_\beta}}{1-2e^{ik_\beta}-e^{ik_\alpha+
ik_\beta}}\Big].
\end{equation}
It is useful to  introduce the rapidities $\{\lambda_\alpha\}$ as 
\begin{equation}
\label{rap}
k_\alpha=\pi-2\arctan(\lambda_\alpha)\quad\mbox{mod}\, 2\pi.
\end{equation}
Taking the logarithm on both sides in~\eref{ba_eq}, and using~\eref{rap}, 
one obtains the Bethe equations in logarithmic form as 
\begin{equation}
\label{ba_eq_log}
\arctan(\lambda_\alpha)=\frac{\pi}{L}J_\alpha+\frac{1}{L}\sum\limits_{
\beta\ne\alpha}\arctan\Big(\frac{\lambda_\alpha-\lambda_\beta}{2}\Big),
\end{equation}
where $-L/2<J_\alpha\le L/2$ are the so-called Bethe quantum numbers. 
Notice that $J_\alpha$ are half-integers and integers for $L-M$ 
even and odd, respectively~\cite{taka-book}. Finally, the total 
eigenstate quasi-momentum $K_T\equiv\sum_\alpha k_\alpha$ is obtained 
from~\eref{ba_eq_log} and~\eref{rap} as 
\begin{eqnarray}
\label{ktot}
K_T=\pi M+\frac{2\pi}{L}\sum_\alpha J_\alpha
\end{eqnarray}

%%%%%%%%%%%%%%%%%%%%%%%%%%%%%%%%%%%%%%%%%%%%%%%%%%%%%%%%%%%%%%%%%%%%%%
\subsection{Real roots of the Bethe equations ($1$-strings)}
\label{real_ex}

In principle any choice of $M$ distinct Bethe quantum numbers $\{J_1,J_2,
\dots,J_M\}$ identifies a set of solutions $\{k_1,k_2,\dots,k_M\}$ 
of~\eref{ba_eq_log} and, using~\eref{ba_eig}, an eigenstate of~\eref{xxx_ham}. 
However, for {\it real} solutions of~\eref{ba_eq_log}, i.e. $\lambda_\alpha
\in{\mathbb R}\,\forall\alpha$, a precise bound  for the Bethe numbers 
can be obtained as~\cite{taka-book}
\begin{equation}
\label{bn_bound}
-J_\infty\le J_\alpha\le J_\infty\quad\mbox{with}\quad J_\infty\equiv
\frac{L-1-M}{2}, 
\end{equation}
where $J_\infty$ is formally the Bethe quantum number associated to an infinite 
rapidity~\cite{taka-book}. It is convenient to introduce the notation $\nm$ 
and $\fm$ for occupied and vacant (``holes'') Bethe quantum numbers, 
respectively. A generic eigenstate of~\eref{xxx_ham}, in the sector with $M$ 
particles, can then be identified by its Bethe quantum numbers as  
\begin{equation}
\label{eig_not}
[\fm^{m_1}\nm^{n_1}\fm^{m_2}\nm^{n_2}\cdots]_1\quad\mbox{with}\quad\nm^m
\equiv\underbrace{\nm\nm\dots\nm}_{m}, 
\end{equation}
where $\sum_{i}n_i=M$ and $\sum_{i}m_i=2J_\infty+1-M$. The brackets 
$[\cdots]_1$ mean that the Bethe quantum numbers are defined  in the interval 
$[-J_\infty,J_\infty]$, whereas the subscript is to stress that real solutions 
of~\eref{ba_eq_log} ($1$-strings, see section~\ref{string_ex}) are considered.  
The total number of real rapidities, according to~\eref{bn_bound}, is 
$(L-M)!/(M!(L-2M)!)$. 

Remarkably, the lowest-energy eigenstate $|\Psi_M^{(0)}\rangle$ of~\eref{xxx_ham} 
in the sector with $M$ particles is given in terms of real rapidities~\cite{yang-1966},  
which are obtained by choosing the Bethe quantum numbers  
\begin{equation}
J_\alpha^{(0)}(M)=-\frac{M+1}{2}+\alpha\qquad\mbox{with}\quad \alpha=1,2,
\dots, M,
\end{equation}
or, in the notation~\eref{eig_not}, 
\begin{equation}
\label{gs_M_bn}
J^{(0)}_\alpha(M)=[\fm^{(L-M)/2}\nm^M\fm^{(L-M)/2}]_1. 
\end{equation}
Notice that~\eref{bn_bound} implies that in the sector with $M=L/2$ there 
is only one set of real roots of the Bethe equations, which corresponds to 
the global ground state $|\Psi^{(0)}\rangle$ of the $XXX$ chain. 

It is natural to interpret the block ``$\nm^M$'' of occupied Bethe numbers 
in~\eref{gs_M_bn} as a ``Fermi sea''. All the excited states above $|\Psi^{(0)}_M
\rangle$, which have the same number of particles, and correspond to real rapidities, 
are obtained by successive exchanges of occupied and empty Bethe quantum numbers 
(``particle-hole'' processes). An example of such processes is shown in 
Table~\ref{table} (first row, p-hole). 

It is useful to classify 
the eigenstates of the $XXX$ chain according to the number of blocks of contiguous 
Bethe quantum numbers. While the ground state $|\Psi_M^{(0)}\rangle$ (cf.~\eref{gs_M_bn}) 
corresponds to a single block, the number of blocks generically increases under p-hole 
processes (for instance from one to three in the process shown in Table~\ref{table}), 
although this is not true in general. A notable exception is provided by sequences 
of p-hole processes that amount to a shift of one or many blocks of quantum  numbers. 
For instance, the process   
\begin{eqnarray}
[\cdots\fm
\nm^{M'}\fm^{h}
\nm^{M''}\fm\cdots]_1\quad\leadsto\quad &  
[\cdots\fm\nm^{M'}\fm^{h'}
\nm^{M''}\fm\cdots]_1,
\end{eqnarray}
where the dots stand for $\fm\cdots\fm$, $M'+M''=M$, and $h\ne h'$ are non-zero 
integers. 
The number of blocks in the Bethe quantum numbers is strikingly reflected in the 
entanglement properties of the excited states of the $XXX$ chain. We 
anticipate here that eigenstates obtained from Bethe quantum number configurations 
with the same number of blocks exhibit similar entanglement entropy (see 
section~\ref{low_density_M2} and~\ref{low_density_M3}). 

A different class of excited states of the Heisenberg chain  
is obtained from roots of~\eref{ba_eq} containing pairs of complex conjugate 
rapidities (strings), which, physically, correspond to groups of bound particles. 
An example of process leading to an eigenstate with a two-particle bound state 
is shown in Table~\ref{table} (see next section for the structure 
of the corresponding Bethe quantum numbers).
\begin{table}[tp]
\begin{center}
\caption{ Excitations processes and Bethe-Takahashi quantum numbers for some 
 eigenstates of the $XXX$ spin chain in the sector with $M$ particles. Only 
 processes preserving the total number of particles are considered. Here $\nm$ 
 ($\fm$) denotes an occupied (vacant) quantum number, $[\cdots ]_\alpha$ is 
 the quantum number configuration in the $\alpha$-string sector, and $s_1\equiv 
 (L-M)/2$, $s_2\equiv (L-2M)/2$, with $L$ the length of the chain. Notice that 
 $1$-strings correspond to real rapidities. $[\fm^{s_1}\nm^M\fm^{s_1}]_1$ 
 corresponds to the lowest-energy state in the sector with $M$ particles. 
 (First row) p-hole excitation: one occupied and empty numbers are exchanged. 
 An example of p-hole process is shown in the second column. The number of 
 blocks of contiguous Bethe quantum numbers increases from one to three. All $1$-strings 
 excited states of the $XXX$ chain can be obtained by successive applications 
 of p-hole. (Second row) Creation of a two-particle bound state ($2-$string). 
 Excited states involving $n$-strings with $n>2$, or more than a single 
 bound states are not shown. 
} 
\label{table}
\vspace{.2cm}
\begin{tabular}{ c c c}
\toprule
\T\B\quad{\bf Type of excitation} & & {\bf Bethe-Takahashi numbers}\\
\hline 
\T\B{\bf p-hole} & $[\fm^{s_1}
\nm^{M}\fm^{s_1}]_1\quad\leadsto\quad$ &$[
\fm^{s_1}\nm\fm
\nm^{M-2}\fm^{s_1-2}\nm
\fm]_1$\B\B\\\\
\T\B
& & $[\fm^{s_1}\nm^{M/2-1}{\color{black}\fm}^2
\nm^{M/2-1}
\fm^{s_1}]_1$\B\B\\
\T\B
{\bf $2$-string} & $[\fm^{s_1}
\nm^{M}\fm^{s_1}]_1
\quad\leadsto\quad$ &$\times$\\
& & $[\fm^{s_2}\nm
\fm^{s_2}]_2$\B\B\\
\bottomrule
\end{tabular}
\end{center}
\end{table} 
%#####################################################################
%

%%%%%%%%%%%%%%%%%%%%%%% STRING SOLUTIONS %%%%%%%%%%%%%%%%%%%%%%%%%%%%%%%%
\subsection{String solutions of the Bethe equations}
\label{string_ex}

Complex solutions of the Bethe equations~\eref{ba_eq} form particular ``string'' 
patterns in the complex plane, at least in the limit of large chains $L\to\infty$
(string hypothesis)~\cite{bethe-1931,taka-book}. Specifically, rapidities 
forming a ``string'' of length $1\le n\le M$ (that we defined here as $n$-string) 
are parametrized as 
\begin{equation}
\label{str_hyp}
\lambda_\gamma^{(n,j)}=\lambda_\gamma^{(n)}-i(n-1-2j),\qquad j=0,1,\dots, n-1, 
\end{equation}
where $\lambda_\gamma^{(n)}$ is the real part of the string (string center), 
and $\gamma$ labels strings with different centers. In this language real 
rapidities are strings of unit length ($1$-strings), i.e., with $n=1$. 
It is natural to classify solutions $\Lambda\equiv\{\lambda_\alpha\}$ ($\alpha
\in[1,M]$) of the Bethe equations~\eref{ba_eq} according to their string content 
$\{N_\alpha\}$, where $N_\alpha$ denotes the number of $\alpha$-strings in 
$\Lambda$. 

\paragraph{Bethe-Takahashi equations.} The string centers $\lambda_\gamma^{(n)}$ 
in~\eref{str_hyp} are obtained by solving the so-called Bethe-Takahashi 
equations~\cite{taka-book}
\begin{equation}
\label{bt_eq}
2L\arctan(\lambda_\gamma^{(n)}/n)=2\pi I_{\gamma}^{(n)}+\sum\limits_{(m,
\beta)\ne(n,\gamma)}\Theta_{m,n}(\lambda_\gamma^{(n)}-\lambda_\beta^{(m)}), 
\end{equation}
where the generalized scattering phases $\Theta_{m,n}$ read 
\begin{eqnarray}
\nonumber\fl\Theta_{m,n}(x)\equiv\left\{\begin{array}{cc}
\vartheta\big(\frac{x}{|n-m|}\big)+\!\!\!\!\!\sum
\limits_{r=1}^{(n+m-|n-m|-1)/2}\!\!\!\!\!2\vartheta\big(\frac{x}
{|n-m|+2r}\big)+\vartheta\big(\frac{x}{n+m}\big) & \quad\mbox{if}
\quad n\ne m\\\fl\sum\limits_{r=1}^{n-1}2\vartheta\big(\frac{x}{2r}\big)+
\vartheta\big(\frac{x}{2n}\big) & \quad\mbox{if}\quad n=m
\end{array}\right.
\end{eqnarray}
and $\vartheta(x)\equiv 2\arctan(x)$. Here $I_{\gamma}^{(n)}$ are the 
Bethe-Takahashi quantum numbers associated with $\lambda_\gamma^{(n)}$. 
It can be shown that $I^{(n)}_{\gamma}$  are integers or half-integers for 
$L-N_n$ odd and even, respectively. Clearly, the constraint $\sum_{\alpha=1}^{M}
\alpha N_\alpha=M$ has to be satisfied. As for the Bethe quantum numbers 
$J_\alpha$ (see~\eref{bn_bound}), an upper bound for the Bethe-Takahashi 
quantum numbers can be derived as~\cite{taka-book}  
\begin{equation}
|I_\gamma^{(n)}|\le I^{(n)}_{MAX}\equiv\frac{1}{2}(L-1-\sum
\limits_{m=1}^Mt_{m,n}N_m),
\label{bt_qn_bound}
\end{equation}
where $t_{m,n}\equiv 2\mbox{min}(n,m)-\delta_{m,n}$. Notice that this implies 
$I^{(n)}_{MAX}\le I^{(n')}_{MAX}$ if $n<n'$. 

It is straightforward to generalize the notation~\eref{eig_not} to the 
case of the Bethe-Takahashi quantum numbers. First, for real solutions  
of the Bethe-Takahashi equations, i.e., $\lambda_\alpha\in {\mathbb 
R},\, \forall \alpha$, one has $\{J_\alpha\}=\{I^{(1)}_\alpha\}$. 
Moreover, given a generic solution of~\eref{bt_eq} with string content 
$\{N_\alpha\}$, the corresponding Bethe-Takahashi numbers can be represented 
as $[{\mathcal C}_1]_1\times[{\mathcal C}_2]_2\times\cdots\times[{\mathcal 
C}_M]_M$, with 
\begin{eqnarray}
\label{eig_not1}
[{\mathcal C}_\alpha]_\alpha\equiv [\fm^{m_1}\nm^{n_1}\fm^{m_2}\nm^{n_2}
\cdots]_\alpha  
\end{eqnarray}
the Bethe-Takahashi numbers corresponding to the $N_\alpha$ $\alpha$-strings. 
Notice that $\sum_i n_i=N_\alpha$ and $\sum_im_i=2I^{(\alpha)}_{MAX}+1-N_\alpha$. 

The quasi-momentum associated to a generic rapidity $\lambda'=\lambda+ip$ 
in a $n$-string, using~\eref{rap}, is  
\begin{equation}
\label{str_qm}
\fl k(\lambda+ip)=\left\{\begin{array}{cc}\pi+\frac{i}{2}\log\Big[
\frac{(1-p)^2+\lambda^2}{(1+p)^2+\lambda^2}\Big]-\arctan\Big[
\frac{\lambda}{1+p}\Big]-\arctan\Big[\frac{\lambda}{1-p}\Big] & 
\quad\mbox{if}\quad |p|>1\\\\\pi+\frac{i}{2}\log\Big[\frac{(1-p)^2+
\lambda^2}{(1+p)^2+\lambda^2}\Big]-\arctan\Big[\frac{\lambda}{1+
|p|}\Big]+\frac{\pi}{2}\textrm{Sign}(\lambda) & \quad\mbox{if}
\quad |p|=1,\end{array}\right. 
\end{equation}
where $\textrm{Sign}(x)$ is the sign function.
By summing over the different string components, it is 
straightforward to obtain the total quasi-momentum $k_s^{(n)}$ of the  
$n$-string as 
\begin{equation}
\label{str_mom}
k^{(n)}_s=\pi+2\arctan\Big(\frac{\lambda_\gamma^{(n)}}{n}\Big)\quad
\mbox{mod}\, 2\pi. 
\end{equation}
The corresponding string energy $E_s^{(n)}$, from~\eref{ba_ener} 
and~\eref{str_qm}, is a function of the string center 
$\lambda_\gamma^{(n)}$ as 
\begin{equation}
\label{str_ener}
E_s^{(n)}=-\frac{2n}{(\lambda^{(n)}_\gamma)^2+n^2}. 
\end{equation}
Interestingly, since $0<|\lambda|<\infty$ one has $E^{(n)}_s\in[-2/n,0]$, 
implying that ``longer'' strings have higher energy. 

It is useful to isolate the imaginary part of the quasi-momentum 
$k(\lambda+ip)$ in~\eref{str_qm}, defining $\ell_b^{(p)}$ as 
\begin{equation}
\label{b_length}
\ell_b^{(p)}\equiv\left[\frac{1}{2}\log\frac{(1+p)^2+
\lambda^2}{(1-p)^2+\lambda^2}\right]^{-1}.  
\end{equation}
Since $\ell_b^{(p)}$ contributes with terms of the form $\exp({-
|x_i-x_j|/\ell_b^{(p)}})$ in the amplitudes~\eref{ba_amp}, it is natural 
to interpret $\ell_b^{(p)}$ as the bound-state extension. Large $\ell_b^{(p)}$ 
corresponds to a weakly-bound cluster of particles, whereas $\ell_b^{(p)}\to 0$ 
signals a strongly-bound one. Finally, we define $\ell_b$ as the length 
associated to the outermost rapidity in a string, i.e., that with the largest 
imaginary part. This can be expressed in terms of the string total energy 
$E^{(n)}_s$ (cf.~\eref{str_ener}) as
\begin{equation}
\label{b_length_e}
\ell_b=\Big[-\frac{1}{2}\log\Big(1+\frac{2n-2}{n}E^{(n)}_s
\Big)\Big]^{-1}
\end{equation}
Since in this work we restrict ourselves to strings with $n=2,3$, we consider 
only $\ell_b$. 

We should stress that the string hypothesis~\eref{str_hyp} holds only in the limit 
$L\to\infty$, whereas for finite chains rapidities of the form~\eref{str_hyp} are 
not solutions of the Bethe equations~\eref{ba_eq_log}. To account for finite chain 
corrections, Equation~\eref{str_hyp} has to be modified as (deviated $n$-string)
\begin{eqnarray}
\label{str_hyp_dev}
\fl\quad 
\lambda_{\gamma}^{(n,j)}=\lambda_{\gamma}^{(n)}+\epsilon^{(n,j)}_{\gamma}+
i(n-1-2j)+i\delta^{(n,j)}_{\gamma}\quad\mbox{with}\quad j=0,1,\dots,n-1,   
\end{eqnarray}
where the string deviations $\epsilon_\gamma^{(n,j)},\delta_{\gamma}^{(n,j)}
\in{\mathbb R}$ are ${\mathcal O}(e^{-L})$, i.e., exponentially vanishing in 
the limit $L\to\infty$. 
One should remark that solutions of the Bethe equations deviating from the string 
picture~\eref{str_hyp_dev} are known. For instance, for two-particle, they have been 
carefully characterized in~\cite{essler-1992} (cf. also~\cite{hagemans-2007}). 
Similar behaviors can be observed for open boundary conditions~\cite{alba-2013}. 
Furthermore, detailed knowledge of the string deviations is important in order to 
extract physical quantities (for instance conformal data) about the Heisenberg 
chain~\cite{devega-1985,alcaraz-1987,devega-1987,alcaraz-1988,destri-1992}. Notice 
that solutions of the Bethe equations are self-conjugate~\cite{vladimirov-1986}, i.e., 
$\{\lambda_\alpha\}=\{\lambda_\alpha\}^*$,  which implies some constraints on the 
string deviations. 

It is interesting to calculate the scattering phases~\eref{s_phases} for deviated 
strings. Given $\lambda_1$ and $\lambda_2$ as $\lambda_1=\lambda+\epsilon+ip+i\delta$ 
and $\lambda_2=\lambda+\epsilon'+iq+i\delta'$, with $p,q\in{\mathbb Z}$ and 
$\epsilon,\epsilon',\delta,\delta'\in{\mathbb R}$, one has using~\eref{s_phases} 
and~\eref{str_hyp_dev}
\begin{eqnarray}
\label{str_phases}
\fl\theta(\lambda_1,\lambda_2)=
\frac{1}{4i}\log\Big[\frac{(p-q+2+\hat\delta)^2+\hat\epsilon^2}
{(p-q-2+\hat\delta)^2+\hat\epsilon^2}\Big]+\\\nonumber
\frac{1}{2}\arctan\Big[\frac{4\hat\epsilon}{(p-q+\hat\delta)^2-4
+\hat\epsilon^2}\Big]+\frac{\pi}{2}\,\textrm{Sign}(\hat\epsilon)
H(4-(p-q+\hat\delta)^2-\hat\epsilon^2), 
\end{eqnarray}
where $\hat\delta\equiv \delta-\delta'$, $\hat\epsilon\equiv\epsilon-
\epsilon'$, and $H(x)$ is the Heaviside step function,  Clearly, in the limit 
$L\to\infty$, equivalently $\hat\epsilon,\hat\delta\to 0$, we have 
$\theta(\lambda_1,\lambda_2)\to\pm i\infty$ for  $p-q=\mp 2$. 
These divergences are reflected in exponentially vanishing or 
diverging amplitudes in~\eref{ba_amp}. The treatment of these singularities 
in constructing the eigenstates of the Heisenberg chain can be tricky and 
it is detailed in~\ref{two_p_cluster},~\ref{three_p_cluster}, 
and~\ref{ba_eig_single_2string} for the string configurations considered 
in this work.

%%%%%%%%%%%%%%%%%%%%%%%%%%%%%%%%%%%%%%%%%%%%%%%%%%%%%%%%%%%%%%%%%%%%%%
\section{Few-particle entanglement: Eigenstates with $M=2$ particles}
\label{low_density_M2}

In this section, as an extreme case of the vanishing density limit (i.e., with 
fixed $M$ and $L\to\infty$), we discuss analytical results for the entanglement 
entropy and entanglement spectrum (ES) of the eigenstates of the $XXX$ chain 
in the sector with two particles $M=2$. We consider both $1$-string and 
$2$-string eigenstates (see sections~\ref{real_ex} and~\ref{string_ex}). 
The former are states of two unbound particles (magnons), whereas 
the latter correspond to bound states of the two particles. The analytical 
results obtained from the Bethe ansatz solution are discussed in~\ref{two_p_ent} 
and~\ref{two_p_cluster}. 

Based on the analytical solution, we present a semiclassical upper 
bound $S_{MAX}(M=2$ for the von Neumann entropy (see~\ref{two_p_ent} and 
\ref{upper_bound}). In the $1$-string sector this is 
saturated in the limit $L\gg 1$. On the other hand, for $2$-string eigenstates 
the entanglement entropy is dramatically reduced, and we find $S_{MAX}(M=1)<
S_A<S_{MAX}(M=2)$. 

We also examine the dependence of the entanglement entropy on the 
eigenstate energy, focusing on the half-chain entropy $S_A(L/2)$. 
In the $1$-string sector this  exhibits a two-band structure. 
In both bands the entropy is approximately constant as a function of energy,  
and in the upper one $S_A(L/2)\approx S_{MAX}(M=2,L/2)$. 
These bands can be understood in terms of the two Bethe-Takahashi quantum 
numbers used to identify the eigenstates of the $XXX$ chain (see 
section~\ref{ba_approach}). Precisely, the entanglement entropy increases 
with the number of blocks of contiguous Bethe quantum numbers: while the lower 
band correspond to two quantum numbers next to each other (i.e., a single block), the upper 
one is obtained for two quantum numbers far apart (two blocks). 
Equivalently, the first situation correspond to the two particles having 
similar quasi-momenta, while the latter is reflected in particles with 
very different ones. 

In the $2$-string sector the entanglement entropy changes continuously (in the 
large $L$ limit) with the eigenstate energy, and no entanglement bands are observed. 
This reflects the behavior of the bound-state length $\ell_b$ (cf.~\eref{b_length}), 
which is a continuous increasing function of the bound-state energy 
(cf.~\eref{b_length_e}). The entanglement entropy generically increases 
upon increasing $\ell_b/L$. For $\ell_b/L\to 0$ (i.e. two strongly-bound 
particles) one has $S_A(L/2)\approx S_{MAX}(M=1,L/2)$.

All these findings are reflected at the level of the half-chain entanglement 
spectrum. In the $1$-string sector the ES levels obtained from 
highly-entangled eigenstates (i.e., saturating the upper bound $S_{MAX}$) 
are simple functions of the ratio $\omega\equiv L_A/L$ that  
can be understood semiclassically (see~\ref{two_p_ent} for analytical results). 
Surprisingly, for $2$-string eigenstates, besides the usual bulk ES levels, 
edge-related ES levels appear, which diverge in the limit $L\gg 1$. The 
corresponding entanglement eigenfunctions are exponentially localized at 
the boundary between the two subsystems (edge-states).

%%%%%%%%%%%%%%%%%%%%%%%%%%%%%%%%%%%%%%%%%%%%%%%%%%%%%%%%%%%%%%%%%%%%%%%
\subsection{Overview: interacting magnons versus bound states}
\label{2p_overview}

%##################################################################
\begin{figure}[t]
\begin{center}
\includegraphics[width=.95\textwidth]{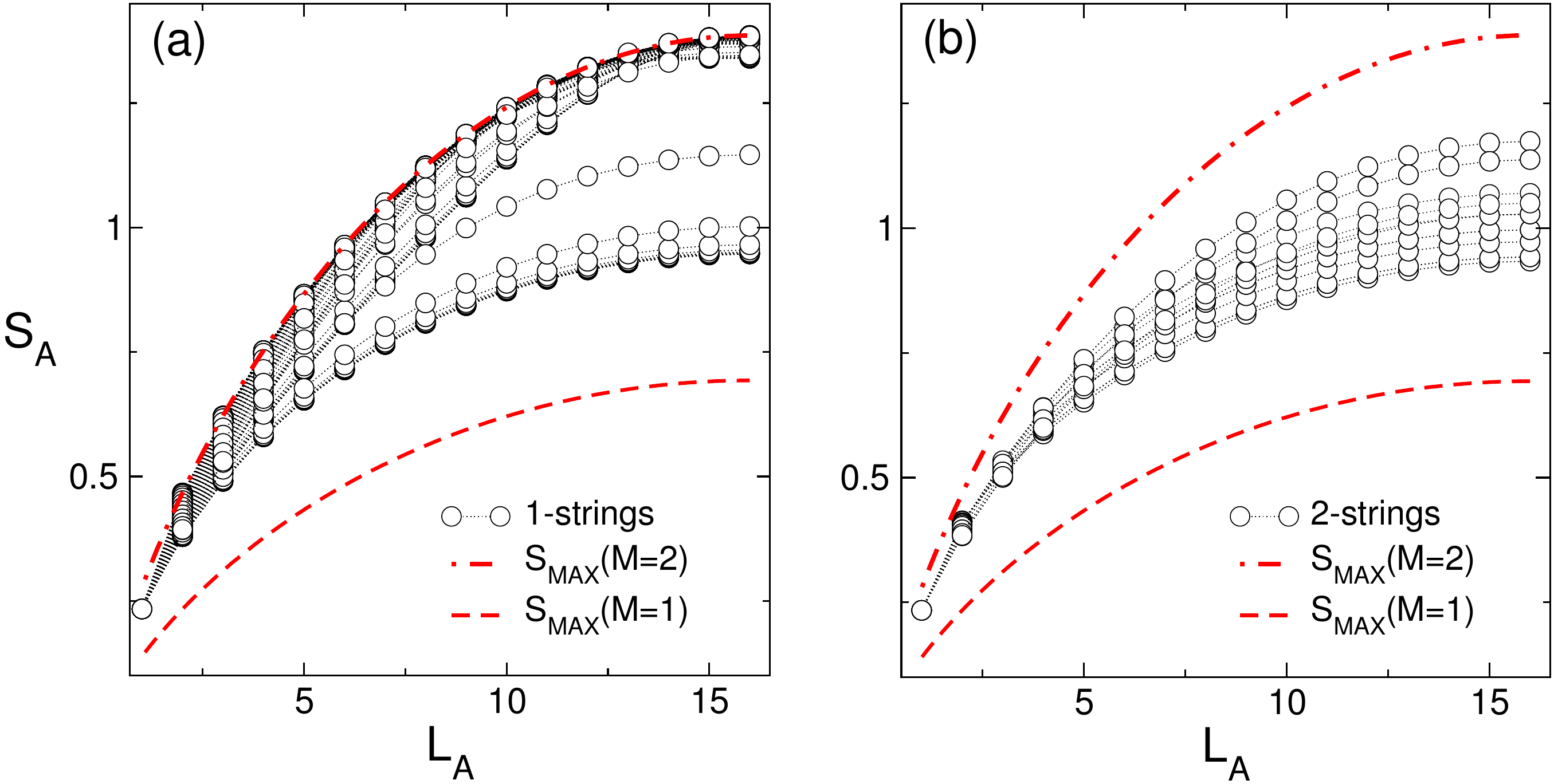}
\end{center}
\caption{ Entanglement of the eigenstates of the Heisenberg spin chain 
 in the sector with two particles ($M=2$): entanglement entropy $S_A$ 
 versus the subsystem size $L_A$ (dotted lines are guides to the eye). Data 
 are for a chain of length $L=32$. (a) Eigenstates corresponding to real 
 solutions of the Bethe equations ($1$-strings). The dashed and 
 dash-dotted lines the (lower and upper) bounds $S_{MAX}(M=1)$ and 
 $S_{MAX}(M=2)$, respectively, with $S_{MAX}(M)=-M(\omega\log(\omega)+
 (1-\omega)\log(1-\omega))$ and $\omega\equiv L_A/L$ (b) Eigenstates 
 corresponding to pairs of complex conjugate solutions of the 
 Bethe-Takahashi equations ($2$-strings): $S_A$ versus $L_A$ (same 
 scale as in (a) on the $y$-axis). Dashed and dash-dotted lines are 
 the same as in (a). Note the lower value of entanglement and 
 the dense structure as compared to the case of $1$-strings in (a). 
}
\label{fig1:ent_2p}
\end{figure}
%##################################################################

The entanglement entropy $S_A(L_A)$ obtained from the eigenstates of the 
Heisenberg chain in the sector with $M=2$ particles is plotted versus the 
block length $L_A$ in Figure~\ref{fig1:ent_2p}. Symbols are exact numerical 
data for $S_A$ obtained from the Bethe ansatz result~\eref{ba_eig}, for a 
chain with $L=32$. All the possible eigenstates are considered 
in the figure. Panels (a) and (b) show the entanglement entropy for 
$1$-string and $2$-string eigenstates, respectively. 

%##################################################################
\begin{figure}[t]
\begin{center}
\includegraphics[width=.8\textwidth]{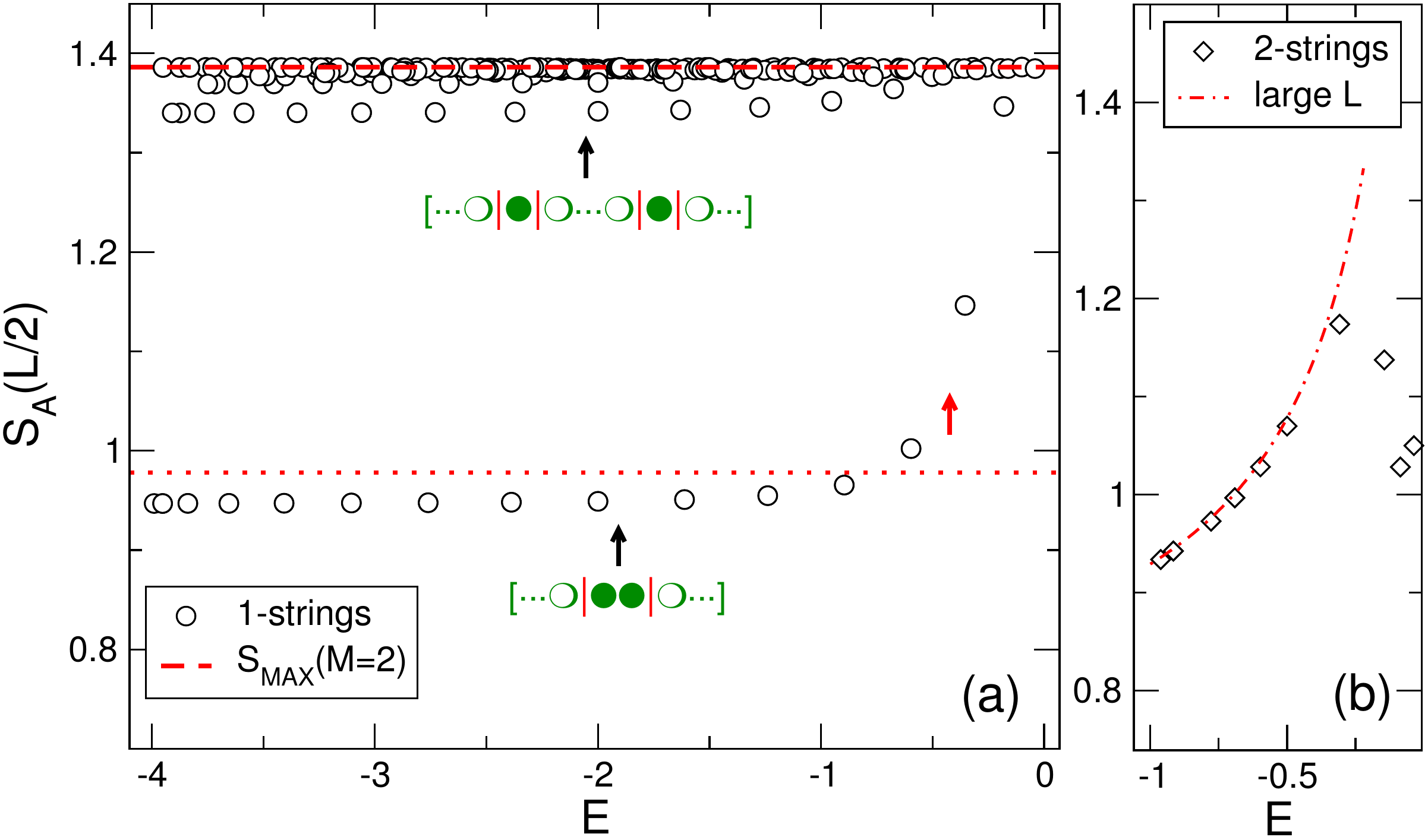}
\end{center}
\caption{ Eigenstates of the Heisenberg spin chain in the sector with 
 two particles ($M=2$): entanglement bands. Half-chain entanglement 
 entropy $S_A(L/2)$ versus eigenstate energy $E$. Data are Bethe ansatz 
 results for a chain with $L=32$. (a) $1$-string eigenstates. The dashed 
 line is $S_{MAX}(M=2)$ for $L_A=L/2$). Two bands with approximately 
 constant entropy (as a function of energy) are visible. The cartoons 
 show typical Bethe-Takahashi number configurations for the eigenstates 
 in the bands. Empty (full) circles denote vacant (occupied) quantum 
 numbers. The vertical lines mark disconnected blocks of occupied quantum 
 numbers. 
 The lower and the upper entanglement band corresponds to 
 configurations with two neighboring and two separated Bethe quantum numbers, 
 respectively. The vertical arrow is to stress the merging of the two bands 
 at high energy. (b) 
 $2-$string eigenstates. No band structure is visible. The 
 dash-dotted line is the analytical result in the large chain limit 
 $L\gg \ell_b$, with $\ell_b$ the size of the bound state. 
}
\label{fig2:energy}
\end{figure}
%##################################################################

We start discussing the $1$-strings. The semiclassical upper-bound~\eref{ub_intro} 
for the von Neumann entropy is reported in Figure~\ref{fig1:ent_2p} (a) for $M=1$ 
and $M=2$ as dashed and dash-dotted line, respectively. Clearly, we have 
$S_{MAX}(M=1)<S_A(L_A)\lesssim S_{MAX}(M=2)$, i.e., the upper bound is almost 
saturated. Deviations from $S_{MAX}(M=2)$ in the upper band are ${\mathcal O}(1/L)$. 

The entanglement entropy obtained from $2$-string eigenstates is shown 
in Figure~\ref{fig1:ent_2p} (b). Dashed and dash-dotted lines are the same 
as in panel (a). A dramatic reduction in the entanglement entropy is observed, 
as compared with the upper-band of the $1$-string eigenstates (panel (a)). Interestingly, 
we have $S_A(L_A)>S_{MAX}(M=1)$, contradicting the naive picture of the bound state 
as an effective single particle. This reflects the effective bound-state length 
$\ell_b$ (cf.~\eref{b_length}) being finite. Only for low-energy $2$-strings 
one has $\ell_b/L\to 0$ and $S_A(L_A)\approx S_{MAX}(M=1)$ in the limit $L\gg 1$. 

%%%%%%%%%%%%%%%%%%%%%%%%%%%%%%%%%%%%%%%%%%%%%%%%%%%%%%%%%%%%%%%%
\subsection{Entanglement versus energy: entanglement bands}

The behavior of the half-chain entanglement entropy $S_A(L/2)$ as a function 
of the eigenstate energy $E$ is illustrated in Figure~\ref{fig2:energy}. 
Data are the same as in Figure~\ref{fig1:ent_2p}. Panel (a) and (b) are for 
the $1$-string and the $2$-string eigenstates, respectively. Notice that 
while for the $1$-strings we have $-4<E<0$, $2$-strings correspond to the higher 
energies $-1<E<0$ (cf.~\eref{str_ener}). 

In the $1$-string sector $S_A(L/2)$ exhibits a two-band structure. The lower 
band with $S_A(L/2)\approx 1$ corresponds to eigenstates obtained from two 
Bethe-Takahashi numbers (see section~\ref{ba_approach}) $\{I^{(1)}_1,I^{(1)}_2\}$ 
next to each other (i.e. $\delta I\equiv |I^{(1)}_2
-I^{(1)}_1|=1$). This is illustrated by the cartoon in the Figure, where full (empty) 
circles denote occupied (vacant) Bethe quantum numbers (see 
section~\ref{string_ex}), whereas vertical lines mark the disconnected blocks of 
occupied quantum numbers. Since $\delta I=1$ implies 
$|k_2-k_1|\sim 2\pi/L$ (at least in the $L\gg 1$ limit, cf.~\ref{low_qm_limit}), the 
lower band corresponds to eigenstates with particles with similar quasi-momenta. In 
particular, the dotted line in the Figure is the analytical result (see~\ref{low_qm_limit}) 
assuming $|k_2-k_1|=2\pi/L$, and it is in agreement with the data,  apart from 
${\mathcal O}(1/L)$ corrections. 

The upper band corresponds to $|I^{(1)}_2-I^{(1)}_1|>1$, i.e., eigenstates 
obtained from two separated Bethe quantum numbers (as shown by the cartoon). The 
substructures (lines) within the band correspond to shifting two equi-distant 
(i.e., with $\delta I=const.$) quantum numbers. 

These substructure are generic for eigenstates with finite $M$ and $L\to\infty$. 
Precisely, we observe that a global shift of the Bethe-Takahashi numbers, which amounts 
to a shift in the eigenstate total quasi-momentum (cf.~\eref{ktot}), does not change 
significantly the entanglement entropy. This is better understood in the non-interacting 
limit (i.e., the $XX$ chain), where $k_\alpha\sim 2\pi J_\alpha/L$~\cite{biegel-2004}, 
implying that a global shift of the Bethe numbers corresponds to shifting 
the quasi-momenta as $k_\alpha\to k_\alpha+\delta k$. As a consequence, the wavefunction 
amplitude $A_M(x_1,x_2,\dots,x_M)$ (cf.~\eref{ba_amp}) is modified as $A_M(x_1,x_2,
\dots,x_M)\to e^{i\delta k\sum_i x_i}A(x_1,x_2,\dots, x_M)$. Since the phase factor 
$\exp({i\delta k\sum_i x_i})=\exp({i\delta k\sum_{j\in A}x_j})\times\exp({i\delta 
k\sum_{j\in B}x_j})$ corresponds to a product of unitary transformations on the 
subsystems Hilbert spaces ${\mathcal H}_A$ and ${\mathcal H}_B$, it does not affect 
the entanglement spectrum and entropy. This observation can be recast in the CFT 
language, which allows to generalize it to other interacting $c=1$ and for finite 
particle density $M/L$. In fact, the total quasi-momentum shift is implemented in 
the CFT language by acting over the ground state with vertex operators $Y_1[\alpha_+,
\alpha_-]=e^{i(\alpha_+\phi+\alpha_-\bar\phi)}$, where $\phi$ and $\bar\phi$ are 
bosonic operators, and $\alpha_\pm$ are related to their scaling dimensions. 
Remarkably, it has been shown in Ref.~\cite{alcaraz-2011} that all the excitations 
obtained in this way have the same entropy as the ground state. 

Interestingly, the results in Figure~\ref{fig2:energy} suggest that, at least in the vanishing 
density regime, the effect of the scattering phases is negligible at low energy. This 
does not hold at high energies ($E\to 0^-$) where the two bands merge. This 
corresponds to $I^{(1)}_1\approx I^{(1)}_2\approx I^{(1)}_{MAX}$ (cf.~\eref{bt_qn_bound}), 
which implies $k_1\approx k_2\approx 0\,\textrm{mod}\, 2\pi$ (cf.~\eref{srq}). Notice 
that for $k_1,k_2\to 0$, since the wavefunction~\eref{ba_eig} becomes ``flat'', the 
semiclassical bound $S_{MAX}$ is exact (see~\ref{upper_bound}).

The entanglement entropy for the bound states of two particles is plotted in 
Figure~\ref{fig2:energy} (b). In contrast with the $1$-strings (panel (a)),  
no band structure is visible. The entanglement entropy increases monotonically up 
to $E\approx -0.3$, when it starts decreasing. Interestingly, one has $S_A(L/2)>S_{MAX}
(M=1,L/2)=\log(2)$, which would be the entropy of a one particle state. This behavior can 
be understood in terms of the bound-state length $\ell_b$ (cf.~\eref{b_length_e}). 
At $E=-1$ we have $\ell_b=0$, implying $S_A(L/2)\approx \log(2)$. On the other hand, 
as $E\to 0^-$, $\ell_b\to\infty$ and the two particles become weakly bound, which 
is associated with a substantial increase in the entanglement. More quantitatively, 
for $\ell_b\lesssim L$ one has (see~\ref{two_p_cluster} for the derivation)
\begin{eqnarray}
\label{two_p_cluster_c}
\fl\qquad S_A(L/2)=\log(2)-\frac{2}{LE}\Big[1-\log(2)+\log(L)+\log(-E)\Big]+
{\mathcal O}(1/L^2), 
\end{eqnarray}
which is shown as dash-dotted line in the Figure. Notice that~\eref{two_p_cluster_c} 
breaks down at $E\approx -0.3$. This can be understood considering the entanglement 
spectrum (ES) of $2$-string eigenstates.

%%%%%%%%%%%%%%%%%%%%%%%%%%%%%%%%%%%%%%%%%%%%%%%%%%%%%%%%%%%%%%%%%%%%%%%%%%%
\subsection{Entanglement spectra}
\label{ent_spectra_2p}

%##################################################################
\begin{figure}[t]
\begin{center}
\includegraphics[width=.85\textwidth]{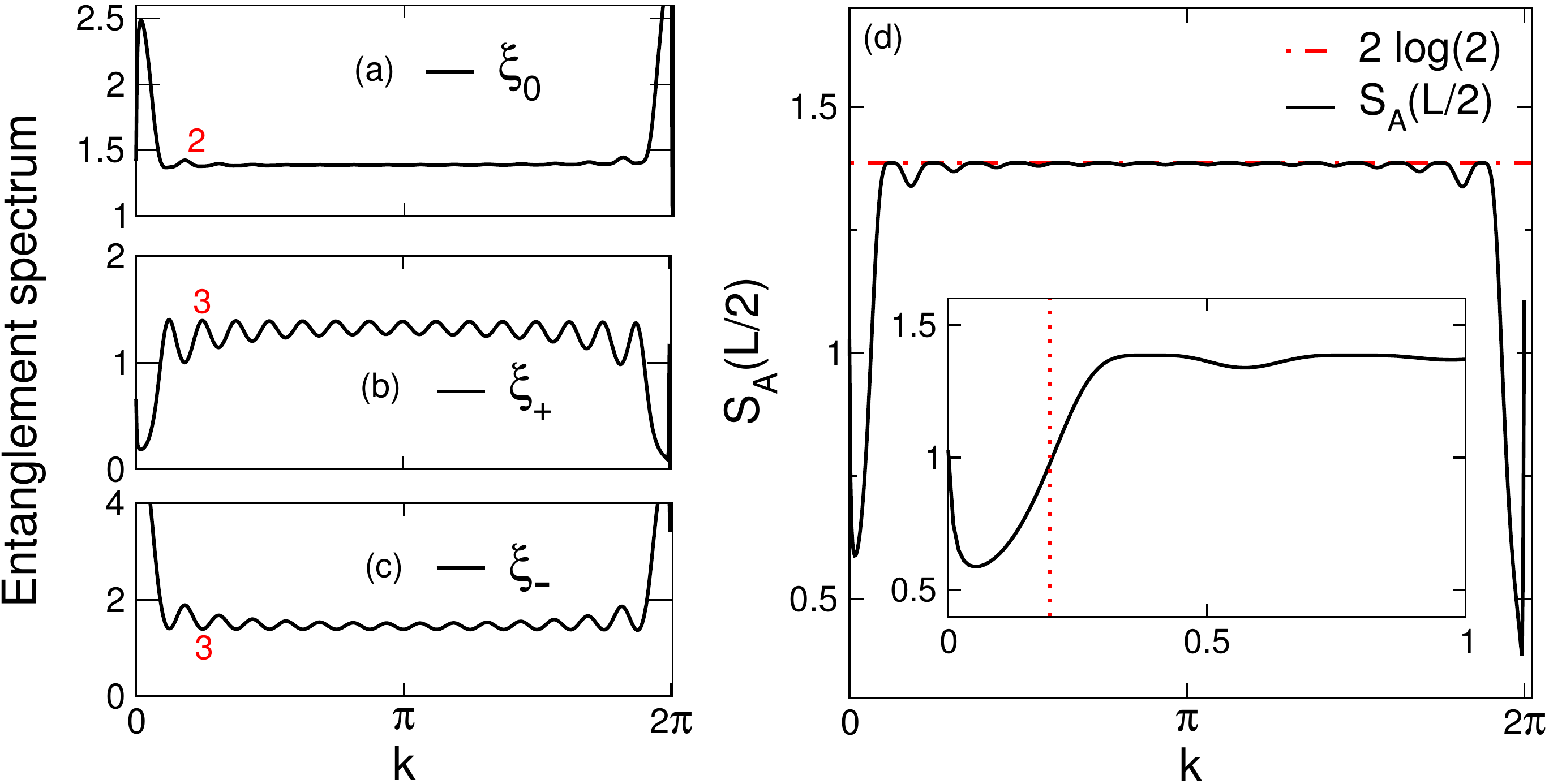}
\end{center}
\caption{ Entanglement spectrum (ES) and entropy for the Heisenberg 
 chain in the sector with two particles $M=2$. (a)(b)(c) ES levels 
 $\xi_+,\xi_-,\xi_0$ plotted versus the relative momentum $k\equiv 
 k_2-k_1$ of the two particles. Data are exact results for a chain 
 with $L=32$. For $0\ll k\ll 2\pi$ we have $\xi_0\approx
 \xi_+\approx\xi_-\approx 2\log(2)$. The accompanying numbers specify the 
 degeneracies of the levels. The resulting half-chain entropy $S_A$ 
 is reported in panel (d) versus $k$ (full line). The dash-dotted 
 line corresponds to the upper bound $S_{MAX}=2\log(2)$. Inset: same as in 
 the main figure zooming around $k\approx 0$. The vertical dotted 
 line is $k=2\pi/L$. 
}
\label{fig3:2p_ES}
\end{figure}
%##################################################################

%##################################################################
\begin{figure}[t]
\begin{center}
\includegraphics[width=.9\textwidth]{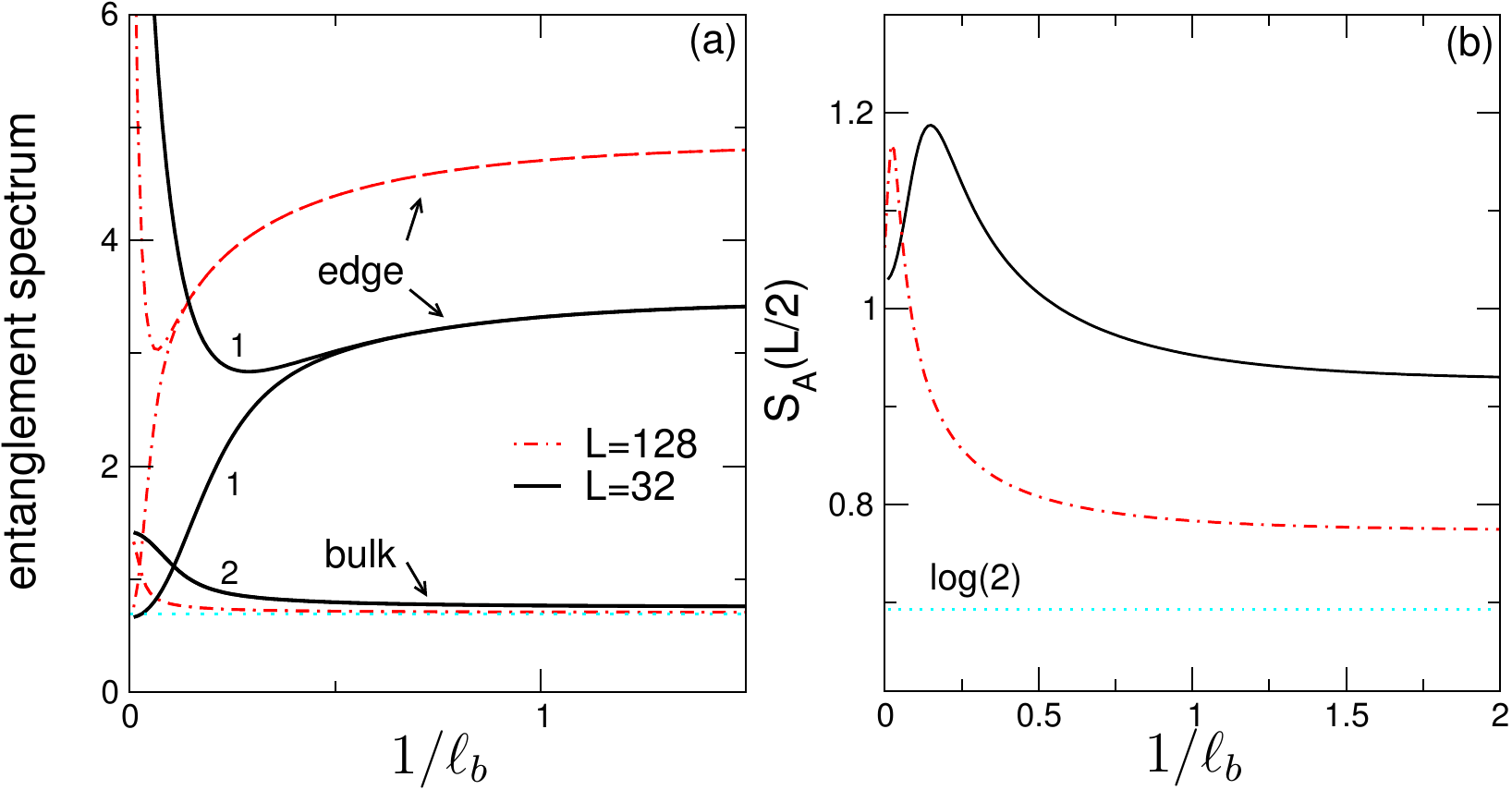}
\end{center}
\caption{Entanglement spectrum and entanglement entropy for $2$-string 
 eigenstates of the $XXX$ chain (bound states of two particles). 
 (a) Entanglement spectrum levels $\xi$ plotted versus 
 $1/\ell_b$, $\ell_b$ being the bound-state length. Full and dash-dotted 
 line are exact numerical results for a chain with $L=32$ and $L=128$, 
 respectively. Accompanying numbers specify the degeneracies (shown only for $L=32$) 
 of the levels. Two edge-related ES levels are present, besides the bulk 
 ones (with $\xi\approx\log(2)$ at $\ell_b\to 0$). At $\ell_b\to\infty$ one 
 of the two edge levels diverges, whereas the remaining is $\sim\log(2)$, i.e., 
 an extra bulk ES level appears. 
 (b) Half-chain entanglement entropy $S_A(L/2)$ obtained from 
 the ES levels in panel (a) plotted versus $1/\ell_b$. Notice that $S_A(L/2)
 \to\log(2)$ at $\ell_b\to 0$ $L\to\infty$, while $S_A(L/2)\to 3/2\log(2)$ for 
 $\ell_b\to 0$. 
}
\label{fig4:2p_bound}
\end{figure}
%##################################################################

In the $1$-string sector, the half-chain ES contains three levels $\xi_0$  
and $\xi_\pm$, which are functions of the relative quasi-momentum 
$k\equiv k_2-k_1$ between the two particles (see~\ref{two_p_ent}). These are 
shown in Figure~\ref{fig3:2p_ES} (panels (a)-(c)) plotted versus $k$, for a chain 
with $L=32$. The accompanying numbers in the panels specify the degeneracies 
of the ES levels. To be general we keep $k$ arbitrary, while in principle its 
value is fixed by using the solutions of the Bethe equations~\eref{ba_eq_log}. 
Moreover, data in Figure~\ref{fig3:2p_ES} are for fixed relative scattering phase 
(cf.~\eref{s_phases}) $\theta\equiv\theta_{2,1}-\theta_{1,2}=\pi$. We verified 
numerically that the main results do not change for different values of $\theta$, 
at least for $0\ll k\ll 2\pi$ (central region in the panels), where the semiclassical 
approximation holds (see~\ref{two_p_ent}). 

As Figure~\ref{fig3:2p_ES} shows, for $0\ll k\ll 2\pi$ one has 
$\xi_0\approx\xi_+\approx\xi_-\approx 2\log(2)$. These results are correct up to terms 
${\mathcal O}(1/L)$, and can be derived rigorously using the Bethe ansatz solution 
(see~\ref{two_p_ent}). This implies, as expected, that in the semiclassical regime 
the effect of the scattering phase $\theta$ (cf.~\eref{s_phases}) can be neglected. 
This scenario breaks down at $k\sim 0\,\mbox{mod}\, 2\pi$, where deviations from the 
semiclassical results are observed for all ES levels (cf.~Figure~\ref{fig3:2p_ES}).
The half-chain entanglement entropy obtained from the ES in (a)-(c) is reported in 
Figure~\ref{fig3:2p_ES} (d) as a function of $k$. While at $k\ne 0$ the entanglement 
entropy exhibits a flat behavior with $S_A(L/2)\approx S_{MAX}(M=2)=2\log(2)$ 
(dash-dotted line in the Figure), at $k\to 0\,\mbox{mod}\, 2\pi$ a strong reduction 
occurs. This is better highlighted in the inset of the Figure zooming in the region 
$0\le k\le 1$. The vertical dotted line is $k=2\pi/L$, the value $S_A(L/2)\approx 1$ is 
the horizontal dotted line shown in Figure~\ref{fig2:energy}. 

We now turn to the entanglement spectra of $2$-string eigenstates. 
The calculation of the ES is outlined in~\ref{two_p_cluster}. 
The ES contains four levels, which are plotted in 
Figure~\ref{fig4:2p_bound} versus the inverse bound-state length $1/\ell_b$ 
(cf.~\eref{b_length} and~\eref{b_length_e}). The full and dash-dotted 
lines are for chains of length $L=32$ and $L=128$, respectively. The 
accompanying numbers specify the degeneracies of the ES levels. 
Clearly, two of the levels are $\sim\log(2)$ as $1/\ell_b\to 0$ (horizontal 
line in the Figure), i.e., for eigenstates containing two strongly-bound 
particles. These are ``bulk'' ES levels, similar to the ES levels for two unbound 
particles (cf.~Figure~\ref{fig3:2p_ES}). Notice that, at fixed $\ell_b$ deviations 
from $\log(2)$ are ${\mathcal O}(1/L)$, as can be confirmed by comparing the curves  
for $L=32$ and $L=128$. Strikingly, the remaining two ES levels diverge in 
the limit $\ell_b,L\to\infty$. These are edge-related levels. The corresponding 
entanglement eigenvectors (i.e., eigenvectors of the reduced density matrix) are 
exponentially localized at the two edges  of subsystem $A$ (edge states) 
(cf.~\ref{two_p_cluster}). Interestingly, at $1/\ell_b\approx 1/2$ one of 
the two edge levels diverges, whereas the other one becomes an 
extra ``bulk'' level. This corresponds to the maximum entropy in 
Figure~\ref{fig2:energy} (b), where~\eref{two_p_cluster_c} breaks down. 
The half-chain entanglement entropy obtained from the ES in panel (a) is 
shown in Figure~\ref{fig3:2p_ES} (b) as a function of $1/\ell_b$. In the limit 
$1\ll \ell_b\ll L$ we have $S_A(L/2)=\log(2)$, i.e. the half-chain entropy 
of a single particle. Notice that for  $\ell_b\to\infty$ we have 
$S_A(L/2)\approx 3/2\log(2)$, due to the extra bulk level appearing 
in the ES (see panel (a)).

%%%%%%%%%%%%%%%%%%%%%%%%%%%%%%%%%%%%%%%%%%%%%%%%%%%%%%%%%%%%%%%%%
\section{Few-particle entanglement : Eigenstates with $M=3$ particles}
\label{low_density_M3}

In this section we focus on entanglement properties in eigenstates of 
the $XXX$ chain in the sector with $M=3$. Besides $1$-string and 
$3$-string eigenstates, which correspond to unbound particles and 
three-particle bound states, the intermediate situation with eigenstates 
that contain a magnon and a two-particle bound state arises. 
These correspond to solutions of the Bethe-Takahashi equations~\eref{bt_eq} 
with string content $\{1,1\}$ ($\{1,1\}$-strings), i.e. one real 
and two complex conjugate rapidities. 

For highly-entangled $1$-string eigenstates we show that $S_A(L_A)\lesssim 
S_{MAX}(M=3)$, confirming what has been found for two particles in 
section~\ref{low_density_M2}. Furthermore, the entanglement entropy is 
progressively reduced in eigenstates that contain bound states (i.e. 
$\{1,1\}$-strings and $3$-strings). We also investigate the behavior of 
the half-chain entanglement entropy as a function of the eigenstate energy. 
For the $1$-strings we observe an entanglement band-like structure with 
three bands. As in the two-particle case (see section~\ref{low_density_M2})), 
higher entanglement bands corresponds to a larger number of blocks 
of contiguous Bethe-Takahashi quantum numbers. In the sector with 
$\{1,1\}$-strings, the half-chain entropy exhibits a dense structure (i.e., 
no entanglement bands), reflecting that the size $\ell_b$ of the two-particle 
bound state varies continuously as a function of the bound-state energy.

%%%%%%%%%%%%%%%%%%%%%%%%%%%%%%%%%%%%%%%%%%%%%%%%%%%%%%%%%%%%%%%%%%%%%%%%%%
\subsection{Overview: unbound particles, bound states, and coexistence 
of the two behaviors}
\label{M3_overview}

%##################################################################
\begin{figure}[t]
\begin{center}
\includegraphics[width=.99\textwidth]{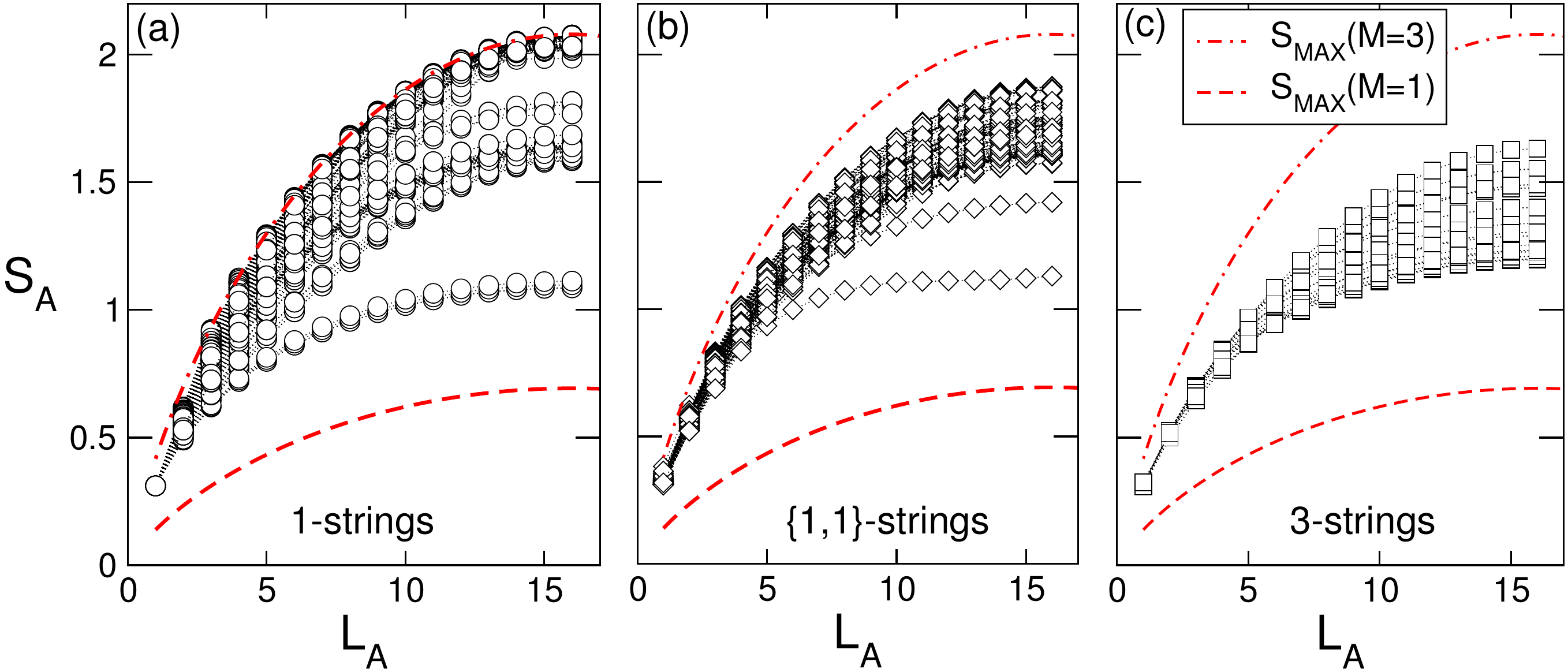}
\end{center}
\caption{ Entanglement entropy of the eigenstates of the $XXX$ chain 
 in the sector with three particles $M=3$. Data  are exact numerical 
 data for $L=32$ sites. (a) $1$-string eigenstates: $S_A$ plotted 
 versus the size of block $A$ $L_A$. The dashed  line is the upper 
 bound $S_{MAX}(M=3)\equiv 3\log(2)$. The dash-dotted line is 
 $S_{MAX}(M=1)\equiv 2\log(2)$ (cf. Figure~\ref{fig1:ent_2p}). (b) 
 Eigenstates corresponding to $\{1,1\}$-strings, i.e. one real and a 
 pair of complex conjugate solutions of the Bethe equations. (c) 
 Same as in (a)(b) but for $3$-string eigenstates (i.e., bound 
 states of the three particles). 
}
\label{fig5:ent_3p}
\end{figure}
%##################################################################

The von Neumann entropy  $S_A(L_A)$ for the eigenstates of the $XXX$ chain 
in the sector with $M=3$ is shown in Figure~\ref{fig5:ent_3p} (panels (a)-(c)) 
for a chain with $L=32$ ($S_A$ plotted versus the subsystem size $L_A$). The 
three panels (a)-(c) correspond to eigenstates obtained from different types of solutions 
of the Bethe-Takahashi equations: $1$-strings, $\{1,1\}$-strings, and $3$-strings  
are shown in panels (a),(b), and (c). The same scale is used on the $y$-axis 
in all the panels. The dashed and dash-dotted lines denote the semiclassical 
bound $S_{MAX}(M)$ (cf.~\eref{ub_intro}) with $M=2$ and $M=3$, respectively. 
The construction of the Bethe wavefunctions, from which the results in the 
Figure are derived, is outlined in~\ref{ba_eig_single_2string} 
and~\ref{three_p_cluster}. 

Clearly, $S_{MAX}(M=3)$ is saturated (apart from ${\mathcal 
O}(1/L)$ terms) in the highest band of the $1$-string sector. A substantial 
decrease in the entanglement entropy occurs for $\{1,1\}$-strings. Large values 
of $S_A(L_A)$ correspond to eigenstates of the $XXX$ chain that contain an unbound 
particle and a pair of weakly bound particles, whereas small ones are obtained 
when two strongly-bound particles are present. Finally, a further reduction 
of the entanglement entropy happens for eigenstates obtained from $3$-strings 
(panel (c)). However, $S_A(L_A)>S_{MAX}(M=1)$, reflecting that the size 
$\ell_b$ of the three-particle bound is finite, similarly to the two-particle 
bound states (see Figure~\ref{fig1:ent_2p}).

%%%%%%%%%%%%%%%%%%%%%%%%%%%%%%%%%%%%%%%%%%%%%%%%%%%%%%%%%%%%%%%%%%%%%
\subsection{Entanglement entropy versus energy: bands and continuous 
structures}
\label{M3_ent_ener}

%##################################################################
\begin{figure}[t]
\begin{center}
\includegraphics[width=.99\textwidth]{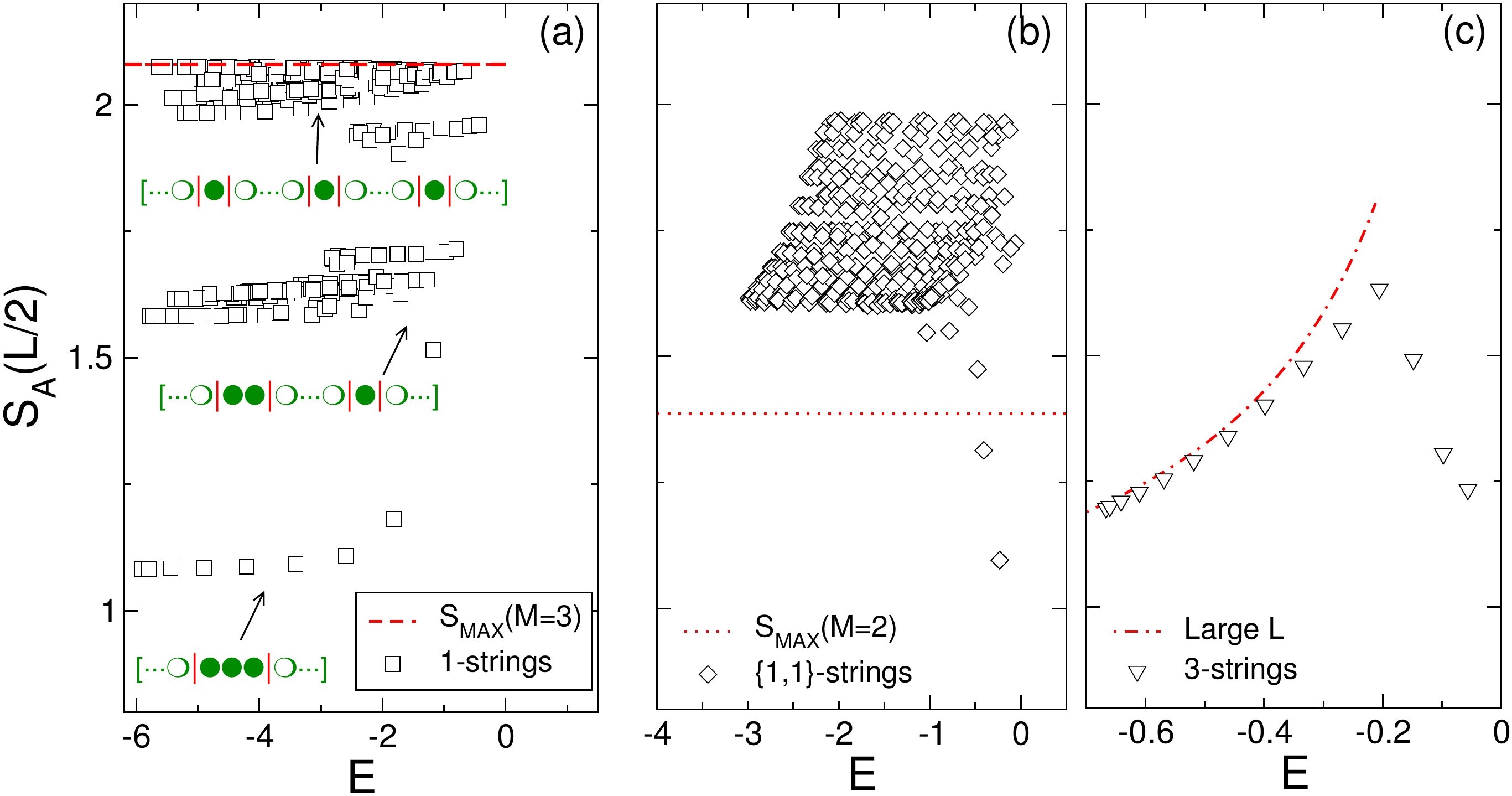}
\end{center}
\caption{ Entanglement entropy for the Heisenberg chain 
 in the sector with three particles $M=3$. Half-chain entropy $S_A(L/2)$ 
 plotted versus the eigenstate energy $E$. Data are exact numerical 
 result from the Bethe ansatz solution of the model for a chain with 
 $L=32$ sites. (a) $1$-string eigenstates:  
 three entanglement bands are visible. The cartoons show typical 
 Bethe-Takahashi number configurations corresponding to eigenstates 
 in a given band. Full and empty symbols denote occupied and vacant 
 Bethe quantum numbers, respectively. The vertical lines mark disconnected 
 blocks of occupied numbers. Different points in a given band are 
 obtained by shifting the blocks independently. 
 The dashed line is the upper bound $S_{MAX}(M=3)=3\log(2)$. (b) The 
 same as in (a) for $\{1,1\}$-string eigenstates. (c) Same as in 
 (a)(b) for the three-particle bound states ($3$-strings). The 
 dash-dotted line is the analytical result in the large $L$ limit.
}
\label{fig6:ent_3p_ener}
\end{figure}
%##################################################################

The half-chain entanglement entropy $S_A(L/2)$ is plotted versus the 
eigenstate energy in Figure~\ref{fig6:ent_3p_ener} for $1$-strings, 
$\{1,1\}$-strings, and $3$-strings. Data are the same as in Figure~\ref{fig5:ent_3p} 
(in all the panels the same scale on the $y$-axis  is used). Notice 
that $1$-string eigenstates have energies $-6\lesssim E\lesssim 0$, 
while $\{1,1\}$-strings and $3$-strings appear at $-3\lesssim E\lesssim 0$, 
and $-2/3\lesssim E\lesssim 0$, respectively. 

In the $1$-string sector (panel (a) in the Figure) $S_A(L/2)$ exhibits 
three bands, which correspond to eigenstates obtained from Bethe-Takahashi 
quantum numbers with different numbers of blocks of contiguous quantum numbers 
(as shown in the cartoons, see also section~\ref{ba_approach}). $S_A(L/2)$ 
increases as a function of the number of Bethe number blocks (from one for the 
lowest band up to three for the highest one). Moreover, $S_A(L/2)\approx 
S_{MAX}(M=3)$ (dashed line in panel (a)) for the eigenstates with three separated 
Bethe numbers. At high energy $E\to 0^-$ all the bands tend to merge, as it  
has been observed for two particles (see Figure~\ref{fig2:energy}). 

The half-chain entropy for $\{1,1\}$-string eigenstates is shown in panel 
(b). A denser (as compared with $1$-strings) structure appears, reflecting 
that  the length $\ell_b$, which is associated with the two-particle bound 
state, is a continuous function of the energy. 

%##################################################################
\begin{figure}[t]
\begin{center}
\includegraphics[width=.9\textwidth]{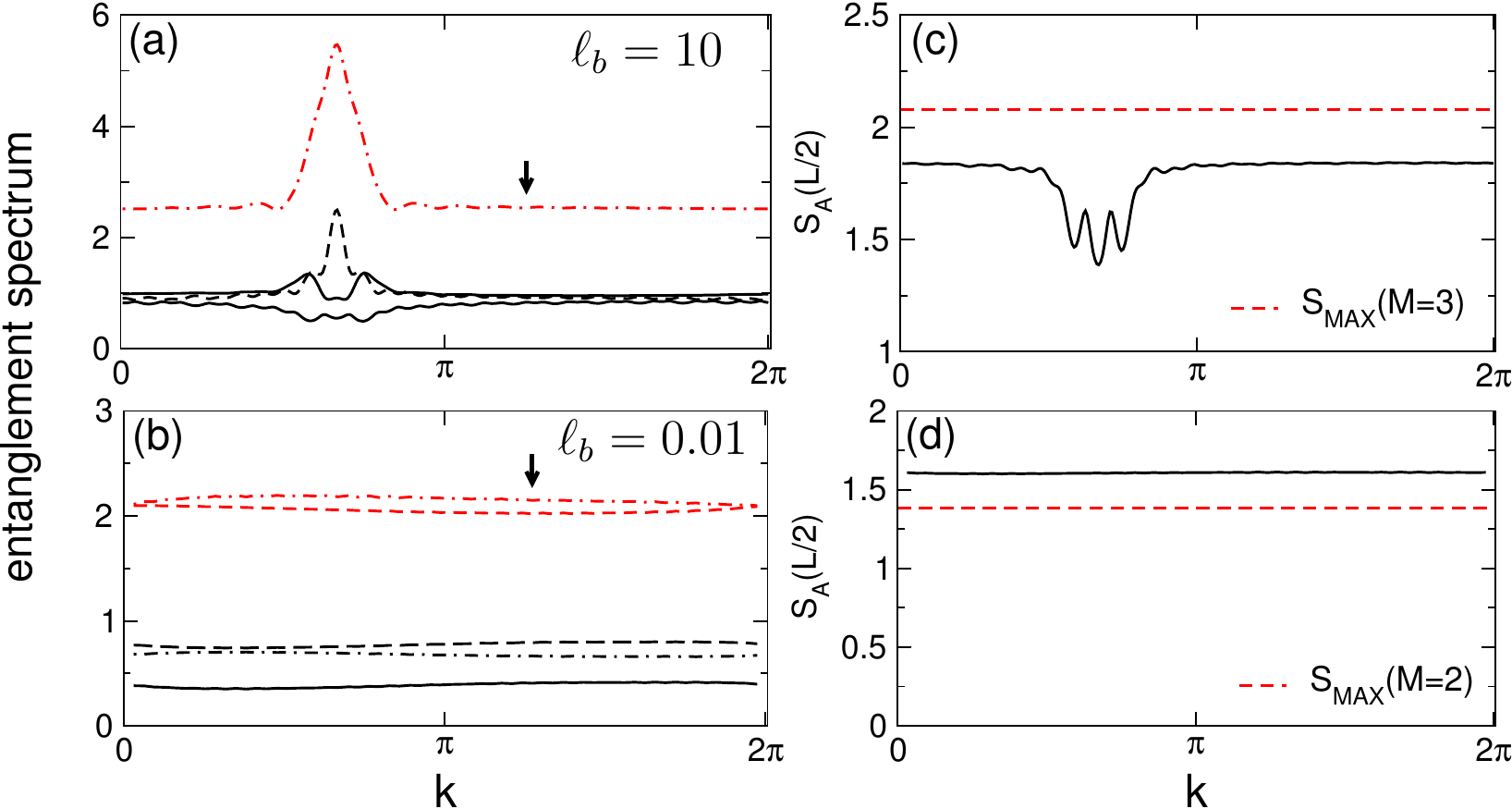}
\end{center}
\caption{ Entanglement entropy and entanglement spectrum (ES) of 
 the Heisenberg spin chain in the sector with $M=3$ particles. Only 
 $\{1,1\}$-string eigenstates are shown in the figure. These 
 comprise a ``free'' particle and a two-particle bound state. 
 Data are exact results from the Bethe ansatz solution for a chain with $L=32$ 
 sites. (a)(b) Half-chain ES for a fixed momentum of the bound state $K_b=
 2\pi/3$ plotted versus the ``free'' particle quasi-momentum $k$. Panels (a) 
 and (b) are for different values of the bound state length $\ell_b=10$ 
 (two weakly-bound particles) and $\ell_b=0.01$ (two strongly-bound 
 particles). The arrow marks edge-related ES levels. (c) Half-chain entropy 
 $S_A(L/2)$ obtained from the ES in panel (a). The dashed line is the 
 semiclassical upper bound $S_{MAX}(M=3)$ (cf.~\eref{ub_intro}). Same as 
 in (c) for the ES in panel (b). The dashed line is now $S_{MAX}(M=2)$. 
}
\label{fig7:ES_1p2}
\end{figure}
%##################################################################

Finally, $S_A(L/2)$ for $3$-string eigenstates is shown in panel (c). The 
dash-dotted line in the Figure is the analytical result in the limit 
$L\gg\ell_b$ (cf.~\eref{M3_bound} and~\eref{M3_bound_e}), 
\begin{eqnarray}
\fl\quad S_A(L/2)=\log(2)-\frac{6}{2LE}\left[1+2\log(2)+
\log(L)+\log(-E)-\log(6)\right]+{\mathcal O}(1/L^2), 
\end{eqnarray}
which accurately describes the behavior of the entropy in the region 
$E\lesssim -0.3$. Notice that deviations from $S_A(L/2)\sim\log(2)$ are 
$\propto 1/(LE)$, as for the two-particle bound state (cf.~\eref{two_p_cluster_c}). 
Interestingly, $S_A(L/2)$ exhibits a maximum with $S_A(L/2)\approx 1.75$ at 
$E\approx -0.2$, similar to the case of a two-particle bound state 
(see Figure~\ref{fig2:energy} and~\ref{fig3:2p_ES}). 

All these findings can be also understood at the level of the half-chain 
entanglement spectrum. In particular, the ES of $1$-string eigenstates 
with maximum number of blocks (three separated Bethe numbers) exhibits $2^M=8$ 
degenerate levels $\xi=\log(8)$, in agreement with the semiclassical result 
in~\ref{upper_bound}. On the other hand, the ES of the three-particle bound 
states is qualitatively similar to what has been found for the two-particles 
(see~\ref{three_p_cluster} for some analytical results). It is interesting, 
however, to consider the entanglement spectrum of $\{1,1\}$-strings.

%%%%%%%%%%%%%%%%%%%%%%%%%%%%%%%%%%%%%%%%%%%%%%%%%%%%%%%%%%%%%%%%%%%%%%%
\subsection{Entanglement spectrum of $\{1,1\}$-string eigenstates}
\label{ES_11strings}

The ES for eigenstates of the $XXX$ chain obtained from $\{1,1\}$-strings 
is shown in Figure~\ref{fig7:ES_1p2}. Data are obtained  
from the Bethe ansatz wavefunction~\eref{ba_eig} for a chain with $L=32$ 
sites. For simplicity we fix in~\eref{ba_eig} the total bound state 
quasi-momentum to $K_b=2\pi/3$, leaving as free parameter the quasi-momentum 
$k$ of the ``free'' particle. Moreover, we treat the bound state length 
$\ell_b$ as an independent variable, although it can be related to the total 
bound-state quasi-momentum $K_b$ via~\eref{b_length} and~\eref{b_length_e}. 
The construction of the Bethe wavefunction for $\{1,1\}$-strings is outlined 
in~\ref{ba_eig_single_2string}. 

Panels (a) and (b) in Figure~\ref{fig7:ES_1p2} show the half-chain ES levels 
as a function of $k$ for $\ell_b=10$ (two weakly-bound particles) and $\ell_b=
10^{-2}$ (two strongly-bound particles), respectively. The corresponding 
entanglement entropies are reported in panels (c) and (d). 
In the case of two weakly-bound particles (panel (a)) the ES exhibits 
four levels. In the limit $L\to\infty$ three ES levels are $\sim\log(2)$ 
(dotted line), whereas the remaining one is diverging 
as $\sim\log(L)$. This is the same edge level found in the ES of the 
two-particle bound states in section~\ref{low_density_M2} (see 
Figure~\ref{fig4:2p_bound}). Notice that the ES changes dramatically 
at $k\approx K_b=2\pi/3$, which is reflected in a 
reduction in the entanglement entropy (see panel (c)). For 
two strongly-bound particles (see panel (b)) two edge levels appear 
(as for two particles, see Figure~\ref{fig4:2p_bound}). Interestingly, 
the ES does not exhibit any significant dependence on $k$, and one has 
$S_A(L/2)\approx S_{MAX}(M=2)$ $\forall k$.

%%%%%%%%%%%%%%%%%%%%%%%%%%%%%%%%%%%%%%%%%%%%%%%%%%%%%%%%%%%%%%%%%%%%%
\section{Entanglement of eigenstates with a finite density of 
particles}
\label{fin_den}

%##################################################################
\begin{figure}[t]
\begin{center}
\includegraphics[width=.8\textwidth]{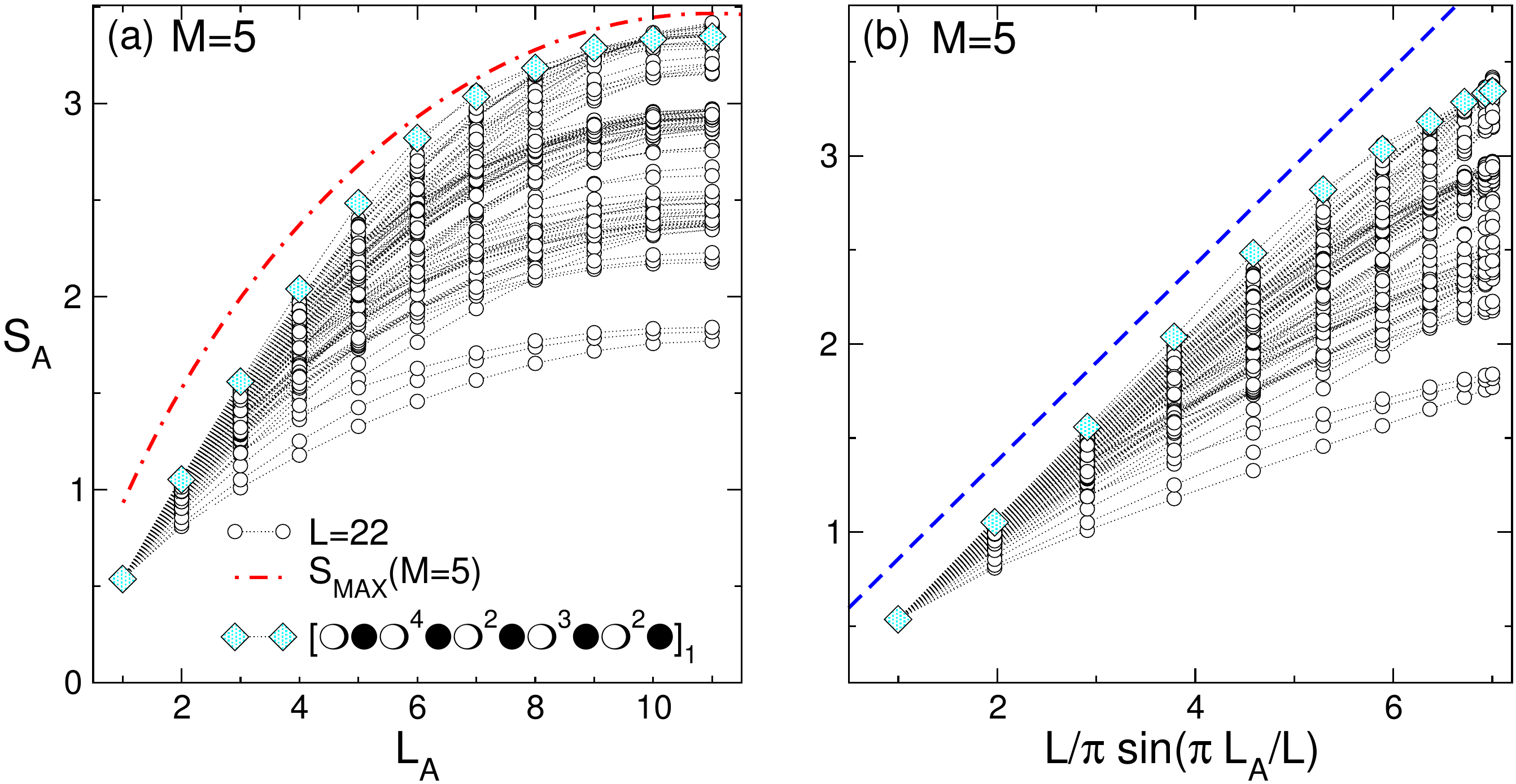}
\includegraphics[width=.8\textwidth]{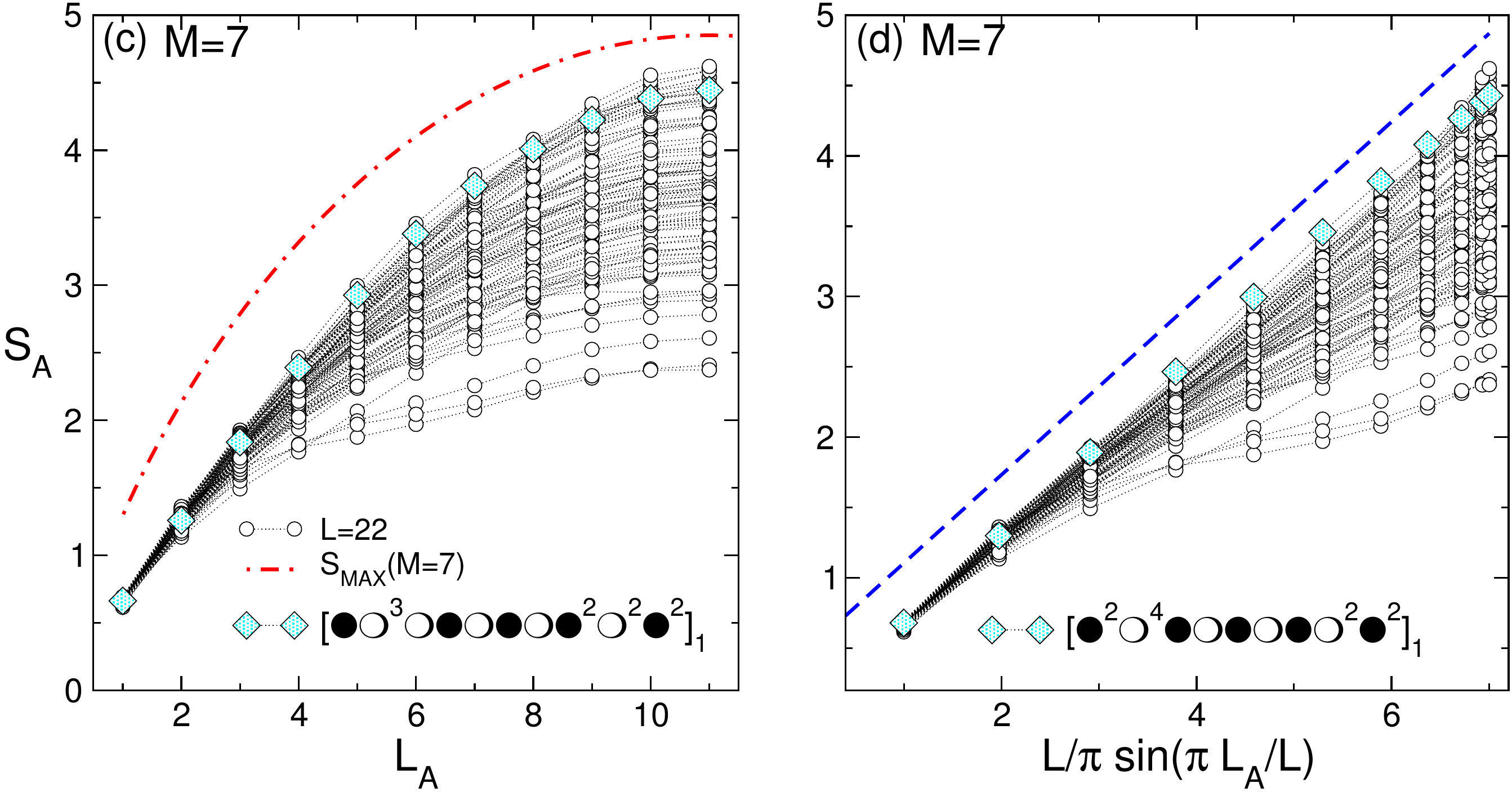}
\end{center}
\caption{ Volume law (i.e. extensive behavior) in the 
 entanglement entropy of the eigenstates of the $XXX$ chain. Here we 
 focus on $1$-string eigenstates. Data are exact numerical results obtained 
 from the Bethe ansatz solution of the model for a chain with $L=22$ 
 sites. (a) Sector with $M=5$: entanglement entropy $S_A$ plotted 
 versus the block length $L_A$. The dash-dotted line is the 
 upper bound $S_{MAX}(M)=-M[\omega\log(\omega)+(1-\omega)\log(1-\omega)]$, 
 with $\omega\equiv L_A/L$. (b) Same data as in (a) plotted versus 
 the chord length $L_c\equiv (L/\pi)\sin(\pi L_A/L)$. 
 The dash-dotted line is a guide to the eye. Full symbols (rhombi) 
 denote $S_A$ obtained from the eigenstate with Bethe-Takahashi 
 quantum numbers $[\fm\nm\fm^4\nm\fm^2\nm\fm\fm^2\nm\fm^2\nm]_1$, 
 where $\nm$ and $\fm$ denote occupied and vacant quantum numbers, 
 respectively. (c)(d) The same as in panels (a)(b) now in the sector 
 with $M=7$. 
}
\label{fig8:log_lin}
\end{figure}
%##################################################################

In this section we focus on the behavior of the entanglement entropy 
for eigenstates of the $XXX$ chain with finite density of particles, 
i.e., with finite $\rho\equiv M/L>0$ in the limit $L\gg 1$. To be specific 
here we present exact numerical data obtained from the Bethe ansatz 
solution of the model (see section~\ref{ba_approach}) for a chain with 
$L=22$, focusing on $M=5$ and $M=7$. We restrict ourselves to $1$-strings 
and $\{M-2,1\}$-strings only (see section~\ref{ba_approach}). Physically, 
the latter corresponds to eigenstates containing a single two-particle 
bound state ($2$-string) and $M-2$ unbound particles. The construction of 
the Bethe ansatz wavefuction~\eref{ba_eig} for $\{M-2,1\}$-strings is 
outlined in~\ref{ba_eig_single_2string}. 

In sharp contrast with the situation with few particles (vanishing density), 
at finite $\rho$ the semiclassical upper bound $S_{MAX}(M)$ (cf.~\eref{ub_intro}) 
is no longer saturated. However, for highly-entangled eigenstates we provide 
numerical evidence that $S_A(L_A)\propto L_c$, with $L_c\equiv (L/\pi)\sin(\pi L_A/L)$ 
the chord length known from conformal field theory. 
Since $L_c\approx L_A$ for $L\gg L_A$ this also signals the volume law $S_A(L_A)
\propto L_A$ (i.e., extensive entanglement). We also 
investigate how  the entanglement entropy depends on the number of blocks of
contiguous Bethe-Takahashi quantum numbers. We numerically observe that 
large number of blocks corresponds to high-entangled eigenstates (as in the 
low-density regime, see sections~\ref{low_density_M2} and~\ref{low_density_M3}). 

Finally, we consider eigenstates in the sector with $M=7$ containing a 
two-particle bound state. We find no significant change in 
the behavior of the von Neumann entropy, as compared to $1$-string 
eigenstates. 

%%%%%%%%%%%%%%%%%%%%%%%%%%%%%%%%%%%%%%%%%%%%%%%%%%%%%%%%%%%%%%%%%%%%%%%%%
\subsection{Extensive entanglement in $1$-string eigenstates} 
\label{volume_law}

%##################################################################
\begin{figure}[t]
\begin{center}
\includegraphics[width=.8\textwidth]{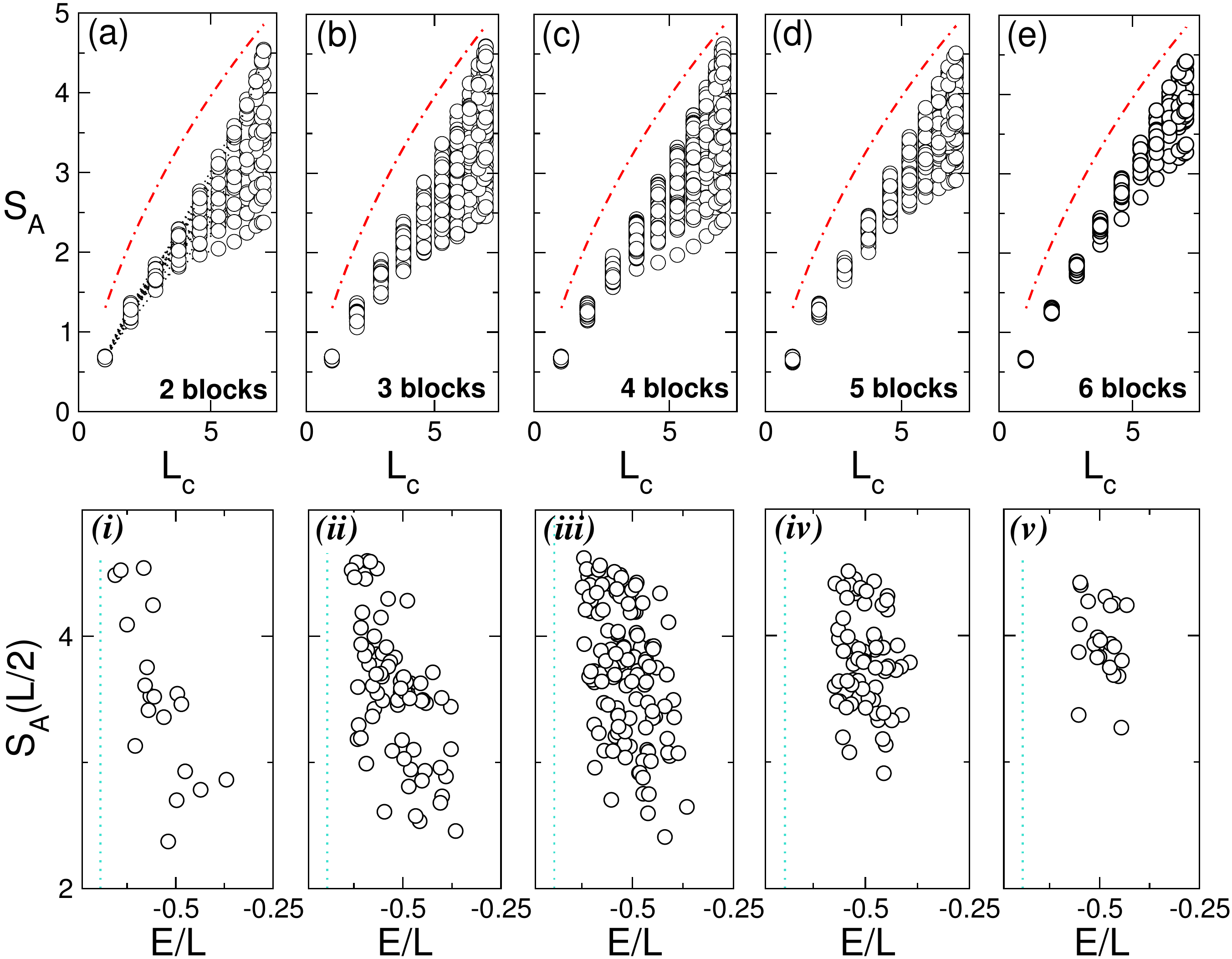}
\end{center}
\caption{ Entanglement entropy of the eigenstates of the $XXX$ chain 
 in the sector with $M=7$. Entanglement entropy as a function of the 
 number of blocks of contiguous occupied Bethe-Takahashi numbers (top panels) 
 and the corresponding eigenstate energies $E$ (bottom panels). Data are exact numerics 
 obtained from the Bethe ansatz solution of the model for a chain of size $L=22$. 
 Here we restrict ourselves to $1$-string eigenstates. (a)-(e) Entanglement 
 entropy $S_A$ as a function of the chord length $L_c\equiv (L/\pi)\sin(\pi 
 L_A/L)$, with $L_A$ the subsystem size. Different panels correspond to 
 different numbers of Bethe number blocks: from two up to six blocks. In all panels the 
 dashed line is the semiclassical upper bound $S_{MAX}(M=7)$. (i)-(v) $S_A(L/2)$ 
 plotted versus the energy density $E/L$. The vertical dotted line denotes the ground 
 state energy density $E_{gs}/L\approx -\log(2)$. Different panels are for different 
 numbers of Bethe number blocks as in (a)-(e). 
}
\label{fig9:dom_wall}
\end{figure}
%##################################################################

Figure~\ref{fig8:log_lin} plots the von Neumann entropy $S_A(L_A)$ for 
eigenstates of the $XXX$ chain with $L=22$ in the sectors with $M=5$ and 
$M=7$ (panels (a)(b) and (c)(d), respectively). Here we restrict ourselves 
to $1$-strings, i.e., real solutions of the Bethe equations~\eref{ba_eq_log}. 
Data were obtained sampling over the Hilbert space of the $XXX$ chain using 
Monte Carlo (see Ref.~\cite{gu-2005} for the details of the Monte Carlo 
algorithm). Only $\sim 100$ eigenstates are shown in the figure. The 
dash-dotted line in panels (a)(c) is the semiclassical upper bound 
$S_{MAX}(M)$ (cf.~\eref{ub_intro}). Already in the sector with $M=5$, 
$S_A(L_A)$ shows deviations from $S_{MAX}(M)$, which become larger in the 
sector with $M=7$ (see Figure~\ref{fig8:log_lin} (c)). 

One should observe that at finite density $\rho$, in the large 
$L_A\ll L$ limit, $S_{MAX}(M)$ reduces to   
\begin{equation}
\label{smax_ext}
S_{MAX}(M)=\rho\Big[1-\log\Big(\frac{L_A}{L}\Big)+{\mathcal O}
(L_A/L)\Big]L_A. 
\end{equation}
The term $\propto\log(L_A/L)$ in the prefactor in~\eref{smax_ext} diverges for  
$L_A/L\ll 1$, signaling that the semiclassical bound $S_{MAX}(M)$ exceeds 
the maximum allowed value $S_A(L_A)=L_A\log(2)$ for the von Neumann entropy. 

Motivated by the result~\eref{vn_cft} for the ground state entanglement entropy 
of the $XXX$ chain~\cite{cc-04}, it is natural to assume that $S_A(L_A)$ is 
a function of the chord length $L_c\equiv (L/\pi)\sin(\pi L_A/L)$,  i.e.,  
\begin{equation}
\label{s_cl}
S_A(L_A)=f(L_c). 
\end{equation}
This is checked numerically in panels (b) and (d) for the eigenstates in the 
sector with $M=5$ and $M=7$, respectively. Clearly, $S_A\propto L_c$ 
for highly-entangled eigenstates. Since $L_c\approx L_A$ for $L_A\ll L$,   
the expected extensive behavior $S_A(L_A)\propto L_A$ is recovered. The full 
symbols (rhombi) in the figure show $S_A(L_A)$ for the eigenstates 
corresponding to Bethe-Takahashi numbers $[\fm\nm\fm^4\nm\fm^2\nm\fm^3\nm
\fm^2\nm]_1$ and $\nm^2\fm^4\nm\fm\nm\fm\nm\fm^2\nm^2]_1$ (see 
section~\ref{ba_approach} for the meaning of the notation). Clearly, a large 
number of blocks in the Bethe numbers is associated with linear behavior of the 
entanglement entropy. 

%##################################################################
\begin{figure}[t]
\begin{center}
\includegraphics[width=.87\textwidth]{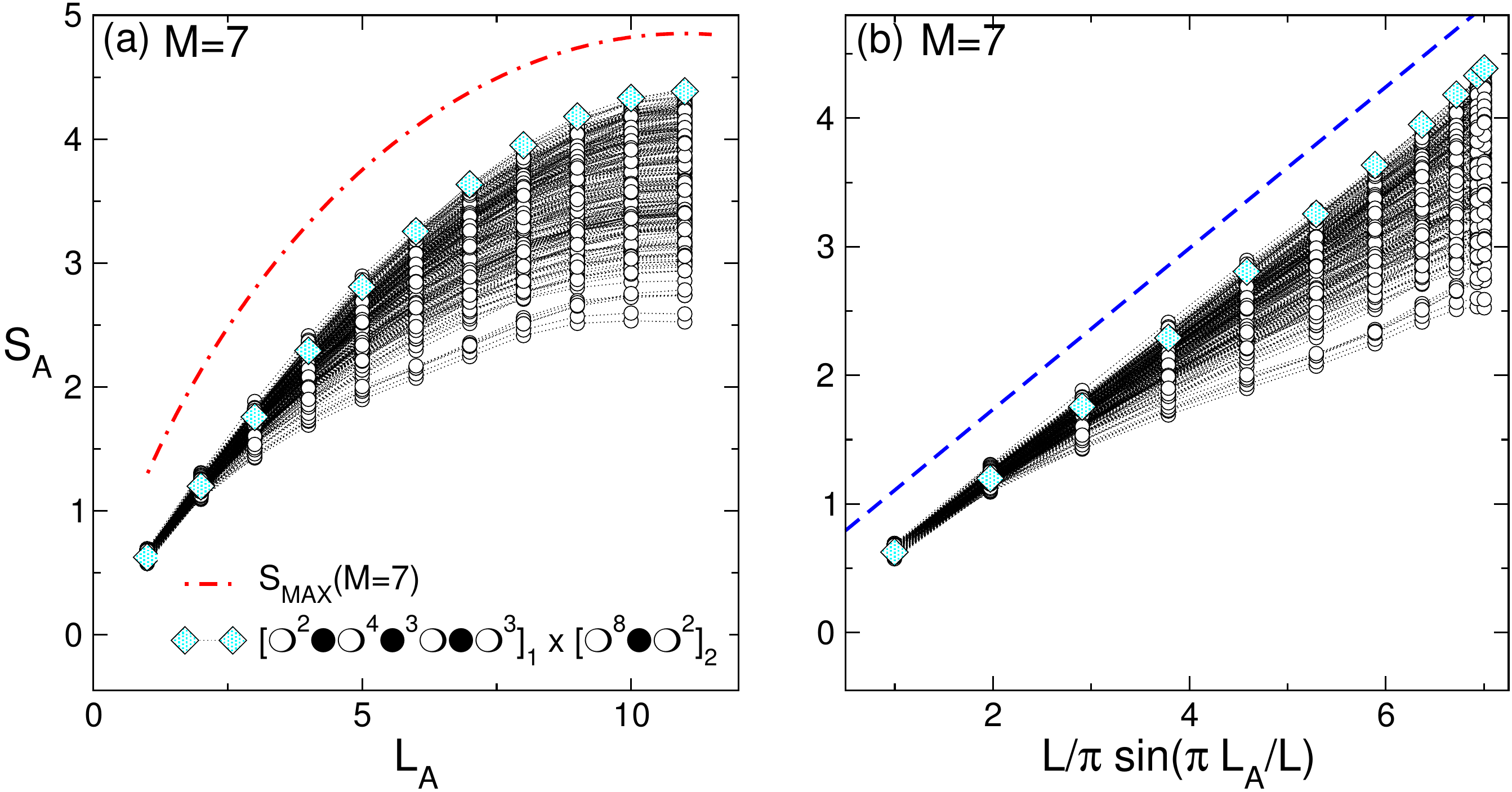}
\end{center}
\caption{ Crossover to the entropy volume law in the sector with $M=7$ 
 particles of the $XXX$  chain.  Here we focus on eigenstates containing 
 only a single $2$-string. Data are exact numerical results (Bethe ansatz) 
 for a chain with $L=22$ sites ($\sim 700$ eigenstates were considered). 
 (a) Entanglement entropy $S_A(L_A)$ plotted versus the block length $L_A$. 
 Full symbols (rhombi) denote $S_A$ for the eigenstate corresponding to 
 $[\fm^2\nm\fm^4\nm^3\fm\nm\fm^3]_1\times [\fm^8\nm\fm^2]_2$, 
 where $\nm$ and $\fm$ denote occupied and vacant Bethe quantum numbers, 
 respectively. Here $[\cdots]_{\alpha}$ denote the Bethe-Takahashi 
 quantum number configurations identifying the $\alpha$-string solutions 
 of the Bethe-Takahashi equations. The dash-dotted line is the 
 semiclassical bound $S_{MAX}(M)$ (same as in Figure~\ref{fig8:log_lin}). 
 (b) Same data as in (a) plotted versus the chord length $L_c\equiv (L/
 \pi)\sin(\pi L_A/L)$. The dash-dotted line is a guide to the eye. 
}
\label{fig10:log_lin_2str}
\end{figure}
%##################################################################

The relation between entanglement and blocks of contiguous Bethe-Takahashi 
numbers is further investigated in Figure~\ref{fig9:dom_wall}, considering 
the sector with $M=7$. The data are the same as in Figure~\ref{fig8:log_lin}. 
Panels (a)-(e) in the figure show the entanglement entropy as a function 
of the chord length $L_c$. Different panels correspond to the different numbers of 
blocks of contiguous Bethe quantum numbers: two blocks in panel 
(a), up to six  in panel (e). The dash-dotted line in all 
panels is the semiclassical bound $S_{MAX}(M=7)$. Interestingly, already 
eigenstates with only two blocks give quite large values for the entanglement 
entropy. However, while $2\lesssim S_A(L/2)\lesssim 4.5$ for eigenstates 
in panel (a), one has $3\lesssim S_A(L/2)\lesssim 4.5$ in panel (e), meaning 
that eigenstates with many blocks typically correspond to highly-entangled 
states. Notice that one should expect a logarithmic entanglement scaling (similar 
to~\eref{vn_cft}) for eigenstates with few blocks, as 
shown in Ref.~\cite{alba-2009}, while volume law should correspond to an extensive 
number of blocks. However, due to finite-size effects 
present for $L=22$, it is not possible to distinguish between the two behaviors 
in Figure~\ref{fig9:dom_wall}. 

Complementary information is shown in panels (i)-(v) where the half-chain 
entropy is plotted versus the eigenstate energy density $E/L\in 
[-\log(2),0]$~\cite{taka-book} (same data as in (a)-(e), same scale is used on both 
axes in all panels). The vertical-dotted line denotes the ground-state energy 
density in the thermodynamic limit $E_{gs}/L= -\log(2)$. 
While eigenstates obtained from Bethe-Takahashi numbers with two blocks 
correspond to energies $-10\lesssim E\lesssim 0$, many blocks correspond to a 
narrower energy window in the ``bulk'' of the spectrum. Also, notice that no clear 
band structure in the half-chain entropy is visible, in contrast with the 
low-density regime (cf. section~\ref{low_density_M2} and~\ref{low_density_M3}). 
This is expected since the number of bands increases exponentially with $M$ (as 
$p(M)$), while the half-chain entropy is at most $S_A\propto M$. 

%%%%%%%%%%%%%%%%%%%%%%%%%%%%%%%%%%%%%%%%%%%%%%%%%%%%%%%%%%%%%%%%%%%%%%%%%
\subsection{Eigenstates containing only a single two-particle bound state}
\label{2strings}

The effects of bound states on the entanglement entropy of eigenstates of the 
$XXX$ chain with finite particle density $\rho$ are investigated in 
Figure~\ref{fig10:log_lin_2str}. We focus on the $XXX$ chain with $L=22$ in 
the sector with $M=7$, restricting ourselves to eigenstates with a single 
two-particle bound state. 

In panel (a) the von Neumann entropy $S_A(L_A)$ is plotted versus the 
block length $L_A$. The dash-dotted line is the semiclassical bound 
$S_{MAX}(M)$ (as in Figure~\ref{fig8:log_lin}). Interestingly, the 
behavior of the entropy is not affected significantly by the two-particle 
bound state, and $2.5\lesssim S_A(L/2)\lesssim 4.5$, as for $1$-string 
eigenstates (compare with Figure~\ref{fig8:log_lin}). The full rhombi show 
the von Neumann entropy for the (highly-entangled) eigenstate obtained 
from Bethe-Takahashi quantum numbers $[\fm^2\nm\fm^4\nm^3\fm\nm\fm^3]_1
\times[\fm^8\nm\fm^2]_2$, with $[\cdots]_1$ and $[\cdots]_2$ denoting the 
quantum numbers for the $1$-strings and the $2$-string, respectively 
(see section~\ref{ba_approach}). The same data are shown in panel (b) versus the 
chord length $L_c\equiv (L/\pi)\sin(\pi L_A/L)$. Highly-entangled 
eigenstates exhibit extensive entanglement entropy, as for the $1$-string 
eigenstates. 

%%%%%%%%%%%%%%%%%%%%%%%%%%%%% CONCLUSIONS %%%%%%%%%%%%%%%%%%%%%%%%%%
\section{Conclusions}
\label{conclusions}

In this paper, by exploiting the Bethe ansatz solution of the 
model~\cite{bethe-1931,taka-book}, we investigated the entanglement entropy 
and entanglement spectra (ES) of the eigenstates of the spin-$\frac{1}{2}$ 
isotropic Heisenberg chain ($XXX$ chain). In contrast with Ref.~\cite{alba-2009}, 
which focused on eigenstates corresponding to real solutions (rapidities) 
of the Bethe equations, here we considered both real ($1$-strings) and 
complex rapidities ($n$-strings, with $n\in{\mathbb N}$, $n>1$). While the 
former are states of interacting magnons, $n$-strings signal the presence 
of many-particle bound states. We addressed both the situations with low and 
finite particle density $\rho\equiv M/L$, $M$ and $L$ being the particle 
number and the chain length, respectively.

We first considered the situation with $M=2$ and $M=3$ (i.e., few particles in 
the chain). For $M=2$, based on the Bethe ansatz results, we derived an upper 
bound $S_{MAX}$ for the entanglement entropy. Using semiclassical arguments this 
was generalized for arbitrary $M$ as 
\begin{eqnarray}
\label{ub_concl}
S_{MAX}(M)=-M(\omega\log(\omega)+(1-\omega)\log(1-\omega)), 
\end{eqnarray}
with $\omega$ the ratio $\omega\equiv L_A/L$, and $L_A$ the length 
of subsystem $A$. For $M=1$,~\eref{ub_concl} is the entanglement entropy 
of a free particle. The linear dependence $S_A(M)\propto M$ reflects that in 
the semiclassical approximation the interactions between particles 
can be neglected. For $L_A/L\ll 1$, $S_{MAX}\approx -(ML_A/L)\log(L_A/L)$, 
which is an ``intermediate'' behavior between logarithmic and extensive. 
Remarkably, for highly-entangled $1$-string eigenstates with $M=2$ 
(and in general in the vanishing density regime, i.e., at fixed $M$ and 
$L\to\infty$) $S_{MAX}(M)$ is saturated, i.e., $S_A\approx S_{MAX}$, apart 
from ${\mathcal O}(1/L)$ terms.  

Focusing on the half-chain entanglement entropy, we investigated the 
relation between entanglement and eigenstate energy. In the vanishing density  
regime, the half-chain entropy exhibits a ``band'' structure, 
when plotted against energy. This is understood in terms of the Bethe-Takahashi 
quantum numbers, which identify the eigenstates of the $XXX$ chain. Specifically, 
higher entanglement bands correspond to states with a larger number of blocks 
of contiguous Bethe-Takahashi numbers. The number of entanglement bands, which 
is obtained by counting the possible ways of grouping the quantum numbers, is 
given by the integer partitions $p(M)$ of $M$. For eigenstates with maximum number 
of Bethe number blocks it is $S_A\approx S_{MAX}$. Interestingly, the corresponding 
entanglement spectrum (ES) levels are simple functions of the ratio 
$\omega\equiv L_A/L$ that are understood in the semiclassical picture. 

For eigenstates that contain bound states, the entanglement entropy is dramatically 
reduced, as compared to the situation with unbound particles only. Since bound 
states can be treated effectively as single particles, this reflects a reduction 
in the number of degrees of freedom. The finite-size behavior of the entanglement 
entropy is quantitatively characterized in terms of the bound-state extension 
$\ell_b$, which is an increasing function of the bound-state energy, and it is 
known analytically from the Bethe ansatz solution. Strongly-bound (equivalently, 
low-energy) and weakly-bound (i.e., high-energy) particles correspond to $\ell_b/
L\ll1$ and $\ell_b/L\gg1$, respectively. Clearly, the entanglement entropy 
increases upon increasing $\ell_b\lesssim L$, i.e., as particles become weakly 
bound. The presence of bound states has striking effects at the level of the 
half-chain entanglement spectrum (ES). Specifically, we demonstrated that the ES 
exhibits edge-related levels. The corresponding entanglement eigenfunctions (i.e., 
the eigenvectors of the reduced density matrix) are exponentially localized at the 
boundary between the two subsystems. 

This semiclassical scenario outlined above breaks down at finite particle density. 
A striking change is that the semiclassical upper bound $S_{MAX}$  
is not saturated. Interestingly, for highly-entangled eigenstates it is now 
$S_A(L_A)\propto L_c$, with $L_c$ the chord length $L_c\equiv (L/\pi)\sin(\pi L_A/L)$. 
Since $L_c\approx L_A$ for small subsystems, this signals the extensive 
entanglement. We also investigated the relation between entanglement and the number of 
blocks of contiguous Bethe-Takahashi numbers. As for vanishing density, eigenstates 
corresponding to a large number of blocks are highly-entangled. Finally, we 
studied the effects of bound states, focusing on eigenstates with $\rho\approx 1/3$ 
and only one two-particle bound state. We provided numerical evidence that the 
presence of the bound state does not affect significantly the entanglement entropy.

\appendix

%%%%%%%%%%%%%%%%%%%%% TWO PARTICLES ENTANGLEMENT %%%%%%%%%%%%%%%%%%%
\section{Entanglement properties of two unbound particles}
\label{two_p_ent} 

In this section we focus on the eigenstates of the $XXX$ chain in the 
sector with two particles, i.e., with $M=2$. We derive analytically the 
entanglement spectrum (ES) and the entanglement entropy for a  
generic wavefunction $|\Psi_2\rangle$ of the form~\eref{ba_eig}.

As a function of the two quasi-momenta $k_1$ and $k_2$, $|\Psi_2\rangle$ 
reads 
\begin{equation}
\label{two_p_bw1}
|\Psi_2\rangle=\frac{1}{\mathcal N}\sum\limits_{x_1<x_2}\left [ 
e^{ik_1 x_1+ik_2 x_2+i\theta_{1,2}}+e^{ik_1 x_2+ik_2 x_1+i\theta_{2,1}}
\right ]|x_1,x_2\rangle,
\end{equation}
where $|x_1,x_2\rangle$ denotes the configuration with the two particles at 
positions $x_1,x_2$ in the chain, and the normalization ${\mathcal N}$ is 
given as  
\begin{equation}
\label{two_p_norm}
{\mathcal N}=\Big[L(L-1)+\frac{-L\cos(k-\theta)+\cos[Lk-\theta]+(L-1)
\cos(\theta)}{\cos(k)-1}\Big]^\frac{1}{2}. 
\end{equation}
Here $k\equiv k_1-k_2$ is the relative quasi-momentum between the two 
particles, and $\theta\equiv\theta_{2,1}-\theta_{1,2}$ the scattering 
phase (cf.~\eref{s_phases}). 

In order to derive entanglement properties it is useful to introduce the 
Schmidt matrix ${\mathbb M}(A|B)$. Given a generic wavefunction $|\Psi\rangle$ 
and a bipartition of a system into two parts $A$ and $B$, ${\mathbb M}(A|B)$ 
is defined such that  
\begin{equation}
\label{schmidt_mat}
|\Psi\rangle=\sum\limits_{i,j} {\mathbb M}_{i,j}(A|B)|\varphi^{(A)}_i
\rangle\otimes|\varphi^{(B)}_j\rangle, 
\end{equation}
where $|\varphi^{(A)}_i\rangle$ and $|\varphi^{(A)}_i\rangle$ form two 
orthonormal bases for part $A$ and $B$, respectively. Notice that the 
Schmidt decomposition~\eref{sch_dec} corresponds to the singular 
value decomposition (SVD) of ${\mathbb M}_{i,j}(A|B)$. In particular, the 
non-zero singular values $\zeta_i$ of ${\mathbb M}(A|B)$ are related to 
the reduced-density-matrix (for part $A$) eigenvalues $\lambda_i$ as 
$\lambda_i=\zeta_i^2$. 

Here for the case with $|\Psi\rangle=|\Psi_2\rangle$ (cf.~\eref{two_p_bw1}), 
the orthonormal basis $\{|\varphi^{(A)}_i\rangle\}$ for $A$ 
can be chosen as 
\begin{eqnarray}
\label{sch_bA}
|\Omega_A\rangle\equiv |\uparrow\uparrow\cdots\uparrow\rangle_A\\
\nonumber
|\varphi_A\rangle=\frac{1}{{\mathcal N}_A}\sum\limits_{x_1<x_2\in A}
\left[e^{ik_1 x_1+ik_2 x_2+i\theta_{1,2}}+e^{ik_1 x_2+ik_2 x_1+
i\theta_{2,1}}\right ]|x_1,x_2\rangle_A\\\nonumber
|\varphi_A'(k_1)\rangle=\frac{1}{\sqrt{L_A}}\sum\limits_{x\in A}e^{ik_1 
x}|x\rangle_A \\\nonumber
|\varphi_A''(k_1,k_2)\rangle=\frac{1}{\sqrt{1-|\Gamma_A|^2}}\Big [|\varphi_A'(k_2)\rangle-\Gamma_A|
\varphi_A'(k_1)\rangle\Big], 
\end{eqnarray}
where $L_A$ is the length of subsystem $A$, and $|x\rangle_A\equiv 
S^-_{x}|\Omega_A\rangle$, $|x_1,x_2\rangle_A\equiv S^-_{x_1}S^-_{x_2}|\Omega_A
\rangle$. In~\eref{sch_bA} ${\mathcal N}_A$ is obtained from~\eref{two_p_norm} 
after replacing $L\to L_A$, and the overlap $\Gamma_A$ reads 
\begin{equation}
\label{over_A}
\Gamma_A\equiv\langle\varphi'_A(k_1)|\varphi'_A(k_2)\rangle=\frac{1}{L_A}
\frac{\sin\frac{k}{2}L_A}{\sin\frac{k}{2}}\,\exp\big[-i\frac{k}{2}(1+L_A)
\big].
\end{equation}
The basis $\{|\varphi^{(B)}_i\rangle\}$ for subsystem $B$ is obtained 
from~\eref{sch_bA} by replacing $A\to B$, and using that 
\begin{equation}
\label{over_B}
\Gamma_B=\frac{1}{L_B}
\frac{\sin\frac{k}{2}L_B}{
\sin\frac{k}{2}}\exp\Big[-i\frac{k}{2}(1+L_B)-ikL_A\Big]. 
\end{equation}
The Bethe wavefunction $|\Psi_2\rangle$ is then rewritten as  
\begin{eqnarray}
\fl\nonumber|\Psi_2\rangle=\frac{1}{\mathcal N}
\Big[{\mathcal N}_A|\varphi_A\rangle\otimes|\Omega_B\rangle+
{\mathcal N}_B|\Omega_A\rangle\otimes|\varphi_B\rangle+
\sqrt{L_AL_B(1-|\Gamma_B|^2)}|\varphi'_A\rangle\otimes
|\varphi''_B\rangle\\\nonumber +e^{i\theta}\sqrt{L_AL_B(1-
|\Gamma_A|^2)}|\varphi''_A\rangle\otimes|\varphi'_B\rangle+
\sqrt{L_AL_B}(\Gamma_B+\Gamma_Ae^{i\theta})|\varphi'_A\rangle
\otimes|\varphi'_B\rangle\Big]. 
\end{eqnarray}
The resulting Schmidt matrix~\eref{schmidt_mat} reads 
\begin{eqnarray}
\fl{\mathbb M}(A|B)=\frac{1}{\mathcal N}
\left[\begin{array}{ccccc}
0               &  {\mathcal N}_B           &  0          &         & 0\\ 
{\mathcal N}_A  &  0                        &  0          &         & 0\\
0               &  0                        &  (L_AL_B)^{1/2}(\Gamma_B+e^{i\theta}\Gamma_A)  &    & (L_AL_B(1-|\Gamma_B|^2))^{1/2}\\
0               &  0                        &  e^{i\theta}(L_AL_B(1-|\Gamma_A|^2))^{1/2}     &    &  0
\end{array}\right].
\label{two_p_schm}
\end{eqnarray}
The two entries in the upper-left $2\times 2$ block in~\eref{two_p_schm} correspond to  
both particles being in subsystem $A$ and subsystem $B$, respectively. The bottom-right block 
corresponds to having one particle in subsystem $A$ and one in $B$. The singular values decompositions 
of~\eref{two_p_schm} is straightforward. ${\mathbb M}(A|B)$ has four 
non-zero singular values. Two singular values are $\zeta_0=
{\mathcal N}_A/{\mathcal N}$ and $\zeta'_{0}={\mathcal N}_B/{\mathcal N}$. 
The remaining ones $\zeta_\pm$ are given as  

\begin{equation*}
\fl\nonumber\zeta_{\pm}=\frac{1}{\mathcal N}\Big[
L_AL_B\Big(1+\textrm{Re}(\Gamma_B^*\Gamma_Ae^{i\theta})\pm 
\sqrt{(1+\textrm{Re}(\Gamma_A^*\Gamma_Be^{-i\theta}))^2-
(|\Gamma_A|^2-1)(|\Gamma_B|^2-1)}\Big )
\Big]^{\frac{1}{2}}.
\end{equation*}

In the limit $L_A,L_B\gg 1$ from~\eref{over_A} and~\eref{over_B} 
one has that $\Gamma_A\propto 1/L_A$ and $\Gamma_B\propto 1/L_B$. The singular 
values of ${\mathbb M}(A|B)$ become $\zeta_0\approx \omega$, 
$\zeta_0'\approx 1-\omega$, and $\zeta_+\approx\zeta_-\approx
\omega^\frac{1}{2}(1-\omega)^\frac{1}{2}$, with $\omega$ being the  
ratio $\omega\equiv L_A/L$. These are related to 
the ``semiclassical'' probabilities of distributing the particles in 
subsystem $A$ and $B$. We mention that similar results are obtained for 
eigenstates with $M>2$ (see~\ref{upper_bound}). Notice that in the 
semiclassical approximation the scattering phase $\theta=\theta_{2,1}-
\theta_{1,2}$ is irrelevant, 
as expected. However, this semiclassical approximation holds at {\it finite} 
(i.e., non-vanishing) relative quasi-momentum $k\equiv k_2-k_1$. For $k\to 
0$, $\theta$ cannot be neglected, and the outlined picture breaks down.
The limit $k\to 0$ is more carefully discussed in the next subsection. 

%%%%%%%%%%%%%%%%%%%% LOW RELATIVE QUASI-MOMENTUM %%%%%%%%%%%%%%%%%%%%
\subsection{Low relative quasi-momentum limit}
\label{low_qm_limit} 

In this section we derive the entanglement entropy and entanglement 
spectrum for the two-particle wavefunction~\eref{two_p_bw1} assuming 
$k\to 0$, i.e., small relative quasi-momentum between the two particles. 
Specifically, we assume $k_2-k_1=2\pi\frac{\alpha}{L}$ and $\alpha={\mathcal 
O}(1)$. Using~\eref{rap} and the logarithmic Bethe 
equations~\eref{ba_eq_log} one obtains  
\begin{equation}
\label{srq}
k_2-k_1=2\pi\frac{\delta J}
{L}+\frac{4}{L}\arctan(\lambda_1-\lambda_2),
\qquad\mbox{with}\quad \delta J\equiv J_2-J_1.  
\end{equation}
Since $\lambda_1\to\lambda_2$ in the limit $L\to\infty$, 
$\alpha\approx\delta J$, which justifies the assumption 
$\alpha={\mathcal O}(1)$. Substituting in~\eref{over_A}\eref{over_B},
one obtains 
\begin{eqnarray}
\label{gamma_app}
\Gamma_A\approx\frac{\sin\pi\alpha\omega}{
\pi\alpha\omega}\, e^{-i\pi\alpha\omega}, & \qquad  
\Gamma_B\approx 
\frac{\sin\pi\alpha(1-\omega)}{\pi\alpha(1-\omega)}\, 
e^{-i\pi\alpha(1+\omega)}.
\end{eqnarray}
where $\omega\equiv L_A/L$, and the limit $L\gg 1$ was taken. 
The two singular values $\zeta_0,\zeta_0'$ of the Schmidt 
matrix ${\mathbb M}(A|B)$ (cf.~\eref{two_p_schm}) become 
\begin{eqnarray}
\label{zeta_0}
\zeta_0=\Big[\frac{2(\pi\alpha\omega)^2-\cos(2\pi\alpha 
\omega+\theta)-\cos(\theta)}{2(\pi\alpha)^2-\cos(2\pi
\alpha+\theta)-\cos(\theta)}\Big]^\frac{1}{2}\\
\zeta_0'=\Big[\frac{2(\pi\alpha(1-\omega))^2-\cos(2
\pi\alpha(1-\omega)+\theta)-\cos(\theta)}{2(\pi\alpha)^2-
\cos(2\pi\alpha+\theta)-\cos(\theta)}\Big]^\frac{1}{2}
\end{eqnarray}
Notice that $\theta$ is not negligible unless the limit $\alpha\to
\infty$ is taken. Similar results can be obtained for $\zeta_\pm$ 
using~\eref{gamma_app}~\eref{over_A}\eref{over_B} and (cf.~\eref{two_p_norm})  
\begin{equation}
{\mathcal N}=L\left[1-\frac{\cos(2\pi\alpha+\theta)-\cos(\theta)}
{(2\pi \alpha)^2}\right]^\frac{1}{2}.
\end{equation}
The entanglement entropy obtained in the low relative quasi-momentum 
limit is reported as dotted line in Figure~\ref{fig2:energy}. 

%%%%%%%%%%%%%%%%%%%%%%%%%%%%%%%%%%%%%%%%%%%%%%%%%%%%%%%%%%%%%%%%%%%%%%%%
\section{Bethe wavefunctions for many-particle bound states}
\label{bw_cluster}

In this section we focus on eigenstates of the $XXX$ chain in the sectors with 
$M=2,3$, focusing on $2$-strings and $3$-strings solutions of the Bethe-Takahashi
equations~\eref{bt_eq}, respectively. Physically, this corresponds to the situation 
in which all the particles form a bound state. Here we first discuss how to treat the 
singularities appearing in the wavefunction amplitudes~\eref{ba_eig}, due to 
the presence of strings. We then derive exact results for the entanglement entropy 
and the entanglement spectrum.

%%%%%%%%%%%%%%%%%%%%%%%%%%%%%%%%%%%%%%%%%%%%%%%%%%%%%%%%%%%%%%%%%%%%%%%%
\subsection{Two-particle bound states}
\label{two_p_cluster}

Here we restrict ourselves to two-particle bound states, i.e., $M=2$. Using the 
string hypothesis~\eref{str_hyp_dev}, the two rapidities $\lambda_1,\lambda_2$ 
identifying the eigenstates of the $XXX$ chain read 
\begin{eqnarray}
\lambda_1=\lambda-i-i\delta,\qquad \lambda_2=\lambda+i+i\delta.
\label{xxx}
\end{eqnarray}
Notice that in~\eref{xxx} we take into account the finite-size deviations 
$\delta={\mathcal O}(e^{-L})$ from the string hypothesis (see section~\ref{ba_approach}). 
In the limit $L\to\infty$ the string center $\lambda$ is determined 
by solving the Bethe-Takahashi equations~\eref{bt_eq}. 
The Bethe ansatz wavefunction for two particles reads (cf.~\eref{ba_eig}) 
\begin{equation}
\label{two_p_bw}
|\Psi_2\rangle=\frac{1}{\mathcal N}\sum\limits_{x_1<x_2}
\left[e^{ik_1 x_1+ik_2 x_2+i\theta_{1,2}}+e^{ik_1 x_2+ik_2 x_1+i
\theta_{2,1}}\right ]|x_1,x_2\rangle, 
\end{equation}
with $k_1,k_2$ the two quasi-momenta obtained from~\eref{two_p_bw} 
using~\eref{str_mom}. From~\eref{str_phases} the scattering phases 
$\theta_{1,2}$ and $\theta_{2,1}$ are given as 
\begin{eqnarray}
\label{xxx1}
\theta_{1,2}=\frac{1}{2i}\log\Big[1+\frac{4}{\hat
\delta}\Big]\to -\frac{1}{2i}\log(\hat\delta)\\
\nonumber\theta_{2,1}=\frac{1}{2i}\log\Big[1+\frac{4}{4-
\hat\delta}\Big]\to \frac{1}{2i}\log(\hat\delta),
\end{eqnarray}
with $\hat\delta=2\delta$. The limit $\delta\to 0$ (equivalently $L\to\infty$) 
was taken in the last step in~\eref{xxx1}. 
Using~\eref{xxx} and~\eref{xxx1} the wavefunction~\ref{two_p_bw} becomes   
\begin{equation}
\label{two_p_bound}
\fl\quad|\Psi_2\rangle=\frac{1}{\mathcal N}\sum\limits_{x_1<x_2}\left[e^{
iK_b(x_1+x_2)}\Big(\frac{1}{\sqrt{\delta}}e^{-(x_2-x_1)/\ell_b}+
\sqrt{\delta}e^{(x_2-x_1)/\ell_b}\Big)\right]|x_1,x_2\rangle,
\end{equation}
with $K_b\equiv k_1+k_2\in{\mathbb R}$ the total momentum of the 
bound state, and $\ell_b$ the bound-state length (~\eref{b_length_e} 
with $n=2$). Notice that, although the second term in~\eref{two_p_bound} 
vanishes exponentially for $L\to\infty$, it cannot be neglected when 
$x_2-x_1\to\infty$. Also, neglecting the second term would lead to the 
inconsistency $\Psi_2(x_1=1,x_2=2)\ne\Psi_2(x_1=1,x_2=L-1)$. However, 
the Bethe equations~\eref{ba_eq} imply that 
\begin{eqnarray}
\label{2p_bound_id}
\theta_{2,1}=\theta_{1,2}-(k_1-k_2)\frac{L}{2}\qquad\mbox{mod}\,
\pi.
\end{eqnarray}
Substituting~\eref{2p_bound_id} in~\eref{two_p_bound}, one obtains 
\begin{eqnarray}
\label{two_p_bw_f}
\fl\qquad|\Psi_2\rangle\propto e^{i\theta_{1,2}}\sum\limits_{x_1<x_2}
e^{iK_b(x_1+x_2)}\left[e^{-(x_2-x_1)/\ell_b}+e^{(x_2-x_1
-L)/\ell_b}\right]|x_1,x_2\rangle.
\end{eqnarray}
Notice that all the terms inside the brackets in~\eref{two_p_bw_f} are now  
regular. The overall divergent factor $e^{i\theta_{1,2}}$ cancels when 
normalizing the wavefunction.

%%%%%%%%%%%%%%%%%%%%%%%%%%%%%%%%%%%%%%%%%%%%%%%%%%%%%%%%%%%%%%%%%%%%%%%%%%%
\paragraph{Entanglement properties.}

In order to derive the entanglement entropy for the two-particle bound 
state, here we consider the simplified situation in which we neglect the 
second term in~\eref{two_p_bw_f}. The consequences of this approximation 
are discussed at the end of the section. The two-particle bound state 
wavefunction~\eref{two_p_bw_f} reads 
\begin{equation}
\label{two_p_b_app}
|\Phi_2\rangle=\frac{1}{\mathcal N}\sum
\limits_{x_1<x_2}e^{iK_b(x_1+x_2)}e^{-(x_2-x_1)/\ell_b}|x_1,x_2\rangle
\end{equation}
The normalization ${\mathcal N}$ does not depend on the total momentum 
$K_b$ and it is given as
\begin{equation}
{\mathcal N}\approx
\frac{(-L+e^{2/\ell_b}(L-1))^{\frac{1}{2}}}
{e^{2/\ell_b}-1},
\end{equation}
where we neglected ${\mathcal O}(e^{-L})$ terms. To construct the Schmidt 
matrix ${\mathbb M}(A|B)$ (cf.~\eref{schmidt_mat}) we choose the orthonormal 
basis for $A$ as 
\begin{eqnarray}
\label{sch_a}
|\Omega_A\rangle\equiv |\uparrow\uparrow\cdots\uparrow\rangle_A\\
\nonumber|\varphi_A\rangle\equiv 
\frac{1}{{\mathcal N}_{A}}\sum\limits_{x_1<x_2\in A}e^{
iK_b(x_1+x_2)}e^{-(x_2-x_1)/\ell_b}|x_1,x_2\rangle_A\\
\nonumber|\varphi'_A\rangle\equiv\frac{1}{{\mathcal 
N}'_{A}}\sum\limits_{x\in A}e^{iK_bx+ x/\ell_b}|x\rangle_A. 
\end{eqnarray}
The basis for the $B$ part can be chosen analogously. The Schmidt matrix then  
reads 
\begin{equation}
\label{sm_2_toy}
{\mathbb M}(A|B)=\frac{1}{\mathcal N}
\left[\begin{array}{ccc}
0 & {\mathcal N}_B & 0\\
{\mathcal N}_A & 0 & 0\\
0 & 0 & {\mathcal N}_A'{\mathcal N}_B'
\end{array}\right],
\end{equation}
where 
\begin{eqnarray}
\label{norms}
\fl\qquad{\mathcal N}_A\approx
\frac{(-L_A+e^{2/\ell_b}(L_A-1))^{\frac{1}{2}}}
{e^{2/\ell_b}-1}, &\qquad
{\mathcal N}_B\approx
\frac{(-L_B+e^{2/\ell_b}(L_B-1))^{\frac{1}{2}}}
{e^{2/\ell_b}-1},\\\nonumber
\fl\qquad{\mathcal N}'_A\approx\frac{e^{L_A/\ell_b}}
{(1-e^{-2/\ell_b})^\frac{1}{2}}, &\qquad
{\mathcal N}'_B\approx\frac{e^{-L_A/\ell_b}}
{(e^{2/\ell_b}-1)^\frac{1}{2}}. 
\end{eqnarray}
In~\eref{norms} ${\mathcal O}(e^{-L/\ell_b})$ terms have been neglected. 
${\mathbb M}(A|B)$ has three non-zero singular values $\zeta_0={\mathcal 
N}_A/{\mathcal N}$, $\zeta_0'={\mathcal N}_B/{\mathcal N}$, and $\zeta
\equiv{\mathcal N}'_A{\mathcal N}'_B/{\mathcal N}$, which read  
\begin{eqnarray}
\label{ES_2p_1b}
\fl\qquad\zeta_0\approx\Big[\frac{-L_A+e^{2/\ell_b}(L_A-1)}{-L+
e^{2/\ell_b}(L-1)}\Big]^\frac{1}{2}, & 
\quad\zeta_0'\approx\Big[\frac{-L_B+e^{2/\ell_b}(L_B-1)}{-L+
e^{2/\ell_b}(L-1)}\Big]^\frac{1}{2},\\\nonumber
\fl\qquad\zeta\approx\Big[\frac{e^{2/\ell_b}}{
-L+e^{2/\ell_b}(L-1)}\Big]^\frac{1}{2}.
\end{eqnarray}
Clearly, $\zeta_0\to \omega^\frac{1}{2}$ and $\zeta'\to (1-\omega)^{
\frac{1}{2}}$ for $L\to\infty$, whereas $\zeta$ is vanishing as 
$\propto 1/L^\frac{1}{2}$. Notice that $\zeta$ corresponds to a diverging 
entanglement spectrum level $\xi\equiv-2\log(\zeta)$. Interestingly, while 
$\zeta_0,\zeta_0'$ are usual ``bulk'' singular values, similar to the 
case with two unbound particles (see~\ref{two_p_ent}), $\zeta$ corresponds to 
an edge-related one. This is clear from the associated singular vector 
$|\varphi'_A\rangle$ (cf.~\eref{sch_a}), which is  exponentially 
localized at the boundary of part $A$. 

It is interesting to rewrite~\eref{ES_2p_1b} in terms of the energy $E$ 
(cf.~\eref{str_ener}). Here we restrict to the case with part $A$ being 
half of the chain. Using~\eref{b_length_e} one obtains  
\begin{eqnarray}
\label{ES_2p_1b_f}
\fl\quad\zeta_0=\zeta_0'=\Big[\frac{2+LE-2(1+E)^{L/2}}
{2+LE-2(1+E)^{L}}\Big]^\frac{1}{2}, & 
\quad\zeta=\frac{-1+(1+E)^{L/2}}{(-1-
LE+(1+E)^L)^\frac{1}{2}}.
\end{eqnarray}
From~\eref{ES_2p_1b} the half-chain entropy reads 
\begin{equation}
\label{2p_b_ent}
\fl\qquad S_A(L/2)=\log(2)-\frac{-1+\log(2)-\log(1-e^{-2/\ell_b})-
\log(L)}{1-e^{-2/\ell_b}}\frac{1}{L}+{\mathcal O}(1/L^2), 
\end{equation}
and, in terms of the energy $E$, one has   
\begin{equation}
\label{2p_b_ent_e}
\fl\qquad S_A(L/2)=\log(2)-\frac{1}{LE}\Big[1-\log(2)+
\log(L)+\log(-E)\Big]+{\mathcal O}(1/L^2).
\end{equation}
Clearly, in the limit $L\to\infty$, $S_A(L/2)\approx\log(2)$. For fixed $L$, 
high-energy bound states exhibit deviations from $S_A(L/2)=\log(2)$. 
This correspond to eigenstates containing two weakly-bound particles 
(cf.~\eref{b_length_e}). In the limit $E\to 0$ (equivalently $\ell_b
\to\infty$), the second term in~\eref{2p_b_ent_e} diverges. Notice that this 
is a spurious divergence  due to the fact that we neglected ${\mathcal O}
(e^{-L/\ell_b})$ terms in~\eref{ES_2p_1b}. 

We should stress that the results outlined above are not strongly affected 
by the second term in~\eref{two_p_bw_f}. The only significant change is that,  
while the Schmidt matrix obtained from~\eref{two_p_b_app} has only one 
edge-related singular value ($\zeta$ in~\eref{ES_2p_1b_f}), an extra one 
$\zeta'\approx\zeta$ appears if one considers~\eref{two_p_bw_f}. This is 
due to the fact that subsystem $A$ has two boundaries. Clearly, $\zeta$ 
and $\zeta'$ correspond to a particle localized at the right and left 
edge of $A$, respectively. Similarly, taking into account the second term 
in~\eref{two_p_bw_f} the expression for the half-chain entropy~\eref{2p_b_ent_e} 
is modified as
\begin{eqnarray}
\label{2p_b_ent_e1}
\fl\qquad S_A(L/2)=\log(2)-\frac{2}{LE}\Big[1-\log(2)+
\log(L)+\log(-E)\Big]+{\mathcal O}(1/L^2).
\end{eqnarray}
Notice the factor two (reflecting the number of boundaries of block $A$) in 
the ${\mathcal O}(1/L)$ term.

%%%%%%%%%%%%%%%%%%%%%%%%%%%%%%%%%%%%%%%%%%%%%%%%%%%%%%%%%%%%%%%%%%%%%%%%
\subsection{Three-particle bound states}
\label{three_p_cluster}

Here we discuss the structure of the Bethe wavefunction~\eref{ba_eig} for 
the $3$-particle bound states. These correspond to three solutions 
$\Lambda=\{\lambda_1,\lambda_2,\lambda_3\}$ of the Bethe equations~\eref{ba_eq} 
forming a (deviated) $3$-string in the complex plane as 
\begin{eqnarray}
\Lambda=\left\{\begin{array}{l}
\label{bs_3p}
\lambda_1=\lambda+\epsilon+2i+i\delta\\
\nonumber\lambda_2=\lambda+\epsilon-2i-i\delta\\
\nonumber\lambda_3=\lambda
\end{array}\right.
\end{eqnarray}
where $\delta,\epsilon$ (string deviations) are ${\mathcal O(e^{-L})}$. 
The string center $\lambda\in {\mathbb R}$ is obtained by solving the 
Bethe-Takahashi equations~\eref{bt_eq}. Using~\eref{str_phases} one 
obtains $\theta_{1,3}\to-i\infty$ and $\theta_{2,3}\to +i\infty$ in the 
limit $\delta\to 0$ (equivalently $L\to\infty$). 
The only permutations of $\{\lambda_1,\lambda_2,\lambda_3\}$ leading to 
singularities in the wavefunction amplitude~\eref{ba_amp} are $\{1,3,2\},
\{2,3,1\},\{2,1,3\}$, and $\{2,3,1\}$. Clearly, the ``most divergent'' 
amplitude in~\eref{ba_eig} is $\propto\exp(i\theta_{1,3}+i\theta_{3,2})$, 
which corresponds to $\{1,3,2\}$. The remaining two permutations 
$\{1,2,3\}$ and $\{3,2,1\}$ contribute with finite terms that can be neglected. 
Using the Bethe equations~\eref{ba_eq} one obtains a relation between  
$\theta_{1,3},\theta_{3,2}$ and $\theta_{3,1},\theta_{2,3}$ (similar 
to~\eref{2p_bound_id})  as 
\begin{equation}
\label{3b_id}
\fl\quad i(k_1-3k_2-k_3)\frac{L}{2}+i\theta_{3,1}+i\theta_{2,3}-4i
\theta_{1,2}-i\theta_{1,3}-i\theta_{3,2}=0\qquad\mbox{mod}\,\pi. 
\end{equation}
Using~\eref{3b_id} the Bethe wavefunction~\eref{ba_eig} reads 
\begin{eqnarray}
\label{3p_bound}
\fl|\Psi_3\rangle\propto \exp(2i\theta_{1,3})\sum
\limits_{x_1<x_2<x_3}\Big\{\\\nonumber\exp\big[iK_b(x_1+x_3)+
ik_3x_2-(x_3-x_1)/\ell_b\big]+\\\nonumber\exp\big[iK_b(x_1+x_2)+
ik_3x_3+(x_2-x_1-L)/\ell_b+iK_T L/4\big]+\\\nonumber\exp
\big[iK_b(x_2+x_3)+ik_3x_1+(x_3-x_2-L)/\ell_b+iK_T L/4\big]\Big\}
|x_1,x_2,x_3\rangle
\end{eqnarray}
with $\ell_b$ the bound-state length obtained 
from~\eref{b_length_e}, $K_b\equiv k_1+k_2\in{\mathbb R}$, and 
$K_T\equiv k_1+k_2+k_3\in{\mathbb R}$ the total momentum of the 
eigenstate. Since $x_1<x_2<x_3$, the first term in~\eref{3p_bound} 
corresponds to a three-particle bound state, whereas the other two 
are superpositions of a two-particle bound state and a free magnon. 
Notice that the overall divergent term $\exp(2i\theta_{1,3})$ 
in~\eref{3p_bound} cancel in the normalized wavefunction. 

%%%%%%%%%%%%%%%%%%%%%%%%%%%%%%%%%%%%%%%%%%%%%%%%%%%%%%%%%%%%%%%%%
\paragraph{Entanglement properties.} 

To discuss the entanglement spectrum and entanglement entropy of the 
three-particle bound state, we consider the simplified situation in 
which we keep only the first amplitude in~\eref{3p_bound}, i.e., 
\begin{eqnarray}
\label{3p_b_toy}
\fl\quad |\Psi_3\rangle=\frac{1}{\mathcal N}\sum
\limits_{x_1<x_2<x_3}e^{-(x_3-x_1)/\ell_b}e^{iK_{b}(x_1+x_3)+
ik_3x_2}|x_1,x_2,x_3\rangle.
\end{eqnarray}
The normalization of~\eref{3p_b_toy} in the large $L$ limit is 
\begin{equation}
{\mathcal N}\approx\frac{(e^{2/\ell_b}(L-2)-L)^\frac{1}{2}
}{(e^{2/\ell_b}-1)^\frac{3}{2}}.
\end{equation}
The Schmidt matrix ${\mathbb M}(A|B)$ from~\eref{3p_b_toy} can be  
given as 
\begin{equation}
\label{S_matr_3b}
{\mathbb M}(A|B)=\frac{1}{\mathcal N}\left[
\begin{array}{cccc}
0 & {\mathcal N}_B & 0 & 0\\
{\mathcal N}_A & 0 & 0 & 0\\
0 & 0 & 0 & {\mathcal N}''_A 
{\mathcal N}'_B\\
0 & 0 & {\mathcal N}'_A
{\mathcal N}''_B & 0
\end{array}\right].
\end{equation}
After neglecting exponentially suppressed ${\mathcal O}(e^{-L})$ terms, 
one obtains 
\begin{eqnarray}
\label{smat_el}
\fl\quad{\mathcal N}_A\approx
\frac{(-L_A+e^{2/\ell_b}(L_A-2))^\frac{1}{2}
}{(e^{2/\ell_b}-1)^\frac{3}{2}}, &\quad  
{\mathcal N}_B\approx
\frac{(-L_B+e^{2/\ell_b}(L_B-2))^\frac{1}{2}}
{(e^{2/\ell_b}-1)^\frac{3}{2}},\\\nonumber 
\fl\quad{\mathcal N}_A''\approx\frac{e^{1/\ell_b}}
{e^{2/\ell_b}-1}e^{L_A/\ell_b}, &\quad 
{\mathcal N}_B''\approx \frac{1}
{e^{2/\ell_b}-1}e^{-L_A/\ell_b},\\\nonumber 
\fl\quad{\mathcal N}_A'\approx\frac{e^{1/\ell_b}}{(
e^{2/\ell_b}-1)^\frac{1}{2}} e^{L_A/\ell_b}, & \quad
{\mathcal N}_B'\approx\frac{1}
{(e^{2/\ell_b}-1)^\frac{1}{2}}e^{-L_A/\ell_b}.
\end{eqnarray}
The singular values of~\eref{S_matr_3b}, using~\eref{smat_el}, 
read   
\begin{eqnarray}
\label{sing_val_3p}
\fl\quad\zeta_0=\frac{{\mathcal N}}{\mathcal N} 
\approx\frac{e^{2/\ell_b}(L_A-2)-L_A}{e^{2/\ell_b}(L-2)-L}, 
&\quad\zeta'=\frac{{\mathcal N}_A''N_B'}{N}\approx
\frac{e^{2/\ell_b}}{e^{2/\ell_b}(L-2)-L},\\
\fl\quad\zeta_0'=\frac{{\mathcal N}_B}{\mathcal N}
\approx\frac{e^{2/\ell_b}(L-L_A-2)-(L-L_A)}{e^{2/\ell_b}
(L-2)-L}, &  \quad\nonumber\zeta''=\frac{{\mathcal N}_A'
{\mathcal N}_B''}{\mathcal N}\approx\frac{e^{2/\ell_b}}
{e^{2/\ell_b}(L-2)-L}. 
\end{eqnarray}
As for two particles (see section~\ref{two_p_cluster}), $\zeta_0$ 
and $\zeta_0'$ are finite in the limit $L\to\infty$, and are bulk-related, 
whereas $\zeta'$ and $\zeta''$, which  are vanishing, have boundary origin. 
It can be checked that the corresponding singular vectors are exponentially 
localized at the boundary of block $A$. 

From~\eref{sing_val_3p} the half-chain entanglement entropy, neglecting 
terms ${\mathcal O}(1/L^2)$, reads 
\begin{equation}
\label{M3_bound}
\fl\quad S_{A}(L/2)=\log(2)+2\frac{-1+\log(2)-\log(1-
e^{-2/\ell_b})-\log(L)}{1-e^{-2/\ell_b}}\frac{1}{L}+
{\mathcal O}(1/L^2),
\end{equation}
and in terms of the bound-state energy $E$ 
(cf.~\eref{b_length_e})
\begin{eqnarray}
\label{M3_bound_e}
\fl\quad S_A(L/2)\approx\log(2)-\frac{3}{2LE}\left[1+2\log(2)+
\log(L)+\log(-E)-\log(6)\right]. 
\end{eqnarray}
As for the two-particle bound state, the correct expression, which takes 
into account all the three components in~\eref{3p_bound}, is obtained as 
\begin{eqnarray}
\fl\quad S_A(L/2)\approx\log(2)-\frac{6}{2LE}\left[1+2\log(2)+
\log(L)+\log(-E)-\log(6)\right]. 
\end{eqnarray}
This is checked numerically in Figure~\ref{fig6:ent_3p_ener} (c).

%%%%%%%%%%%%%%%%%%%%%%%%%%%%%%%%%%%%%%%%%%%%%%%%%%%%%%%%%%%%%%%%%%
\section{Eigenstates containing a single $2$-particle bound state}
\label{ba_eig_single_2string}

In this section we discuss the structure of the Bethe 
wavefunction~\eref{ba_eig} for $M$-particle eigenstates of the $XXX$ chain 
with string content $\{M-2,1\}$. These correspond to solutions of the 
Bethe-Takahashi equations~\eref{bt_eq} containing $M-2$ $1$-strings and 
a single $2$-string. Physically, the resulting wavefunctions contain 
magnons and a two-particle bound state. Due to the presence of the 
$2$-string, divergent terms in~\eref{ba_amp} appear. 

To illustrate the construction of the wavefunction we first focus on 
the elementary case with $M=3$ (i.e., string content $\{2,1\}$), 
considering $\Lambda=\{\lambda_1,\lambda_2,\lambda_3\}$ roots of 
the Bethe equations~\eref{ba_eq_log} of the form  
\begin{eqnarray}
\Lambda=\left\{\begin{array}{l}
\lambda_1=\lambda\\
\lambda_2=\lambda'+i+i\delta\\
\lambda_3=\lambda'-i-i\delta
\end{array}\right.
\end{eqnarray}
Notice that we assume $\textrm{Re}(\lambda_2)=\textrm{Re}(\lambda_3)\ne
\textrm{Re}(\lambda_1)$. Using~\eref{str_phases} one obtains that the only 
diverging scattering phases are $\theta_{3,2}=-\theta_{2,3}\to -i\log(
\delta)$. It is straightforward, using the Bethe equations~\eref{ba_eq}, 
to derive the relation 
\begin{equation}
\label{s_2str_id}
\theta_{3,2}=\theta_{2,3}+\theta_{1,3}-\theta_{1,2}-(k_2-k_3)
\frac{L}{2}\quad\mbox{mod}\,\pi.
\end{equation}
Using~\eref{s_2str_id} the Bethe wavefunction~\eref{ba_eig} reduces to  
\begin{eqnarray}
\label{bw_1p2}
\fl|\Psi_{\{2,1\}}
\rangle\propto\exp(i\theta_{2,3})\sum\limits_{x_1<x_2<x_3}
\Big\{\\\nonumber\exp\left[iK_b(x_2+x_3)+ik_1x_1+\theta_{1,2}+
\theta_{1,3}-(x_3-x_2)/\ell_b\right]+\\\nonumber
\exp\left[iK_b(x_1+x_3)+ik_1x_2+\theta_{2,1}+\theta_{1,3}-
(x_3-x_1)/\ell_b\right]+\\\nonumber
\exp\left[iK_b(x_2+x_3)+ik_1x_3+\theta_{2,1}+\theta_{3,1}-
(x_2-x_1)/\ell_b\right]+\\\nonumber
\exp\left[iK_b(x_2+x_3)+ik_1x_1+2\theta_{1,3}+\theta_{1,2}-
(x_2-x_3+L)/\ell_b\right]+\\\nonumber
\exp\left[iK_b(x_1+x_2)+k_1x_3+2\theta_{2,1}-
(x_1-x_2+L)/\ell_b\right]+\\\nonumber
\exp\left[iK_b(x_1+x_3)+ik_1x_2-(x_1-x_3+L)/\ell_b\right]\Big\}
|x_1,x_2,x_3\rangle, 
\end{eqnarray}
where $K_b\equiv k_2+k_3\in{\mathbb R}$, $\ell_b$ is the bound-state 
length (cf.~\eref{b_length_e}), and the subscript in $|\Psi_{\{2,1\}}
\rangle$ is to stress that we are considering string 
content $\{2,1\}$. Notice the divergent factor $\exp(i\theta_{2,3})$ 
in~\eref{bw_1p2}. 

A similar procedure can be used in the generic case with $M$ particles 
and string content $\{M-2,1\}$. First, the generalization of~\eref{s_2str_id}, 
assuming that $\lambda_{M-1},\lambda_{M}$ form a $2$-string, is obtained as  
\begin{equation}
\label{M_part_sing}
\fl\quad\theta_{M,M-1}=\theta_{M-1,M}-(k_{M-1}-k_M)\frac{L}{2}-\sum
\limits_{j<M-1}(\theta_{j,M-1}+\theta_{j,M})\quad\mbox{mod}\,\pi.
\end{equation}
The Bethe wavefunction is obtained from the general expression~\eref{ba_eig} 
substituting every occurrence of $\theta_{M-1,M}$ using~\eref{M_part_sing}.

%%%%%%%%%%%%%%%%%%%%%%%%%%%%%%%%%%%%%%%%%%%%%%%%%%%%%%%%%%%%%%%%%%%%%%
\section{A semiclassical  upper bound for the entanglement entropy}
\label{upper_bound}

In this section we derive the upper bound~\eref{ub_intro} for the entanglement 
entropy. 

We start considering a bipartition of a chain of length $L$ into two parts 
$A$ and $B$ of length $L_A$ and $L_B$, respectively. Here we define the  
ratio $\omega$ as $\omega\equiv L_A/L$. Given the generic 
eigenstate~\eref{ba_eig} of the $XXX$ chain in the sector with $M$ particles, 
the Schmidt matrix ${\mathbb M}(A|B)$ (cf.~\eref{schmidt_mat} for the definition) 
exhibits a block structure, each block ${\mathbb M}^{(k)}(A|B)$ corresponding  
to a different number of particles $k$ in subsystem $A$. 

The number of non-zero singular values in ${\mathbb M}^{(k)}(A|B)$ is given as 
the number of ways of assigning $k$ quasi-momenta from the set $\{k_1,k_2,\dots, k_M\}$ 
(cf.~\eref{ba_eig}) to the $k$ particles in $A$. This is given by the binomial 
coefficient $C(M,k)\equiv M!/(k!(M-k)!)$.  

Assuming that every particle with given momentum is fully delocalized in the 
chain (``semiclassical'' approximation), the singular values $\{\zeta_i\}$ 
with $i=1,2\dots,C(M,k)$ in ${\mathbb M}^{(k)}(A|B)$ are all degenerate,  
and are given as $\zeta_i=(\omega^{k}(1-\omega)^{M-k})^{1/2}\,\forall i$. 
Notice that $\{\zeta_i^2\}$, which would be the eigenvalues  of the reduced 
density matrix $\rho_A$ for block $A$, correspond to the semiclassical 
probabilities of finding $k$ particles in $A$ and $M-k$ in $B$.  

Finally, the upper bound $S_{MAX}(M,L_A)$ for the entanglement entropy is 
given as  
\begin{equation}
\label{upp_begin}
S_{MAX}(M,L_A)=-\sum\limits_{k=0}^{M}
\frac{M!}{k!(M-k)!}\omega^k(1-\omega)^{M-k}\log\left[
\omega^k(1-\omega)^{M-k}\right].
\end{equation}
It is convenient to consider the large $M$ limit. Thus, approximating 
the binomial coefficient with a gaussian, one obtains 
\begin{eqnarray}
\label{sc_ent}
\fl S_{MAX}(M,L_A)=-\frac{1}{\sqrt{2\pi M\omega(1-\omega)}}\sum
\limits_{k=0}^M e^{-\frac{(k-M\omega)^2}{2M\omega(1-\omega)}} (k
\log\omega+(M-k)\log(1-\omega)). 
\end{eqnarray}
Turning the sum over $k$ into an integral, and taking the integration 
in the full interval $k\in (-\infty,\infty)$ (instead of $[0,\infty
)$),~\eref{sc_ent} becomes 
\begin{equation}
\label{S_ub}
S_{MAX}(M,L_A)=-M[\omega\log\omega+(1-\omega)
\log(1-\omega)].
\end{equation}
Interestingly, $S_{MAX}(M,L_A)$ is proportional to the total number of 
particles in the system. Notice the symmetry under the exchange of the two 
subsystems $L_A\leftrightarrow L_B$, i.e.,  $\omega\leftrightarrow 1-
\omega$.

%%%%%%%%%%%%%%%%%%%%%% REFERENCES %%%%%%%%%%%%%%%%%%%%%%%%%%%%%%%%%%%%%%%%


\section*{References}


\begin{thebibliography}{99}
\bibitem{amico-2008}
L. Amico, R. Fazio, A. Osterloh, and V. Vedral, Entanglement in Many-Body 
Systems, Rev. Mod. Phys. {\bf 80}, 517 (2008).


\bibitem{eisert-2009}
J. Eisert, M. Cramer, and M. B. Plenio, Area laws for the entanglement 
entropy - a review, Rev. Mod. Phys. {\bf 82}, 277 (2009).


\bibitem{calabrese-2009}
P.~Calabrese, J.~Cardy, and B. Doyon Eds., Special issue: Entanglement 
entropy in extended systems, J.\ Phys.\ A {\bf 42}, 50 (2009).


\bibitem{cc-rev}
P.~Calabrese and J.~Cardy, Entanglement entropy and conformal field theory, 
J.\ Phys.\ A {\bf 42} 504005 (2009).


\bibitem{kauke-1999}
M.~Kaulke and I.~Peschel, A DMRG study of the q-symmetric Heisenberg chain, 
Eur.\ Phys.\ J.\ B {\bf 5}, 727 (1998). 


\bibitem{li-2008}
H.~Li, and F.~D.~M.~Haldane, Entanglement Spectrum 
as a Generalization of Entanglement Entropy: Identification of Topological 
Order in Non-Abelian Fractional Quantum Hall Effect States, 
Phys.\ Rev.\ Lett.\ {\bf 101}, 010504 (2008).


\bibitem{calabrese-2008}
P.~Calabrese, A.~Lefevre, Entanglement spectrum in one-dimensional systems, 
Phys.\ Rev.\ A {\bf 78}, 032329 (2008).


\bibitem{regnault-2009}
N.~Regnault, B.~A.~Bernevig, F.~D.~M.~ Haldane, Topological Entanglement and 
Clustering of Jain Hierarchy States, Phys.\ Rev.\ Lett. {\bf 103}, 016801 
(2009).


\bibitem{nienhuis-2009}
B.~Nienhuis, M.~Campostrini, and P.~Calabrese, Entanglement, combinatorics and 
finite-size effects in spin-chains, J.\ Stat.\ Mech.\ (2009) P02063. 


\bibitem{bray-ali-2010}
N.~Bray-Ali, L.~Ding, and S.~Haas, Topological order in paired states of fermions 
in two dimensions with breaking of parity and time-reversal symmetries,
Phys.\ Rev.\ B {\bf 80}, 180504(R) (2009). 


\bibitem{fidkowski-2010}
L.~Fidkowski, Entanglement Spectrum of Topological Insulators and Superconductors,
Phys.\ Rev.\ Lett.\ {\bf 104}, 130502 (2010).


\bibitem{lauchli-2010}
A.~M.~L\"auchli, E.~J.~Bergholtz, J.~Suorsa, and M.~Haque, Disentangling Entanglement 
Spectra of Fractional Quantum Hall States on Torus Geometries, Phys.\ Rev.\ Lett.\ 
{\bf 104}, 156404 (2010).


\bibitem{thomale-2010}
R.~Thomale, A.~Sterdyniak, N.~Regnault, and B.~A.~Bernevig, Entanglement Gap and 
a New Principle of Adiabatic Continuity, Phys.\ Rev.\ Lett.\ {\bf 104}, 180502 (2010).


\bibitem{yao-2010}
H.~Yao and X.~L.~Qi, Entanglement Entropy and Entanglement Spectrum of the Kitaev
Model, Phys.\ Rev.\ Lett.\ {\bf 105}, 080501 (2010).


\bibitem{prodan-2010}
E.~Prodan, T.~L.~Hughes, and B.~A.~Bernevig, Entanglement Spectrum of a Disordered 
Topological Chern Insulator, Phys.\ Rev.\ Lett.\ {\bf 105}, 115501 (2010).


\bibitem{pollmann-2010}
F.~Pollmann, A.~M.~Turner, E.~Berg, M.~Oshikawa, Entanglement spectrum of a 
topological phase in one dimension, Phys.\ Rev.\ B {\bf 81}, 064439 (2010).


\bibitem{kargarian-2010}
M.~Kargarian and G.~A.~Fiete, Topological phases and phase transitions on the 
square-octagon lattice, Phys.\ Rev.\ B {\bf 82}, 085106 (2010).


\bibitem{turner-2010}
A.~M.~Turner, Y.~Zhang, A.~Vishwanath, Entanglement and inversion symmetry in 
topological insulators, Phys.\ Rev.\ B {\bf 82}, 241102R (2010).


\bibitem{papic-2011}
Z.~Papic, B.~A.~Bernevig, and N.~Regnault, Topological Entanglement in Abelian 
and Non-Abelian Excitation Eigenstates, Phys.\ Rev.\ Lett.\ {\bf 106}, 056801 
(2011).


\bibitem{fidkowski-2011}
L.~Fidkowski, T.~S.~Jackson and I.~Klich, Model Characterization of Gapless Edge 
Modes of Topological Insulators Using Intermediate Brillouin-Zone Functions, 
Phys.\ Rev.\ Lett.\ {\bf 107}, 036601 (2011).


\bibitem{deng-2011}
X.~Deng and L.~Santos, Entanglement spectrum of one-dimensional extended 
Bose-Hubbard models, Phys.\ Rev.\ B {\bf 84}, 085138 (2011). 


\bibitem{dubail-2011}
J.~Dubail, and N.~Read, Entanglement Spectra of Complex Paired Superfluids, 
Phys.\ Rev.\ Lett.\ {\bf 107}, 157001 (2011). 


\bibitem{schliemann-2011}
J.~Schliemann, Entanglement spectrum and entanglement thermodynamics of 
quantum Hall bilayers at $\nu=1$, Phys.\ Rev.\ B {\bf 83}, 115322 (2011). 


\bibitem{hughes-2011}
T.~L.~Hughes, E.~Prodan, B.~A.~Bernevig, Inversion-symmetric topological 
insulators, Phys.\ Rev.\ B {\bf 83}, 245132 (2011).


\bibitem{regnault-2011}
N.~Regnault and B.~A.~Bernevig, Fractional Chern Insulator, 
Phys.\ Rev.\ X {\bf 1}, 021014 (2011).


\bibitem{alba-2011} V.~Alba, M.~Haque and A.~M.~L\"auchli, Boundary-Locality 
and Perturbative Structure of Entanglement Spectra in Gapped Systems, 
Phys.\ Rev.\ Lett.\ {\bf 108}, 227201 (2012).


\bibitem{qi-2012}
X.~L.~Qi, H.~Katsura, and A.~W.~W.~Ludwig, General Relationship between the 
Entanglement Spectrum and the Edge State Spectrum of Topological Quantum 
States, Phys.\ Rev.\ Lett.\ {\bf 108}, 196402 (2012).


\bibitem{lundgren-2012}
R.~Lundgren, V.~Chua, and G.~Fiete, Entanglement Entropy and Spectra of the 
One-dimensional Kugel-Khomskii Model, Phys.\ Rev.\ B {\bf 86}, 224422 (2012).


\bibitem{poilblanc-2012}
D. Poilblanc, N. Schuch, D. Perez-Garcia, and J.I. Cirac, 
Topological and entanglement properties of resonating valence bond wave functions, 
Phys. Rev. B {\bf 86}, 014404 (2012). 


\bibitem{dechiara-2012}
G. De Chiara, L. Lepori, M. Lewenstein, and A. Sanpera, Entanglement Spectrum, 
Critical Exponents, and Order Parameters in Quantum Spin Chains, Phys.\ Rev.\ Lett.\  
{\bf 109}, 237208 (2012).


\bibitem{alba-2012}
V.~Alba, M.~Haque, and A.~M.~L\"auchli, Entanglement spectrum of the Heisenberg 
XXZ chain near the ferromagnetic point, J.\ Stat.\ Mech.\ (2012) P08011. 



\bibitem{lepori-2013}
L.~Lepori, G.~De~Chiara, and A.~Sanpera, Entanglement Spectrum, Critical Exponents, 
and Order Parameters in Quantum Spin Chains, Phys.\ Rev.\ B {\bf 87} 235107 (2013).


\bibitem{laeuchli-2013}
A.~M.~L\"auchli, Operator content of real-space entanglement spectra at conformal 
critical points, arXiv:1303.0741 (2013) unpublished.


\bibitem{chung-2013}
C.-M.~Chung, L.~Bonnes, P.~Chen, and A.~L\"auchli, Entanglement spectroscopy using 
Quantum Monte Carlo, Phys.\ Rev.\ B {\bf 89}, 195147 (2014).  


\bibitem{lundgren-2013}
R.~Lundgren, Y.~Fuji, S.~Furukawa, and M.~Oshikawa, 
Entanglement spectra between coupled Tomonaga-Luttinger liquids: Applications 
to ladder systems and topological phases, Phys.\ Rev.\ B {\bf 88}, 245137 (2013).

\bibitem{kolley-2013}
F.~Kolley, S.~Depenbrock, I.~P.~McCulloch, U.~Schollw\"ock, and 
V.~Alba, Entanglement spectroscopy of SU(2)-broken phases in 
two dimensions, Phys.\ Rev.\ B {\bf 88}, 144426 (2013). 


\bibitem{chandran-2013}
A.~Chandran, V.~Khemani, and S.~L.~Sondhi, How universal is the entanglement 
spectrum?, arXiv:13112964. 


\bibitem{grover-2013}
T.~Grover, Entanglement of Interacting Fermions in Quantum Monte Carlo 
Calculations, Phys.\ Rev.\ Lett.\  {\bf 111}, 130402 (2013). 


\bibitem{assaad-2013}
F.~F.~Assaad, T.~C.~Lang, and F.~Parisen Toldin, Entanglement Spectra of Interacting 
Fermions in Quantum Monte Carlo simulations,  Phys.\ Rev.\ B {\bf 89}, 125121 (2013). 


\bibitem{petrescu-2014}
A.~Petrescu, H.~F.~Song, S.~Rachel, Z.~Ristivojevic, C.~Flindt, N.~Laflorencie, I.~Klich, 
N.~Regnault, and K.~Le~Hur, Fluctuations and Entanglement spectrum in quantum Hall states, 
arXiv:1405.7816 (2014). 


\bibitem{luitz-2014}
D.~J.~Luitz, X.~Plat, N.~Laflorencie, and F.~Alet, 
Improving entanglement and thermodynamic R\'enyi entropy measurements in Quantum Monte 
Carlo, arXiv:1405.7391. 


\bibitem{luitz-2014a}
D.~J.~Luitz, N.~Laflorencie, and F.~Alet, Participation spectroscopy and 
entanglement Hamiltonian of quantum spin models, arXiv:1404.3717. 


\bibitem{luitz-2014b} 
D.~J.~Luitz, F.~Alet, and N.~Laflorencie, Shannon entropy and participation 
spectra across the $3d$ $O(3)$ criticality, Phys.\ Rev.\ B {\bf 89}, 
165106 (2014). 


\bibitem{lundgren-2014} 
R.~Lundgren, J.~Blair, M.~Greiter, A.~L\"auchli, G.~Fiete, and R.~Thomale, 
Momentum Space Entanglement Spectrum of Bosons and Fermions with Interactions, 
arXiv:1404.7545. 


\bibitem{schliemann-2014} 
J.~Schliemann, Entanglement thermodynamics, arXiv:1405.2150. 


\bibitem{udagawa-2014}
M.~Udagawa, Y.~Motome, Entanglement Spectrum in Cluster Dynamical Mean-Field 
Theory, arXiv:14065960.


\bibitem{Holzhey} C. Holzhey, F. Larsen, and F. Wilczek,
Geometric and Renormalized Entropy in Conformal Field Theory,
Nucl. Phys. B {\bf 424}, 443 (1994).


\bibitem{Vidal}
G.~Vidal, J.~I.~Latorre, E.~Rico, and A.~Kitaev,
Entanglement in quantum critical phenomena, Phys.\ Rev.\ Lett.\ 
{\bf 90}, 227902 (2003). J.~I.~Latorre, E.~Rico, and G.~Vidal,
Ground state entanglement in quantum spin chains, Quant.\ Inf.\ 
and\ Comp.\ {\bf 4}, 048 (2004).


\bibitem{cc-04}
P.~Calabrese and J.~Cardy, Entanglement entropy and quantum field theory,
J.\ Stat.\ Mech.\ (2004) P06002. 
P.~Calabrese and J.~Cardy, Entanglement entropy and quantum field theory: 
a non-technical introduction, Int. J.\ Quant.\ Inf.\ {\bf 4}, 429 (2006).


\bibitem{fps-08}
S.~Furukawa, V.~Pasquier, and J.~Shiraishi,
Mutual information and compactification radius in a c=1 critical phase in
one dimension, Phys.\ Rev.\ Lett.\ {\bf 102}, 170602 (2009).



\bibitem{cg-08}
M.~Caraglio and F.~Gliozzi, Entanglement entropy and twist fields,
JHEP 0811: 076 (2008).



\bibitem{cct-09}
P.~Calabrese, J.~Cardy, and E.~Tonni,
Entanglement entropy of two disjoint intervals in conformal field theory,
J.\ Stat.\ Mech. P11001 (2009).



\bibitem{ip-09}
F.~Igloi and I.~Peschel, On reduced density matrices for disjoint subsystems, 
EPL {\bf 89} 40001 (2010).


\bibitem{fc-10}
M.~Fagotti and P.~Calabrese,
Entanglement entropy of two disjoint blocks in XY chains,
J.\ Stat.\ Mech.\ (2010) P04016.

\bibitem{cct-11}
P.~Calabrese, J.~Cardy, and E.~Tonni,
Entanglement entropy of two disjoint intervals in conformal field theory II,
J.\ Stat.\ Mech.\ (2011) P01021.


\bibitem{ch-04}
H.~Casini and M.~Huerta,
A finite entanglement entropy and the c-theorem,
Phys.\ Lett.\ B {\bf 600}  142 (2004).
H.~Casini, C.~D.~Fosco, and M.~Huerta,
Entanglement and alpha entropies for a massive Dirac field in two dimensions,
J.\ Stat.\ Mech.\ P05007 (2005).
H.~Casini and M.~Huerta,
Remarks on the entanglement entropy for disconnected regions,
JHEP 0903: 048 (2009).
H.~Casini and M.~Huerta,
Reduced density matrix and internal dynamics for multicomponent regions, 
Class.\ Quant.\ Grav.\ {\bf 26}, 185005 (2009).
H.~Casini, Entropy inequalities from reflection positivity,
J.\ Stat.\ Mech.\  P08019 (2010).
D.~D.~Blanco, H.~Casini, Entanglement entropy for non-coplanar regions 
in quantum field theory, Class.\ Quant.\ Grav.\ 28: 215015 (2011). 


\bibitem{ffip-08}
P.~Facchi, G.~Florio, C.~Invernizzi, and S.~Pascazio,
Entanglement of two blocks of spins in the critical Ising model,
Phys.\ Rev.\ A {\bf 78}, 052302 (2008).


\bibitem{rt-06}
S.~Ryu and T.~Takayanagi, 
Holographic derivation of entanglement entropy from AdS/CFT,
Phys.\ Rev.\ Lett.\ {\bf 96}, 181602 (2006). 
S.~Ryu and T.~Takayanagi, 
Aspects of holographic entanglement entropy,
JHEP 0608: 045 (2006). 
V.~E.~Hubeny and M.~Rangamani,
Holographic entanglement entropy for disconnected regions,
JHEP 0803: 006 (2008). 
M.~Headrick and T.~Takayanagi,
A holographic proof of the strong subadditivity of entanglement entropy,
Phys.\ Rev.\ D {\bf 76}, 106013 (2007). 
T.~Nishioka, S.~Ryu, and T.~Takayanagi, Holographic entanglement entropy: 
an overview, J.\ Phys.\ A {\bf 42} 504008 (2009). 
E.~ Tonni, Holographic entanglement entropy: near horizon geometry and 
disconnected regions, JHEP 1105: 004 (2011). 
A.~Allais, E.~Tonni, Holographic evolution of the mutual information,
JHEP 1201: 102 (2012). 



\bibitem{atc-09}
V.~Alba, L.~Tagliacozzo, and P.~Calabrese,
Entanglement entropy of two disjoint blocks in critical Ising models, 
Phys.\ Rev.\ B {\bf 81}, 060411 (2010).



\bibitem{atc-11}
V.~Alba, L.~Tagliacozzo, and P.~Calabrese,
Entanglement entropy of two disjoint intervals in 
critical c=1 theories, J.\ Stat.\ Mech.\ (2011) P06012.


\bibitem{f-12}
M.~Fagotti, New insights in the entanglement of two
disjoint blocks, EPL {\bf 97} 17007 (2012).

\bibitem{coser-2013} 
A.~Coser, L.~Tagliacozzo, and E.~Tonni, On R\'enyi entropies of disjoint 
intervals in conformal field theory, J.\ Stat.\ Mech.\ (2014) P01008. 


\bibitem{ccen-10} P.~Calabrese, M.~Campostrini, F.~H.~L.~Essler, 
and B.~Nienhuis, Parity effects in the scaling of block entanglement in 
gapless spin chains, Phys.\ Rev.\ Lett.\ {\bf 104}, 095701 (2010).



\bibitem{ce-10}
P.~Calabrese and F.~H.~L.~Essler,
Universal corrections to scaling for block entanglement in spin-1/2 XX chains,
J.\ Stat.\ Mech.\ (2010) P08029.



\bibitem{cc-10}
J.~Cardy and P.~Calabrese, Unusual corrections to scaling in 
entanglement entropy, J.\ Stat.\ Mech.\ (2010) P04023.



\bibitem{ccp-10}
P.~Calabrese, J.~Cardy, and I.~Peschel, Corrections to scaling for block 
entanglement in massive spin-chains, J.\ Stat.\ Mech.\ (2010) P09003.



\bibitem{fc-10b}
M.~Fagotti and P.~Calabrese,
{Universal parity effects in the entanglement entropy of XX chains with open 
boundary conditions}, J.\ Stat.\ Mech.\ (2011) P01017.

\bibitem{xa-11}
J.~C.~Xavier and F.~C.~Alcaraz,
R\'enyi Entropy and Parity Effect of the Anisotropic Spin-s Heisenberg 
Chains with a Magnetic Field, Phys.\ Rev.\ B {\bf 83}, 214425 (2011).


\bibitem{cmv-11}
P.~Calabrese, M.~Mintchev, and E.~Vicari, The entanglement entropy of 1D 
systems in continuous and homogenous space, J.\ Stat.\ Mech.\ (2011) P09028.


\bibitem{alba-2009}
V.~Alba, M.~Fagotti, and P.~Calabrese, Entanglement entropy of excited states, 
J.\ Stat.\ Mech.\ (2009) P10020. 



\bibitem{alcaraz-2011}
F.~C.~Alcaraz, M.~I.~Berganza, and G.~Sierra, Entanglement of low-energy 
excitations in conformal field theory, Phys.\ Rev.\ Lett.\ {\bf 106}, 201601 
(2011).


\bibitem{pizorn-2012}
I.~Pizorn, Universality in entanglement of quasiparticle excitations, 
arXiv:1202.3336. 


\bibitem{berganza-2012}
M.~I.~Berganza, F.~C.~Alcaraz, and G.~Sierra, Entanglement of excited states 
in critical spin chains, J.\ Stat.\ Mech.\ (2012) P01016. 


\bibitem{wong-2013}
G.~Wong, I.~Klich, L.~A.~P.~Zayas, and D.~Vaman, Entanglement Temperature and 
Entanglement Entropy of Excited States, JHEP {\bf12} (2013) 020. 


\bibitem{storms-2013}
M.~Storms, and R.~R.~P.~Singh, Entanglement in ground and excited states 
of gapped fermion systems and their relationship with fermi surface and 
thermodynamic equilibrium properties, Phys.\ Rev.\ E {\bf 89}, 012125 
(2014). 


\bibitem{berkovits-2013}
R.~Berkovits, Two-particle excited states entanglement entropy in a 
one-dimensional ring, Phys.\ Rev.\ B {\bf 87}, 075141 (2013). 


\bibitem{essler-2013}
F.~H.~L.~Essler, A.~M.~L\"auchli, and P.~Calabrese, Shell-Filling Effect in the 
Entanglement Entropies of Spinful Fermions, Phys.\ Rev.\ Lett.\ {\bf 110}, 115701 
(2013).  


\bibitem{nozaki-2014}
M.~Nozaki, T.~Numasawa, T.~Takayanagi, Quantum Entanglement of Local Operators 
in Conformal Field Theories,  Phys.\ Rev.\ Lett.\ {\bf 112}, 111602 (2014).


\bibitem{ramirez-2014}
G.~Ramirez, J.~Rodriguez-Laguna, and G.~Sierra, Entanglement in low-energy 
states of the random-hopping model,  arXiv:1402.5015.


\bibitem{ares-2014}
F.~Ares, J.~G.~Esteve, F.~Falceto, and E.~S\'anchez-Burillo, 
Excited states entanglement in homogeneous fermionic chains, 
arXiv:1401.5922.


\bibitem{huang-2014}
Y.~ Huang, and J.~Moore,  Excited-state entanglement and thermal mutual 
information in random spin chains, arXiv:1405.1817.


\bibitem{palmai-2014}
T.~P\'almai, Excited state entanglement in conformal field theory: extensivity 
and the role of microscopic details, arXiv:1406.3182.


\bibitem{lieb-1961} 
E.~Lieb, T.~Schultz, and D.~Mattis,  Ann.\ Phys.\ {\bf 16}, 407 (1961). 

\bibitem{barouch-1970}
E.~Barouch, B.~M.~McCoy, and M.~Dresden, Phys.\ Rev.\ A {\bf 2}, 
1075 (1970).

\bibitem{barouch-1971}
E.~Barouch and B.~McCoy, Phys.\ Rev.\ A {\bf 3}, 786 (1971).

\bibitem{barouch-1971a}
E.~Barouch and B.~M.~McCoy, Phys.\ Rev.\ A {\bf 3}, 2137 (1971).

\bibitem{mccoy-1971}
B.~M.~McCoy, E.~Barouch, and D.~B.~Abraham, Phys.\ Rev.\ A {\bf 4}, 2331 
(1971).


\bibitem{kor-book}
V.~E.~Korepin, N.~M.~Bogoliubov, A.~G.~Izergin, {\it Quantum Inverse 
Scattering Method and Correlation Functions}, Cambridge University Press, 
Cambridge, 1997.


\bibitem{barmettler-2010}
P.~Barmettler, M.~Punk, V.~Gritsev, E.~Demler, and E.~Altman, 
Quantum quenches in the anisotropic spin-$1/2$ Heisenberg chain: 
different approaches to many-body dynamics far from equilibrium, 
New\ J.\ Phys.\ {\bf 12}, 055017 (2010).


\bibitem{gritsev-2010}
V.~Gritsev, T.~Rostunov, and E.~Demler, Exact methods in analysis of 
nonequilibrium dynamics of integrable models: application to the study 
of correlation function in nonequilibrium $1D$ Bose gas, 
J.\ Stat.\ Mech.\ (2010) P05012. 


\bibitem{vidal-2004}
G.~Vidal, Efficient classical simulation of one-dimensional quantum many-body 
systems, Phys.\ Rev.\ Lett.\ {\bf 93}, 040502 (2004).


\bibitem{daley-2004}
A.~J.~Daley, C.~Kollath, U.~Schollw\"ock, and G.~Vidal, Time-dependent 
density-matrix renormalization-group using adaptive effective Hilbert spaces, 
J.\ Stat.\ Mech.\ (2004) P04005. 


\bibitem{feiguin-2004} 
S.~White and A.~Feiguin, Real-Time Evolution Using the Density Matrix 
Renormalization Group, Phys.\ Rev.\ Lett.\ {\bf 93}, 076401 (2004).


\bibitem{schollwoeck-2005}
U.~Schollw\"ock, The density-matrix renormalization group, 
Rev.\ Mod.\ Phys.\ {\bf 77}, 259 (2005). 


\bibitem{schollwoeck-2011}
U.~Schollw\"ock, The density-matrix renormalization group in the age 
of matrix product states, Annals of Physics {\bf 326}, 96 (2011).


\bibitem{bethe-1931}
H.~Bethe, Zur Theorie der Metalle. I. Eigenwerte und Eigenfunktionen 
der linearen Atomkette, Z.\ Phys.\ {\bf 71}, 205 (1931).


\bibitem{taka-book}
M.~Takahashi, {\it Thermodynamics of one-dimensional solvable models}, 
Cambridge University Press, Cambridge, 1999.



\bibitem{hanus-1963} 
J.~Hanus, Bound states in the Heisenberg ferromagnet, Phys.\ Rev.\ Lett.\ 
{\bf 11}, 336 (1963). 


\bibitem{wortis-1963} 
M.~Wortis, Bound states of two spin waves in the Heisenberg ferromagnet, 
Phys.\ Rev.\ {\bf 132}, 85 (1963). 


\bibitem{fogedby-1980} 
H.~C.~Fogedby, The spectrum of continuous isotropic quantum Heisenberg chain: 
quantum solitons as magnon bound states, J.\ Phys.\ C {\bf 13}, L195 (1980). 


\bibitem{schneider-1981} 
T.~Schneider, Solitons and magnon bound states in ferromagnetic 
Heisenberg chains, Phys.\ Rev.\ B {\bf 24}, 5327 (1981). 


\bibitem{kohno-2009}
M.~Kohno, Dynamically Dominant Excitations of String Solutions in the 
Spin-$1/2$ Antiferromagnetic Heisenberg Chain in a Magnetic Field, 
Phys.\ Rev.\ Lett.\ {\bf 102}, 037203 (2009).


\bibitem{ganahl-2012} M.~Ganahl, E.~Rabel, F.~H.~L.~Essler, H.~G.~Evertz, 
Observing complex bound states in the spin-$1/2$ Heisenberg XXZ chain, 
Phys.\ Rev.\ Lett.\ {\bf 108} 077206 (2012).


\bibitem{haller-2009} 
E.~Haller, M.~Gustavsson, M.~J.~Mark, J.~G.~Danzl, R.~Hart, 
G.~Pupillo, and H-C.~N\"agerl, Relaization of an Excited, Strongly 
Correlated Quantum Gas Phase, Science {\bf 325}, 1224 (2009). 


\bibitem{fukuhara-2013}
T.~Fukuhara, P.~Schauss, M.~Endres, S.~Hild, M.~Cheneau, I.~Bloch, and C.~Gross, 
Microscopic observation of magnon bound states and their dynamics, 
Nature {\bf 502}, 76 (2013). 


\bibitem{popkov-2005}
V.~Popkov and M.~Salerno, Logarithmic divergence of the block entanglement 
entropy for the ferromagnetic Heisenberg model, Phys.\ Rev.\ A {\bf 71}, 012301 
(2005).

\bibitem{popkov-2010}
V.~Popkov and M.~Salerno, Reduced-density-matrix spectrum and block entropy of 
permutationally invariant many-body systems, Phys.\ Rev.\ E {\bf 82}, 011142 (2010).

\bibitem{castro-alvaredo-2011} O.~A.~Castro-Alvaredo and B.~Doyon, 
Permutation operators, entanglement entropy, and the XXZ spin chain in the limit 
$\Delta\to -1$, J.\ Stat.\ Mech. (2011) P02001.

\bibitem{castro-alvaredo-2012}
O.~A.~Castro-Alvaredo and B.~Doyon, Entanglement entropy of highly degenerate 
states and fractal dimensions, Phys.\ Rev.\ Lett.\ {\bf 108}, 10401 (2012). 

\bibitem{karbach-1997}
M.~Karbach, G.~Muller, Introduction to the Bethe ansatz I, Computers\ 
in\ Physics {\bf 11}, 36 (1997). M.~Karbach, K.~Hu, and G.~Muller, 
Introduction to the Bethe ansatz II, Computers\ in\ 
Physics {\bf 12}, 565 (1998). M.~Karbach, K.~Hu, and G.~Muller, 
Introduction to the Bethe ansatz III, arxiv:0008018.


\bibitem{yang-1966}
C.~N.~Yang, and C.~P.~Yang, One-Dimensional Chain of Anisotropic Spin-Spin 
Interactions. I. Proof of Bethe's Hypothesis for Ground State in a Finite System, 
Phys.\ Rev.\ {\bf 150}, 321 (1966). 


\bibitem{essler-1992}
F.~H.~L.~Essler, V.~E.~Korepin, and K.~Schoutens, Fine structure of the Bethe 
ansatz for the spin-$\frac{1}{2}$ Heisenberg $XXX$ model, J.\ Phys.\ A:\ Math.\ 
Theor.\ {\bf 25} 4115 (1992).

\bibitem{hagemans-2007}
R.~Hagemans and J.-S.~Caux, Deformed strings in the Heisenberg model, 
J.\ Phys.\ A:\ Math.\ Theor.\ {\bf 40} 14605 (2007).

\bibitem{alba-2013}
V.~Alba, K.~Saha, and M.~Haque, Bethe ansatz description of edge-localization 
in the open-boundary $XXZ$ spin chain,  J. Stat. Mech. (2013) P10018.  

\bibitem{devega-1985}
H.~J.~de~Vega and F.~Woynarovich, Method for calculating finite size corrections 
in Bethe ansatz systems: Heisenberg chain and six-vertex model, 
Nucl.\ Phys.\ B {\bf 251}, 439 (1985). 

\bibitem{alcaraz-1987}
F.~C.~Alcaraz, M.~N.~Barber, and M.~T.~Batchelor, Conformal invariance and the 
spectrum of the $XXZ$ chain, Phys.\ Rev.\ Lett.\ {\bf 58}, 771 (1987). 

\bibitem{devega-1987}
H.~J.~de~Vega, Finite-size corrections for nested Bethe ansatz models and 
conformal invariance  J.\ Phys.\ A:\ Math.\ Gen.\ {\bf 20}, 6023 (1987). 

\bibitem{alcaraz-1988}
F.~C.~Alcaraz and M.~J.~Martins, Conformal invariance and critical exponents 
of the Takhtajan-Babujian modelsJ.\ Phys.\ A:\ Math.\ Gen.\ {\bf 21}, 4397 
(1988).


\bibitem{destri-1992}
C.~Destri and H.~J.~de~Vega, New thermodynamic Bethe ansatz equations without 
strings, Phys.\ Rev.\ Lett.\ {\bf 69}, 2313 (1992). 

\bibitem{vladimirov-1986} A.~A.~Vladimirov, Proof of the invariance of the 
Bethe-ansatz solutions under complex conjugation, Teor.\ Mat.\ Fiz.\ {\bf 66}, 
154 (1986).


\bibitem{biegel-2004}
D.~Biegel, M.~Karbach, G.~M\"uller, K.~Wiele, Spectrum and transition rates of 
the $XX$ chain analyzed via Bethe ansatz, Phys.\ Rev.\ B {\bf 69}, 174404 (2004). 


\bibitem{gu-2005}
S.-J.~Gu, N.~M.~R.~Peres, and Y.-Q.~Li, Numerical and Monte Carlo Bethe 
ansatz method: 1D Heisenberg model, Eur.\ Phys.\ J.\ B {\bf 48}, 157 
(2005). 



\end{thebibliography}
\end{document}